# Toward Optimal Approximations for Resource-Minimization for Fire Containment on Trees and Non-Uniform $k$-Center

Jannis Blauth[1]   Christian Nöbel[1]   Rico Zenklusen[1]

**Abstract**

One of the most elementary spreading models on graphs can be described by a fire spreading from a burning vertex in discrete time steps. At each step, all neighbors of burning vertices catch fire. A well-studied extension to model fire containment is to allow for fireproofing a number $B$ of non-burning vertices at each step. Interestingly, basic computational questions about this model are computationally hard even on trees. One of the most prominent such examples is Resource Minimization for Fire Containment (RMFC), which asks how small $B$ can be chosen so that a given subset of vertices will never catch fire. Despite recent progress on RMFC on trees, prior work left a significant gap in terms of its approximability. We close this gap by providing an optimal 2-approximation and an asymptotic PTAS, resolving two open questions in the literature. Both results are obtained in a unified way, by first designing a PTAS for a smooth variant of RMFC, which is obtained through a careful LP-guided enumeration procedure.

Moreover, we show that our new techniques, with several additional ingredients, carry over to the non-uniform $k$-center problem (NUkC), by exploiting a link between RMFC on trees and NUkC established by Chakrabarty, Goyal, and Krishnaswamy. This leads to the first approximation algorithm for NUkC that is optimal in terms of the number of additional centers that have to be opened.

[1]Department of Mathematics, ETH Zurich, Zurich, Switzerland. Email: jblauth@ethz.ch, cnoebel@amazon.lu, ricoz@ethz.ch

# 1 Introduction

To find the most efficient way to inhibit harmful spreading processes is a central question reaching into various fields. One of the most basic and most extensively studied spreading models on graphs to capture such problems, including the spreading of viruses (see, e.g., [14, 26] and references therein), is the firefighter model, introduced by Hartnell [15]. It is typically explained using the example of a fire that spreads, and we follow this tradition.

In this model, a fire starts at one vertex $r$, also called the *root*, of a graph $G = (V, E)$, and propagates in discrete time steps. At every time step, all vertices adjacent to at least one burning vertex catch fire. The way this process can be inhibited is by fireproofing a number $B$ of non-burning vertices at each step before the fire spreads, which prevents them from ever catching (and therefore also propagating) fire. $B$ is called the *budget* or the number of *firefighters*. One of the most basic objectives is to decide, for a given set $S \subseteq V$ of vertices, how large $B$ has to be to prevent any vertex in $S$ from catching fire. This is known as the *Resource Minimization for Fire Containment* (RMFC) problem, and is the focus of this paper. Another objective that has been studied is to maximize the number of vertices that can be saved from burning, which is known as the *Firefighter problem*.

Since its introduction, there has been significant work on the firefighter model. We refer the interested reader to the classical survey of Finbow, King, MacGillivray, and Rizzi [12], a more recent one by Wagner [25], as well as recent work on the subject [1, 2, 8] and references therein. In particular, it has been studied in various graph classes (see, e.g., [1, 5, 8, 12, 13, 21, 22]), and in other computational models, like parameterized complexity [3, 4, 10] or online optimization [9]. Despite the considerable attention it received, basic foundational computational questions about the firefighter model remain open.

This is the case even for one of the most studied and most basic graph classes, namely when $G$ is a spanning tree. Interestingly, even for trees of maximum degree three, it is NP-hard to decide whether a budget of $B = 1$ is sufficient for RMFC, or whether one needs at least $B = 2$ [21].[2] Consequently, unless P = NP, the best approximation guarantee one can hope for is 2. The first non-trivial approximation for RMFC on trees (RMFC-T) was by Chalermsook and Chuzhoy [8], who presented a log* $n$-approximation algorithm, where $n = |V|$ is the number of vertices of the tree $G = (V, E)$.[3] Their algorithm is based on the natural linear program of RMFC-T, and they show that $\Theta(\log^* n)$ is indeed the integrality gap of this program. The lack of a strong natural LP relaxation for RMFC-T further underlines the difficulty of the problem. Only quite recently, the first constant-factor approximation algorithm for RMFC-T was obtained by Adjiashvili, Baggio, and Zenklusen [1], who presented a 12-approximation. This was later improved by Rahgoshay and Salavatipour [23], who showed how to compute a solution of budget $\lceil (5 + \varepsilon) B^* \rceil$ for any $\varepsilon > 0$, where $B^*$ is the optimal solution value. Rahgoshay and Salavatipour [23] also presented an asymptotic quasi-polynomial time approximation scheme (a QPTAS), which shows that, for any fixed $\varepsilon > 0$, and any instance with an optimal budget of $B^*$ that is sufficiently big, one can find a $(1 + \varepsilon)$-approximation in time $n^{O(\frac{\log \log n}{\varepsilon})}$.

Despite this recent progress, the following central questions remained open:

- Is it possible to achieve an approximation factor of 2 that matches the hardness result?
- Is there an asymptotic PTAS for RMFC-T, and if so, how to get around the current barriers that only lead to quasi-polynomial running time?

One main goal of this paper is to settle these questions.

---

[2] See also [2] for further approximation hardness results for the Firefighter problem on trees.
[3] The term log* $n$ denotes the number of times we need to apply the (natural) logarithm function until the result is at most 1. Formally, for $k \in \mathbb{Z}_{\geq 0}$, let $\log^{(k)} n$ be the $k$-th iterated logarithm of $n$, i.e., $\log^{(0)} n = n$ and $\log^{(k)} n = \log \log^{(k-1)} n$ for $k \in \mathbb{Z}_{\geq 1}$. Then, log* $n$ denotes the minimum number $k \in \mathbb{Z}_{\geq 0}$ such that $\log^{(k)} n \leq 1$.

An additional motivation to study the approximability of RMFC-T and to answer the above questions comes from a connection between RMFC-T and the *Non-Uniform k-Center* (NUkC) problem, as revealed by Chakrabarty, Goyal, and Krishnaswamy [7]. NUkC is a generalization of $k$-center where one is given a collection of balls with unrelated radii. The goal is to place these balls in a metric space and increase their radii by a factor $\beta$ chosen as small as possible, so that all points in the metric space are covered by the $\beta$-dilated balls. Hence, if all radii are the same, we obtain the classical $k$-center problem. Another well-known special case of it is $k$-center with outliers, which can be modeled as NUkC with two different radii. Contrary to these special cases, NUkC does not allow a constant-factor approximation unless $\mathrm{P} = \mathrm{NP}$ [7]. Interestingly, Chakrabarty, Goyal, and Krishnaswamy [7] show that $(O(1), O(1))$-bicriteria approximations are possible for NUkC, where an $(\alpha, \beta)$-approximation opens a factor of $\alpha$ more balls of each radius, and $\beta$ is the dilation of the radii. They identify and crucially exploit a strong connection between NUkC and RMFC-T. This is not a black-box reduction, but they showed an intimate relation between the corresponding linear programs that allowed for adapting the RMFC-T algorithms of [1] to NUkC. This way, they obtain a $(24, 10)$-bicriteria approximation.

Nevertheless, it remains badly understood what $(\alpha, \beta)$-approximations for NUkC are possible, especially in terms of the parameter $\alpha$. The hardness result of [7] implies that no $(\alpha, \beta)$-approximation is possible for any $\alpha < 2$ and $\beta \leq 2^{\mathrm{poly}(n)}$, unless $\mathrm{P} = \mathrm{NP}$.[4] This leads to the following natural question:

- Is there a $(2, O(1))$-approximation for NUkC?

Improvements on $\alpha$ have only been obtained for the special case of NUkC with up to four different radii. In these cases, the hardness result does not apply, and $(1, O(1))$-approximations have been found. First Chakrabarty, Goyal, and Krishnaswamy [7] presented a $\left(1, 1+\sqrt{5}\right)$-approximation for two different radii, which was later extended to a $(1, 22)$-approximation for three different radii by Jia, Rohwedder, Sheth, and Svensson [20], and then to a $(1, O(1))$-approximation for four different radii by Inamdar and Varadarajan [18].

## 1.1 Our results

We answer all three questions raised in the introduction in the affirmative. In particular, we obtain the following results for RMFC-T, which essentially settles its approximability.

**Theorem 1.1.** *There is a $2$-approximation algorithm for RMFC-T.*

**Theorem 1.2.** *There is an asymptotic PTAS for RMFC-T.*

We prove the above theorems through a single unified approach. More precisely, we will introduce a *smooth* RMFC problem on trees, and show how to obtain a PTAS for it. This smooth version eliminates the artifact that small budgets are harder to approximate, and we believe that it is a natural problem in its own right. As we will see, a PTAS for the smooth version will easily imply both Theorem 1.1 and Theorem 1.2 for classical (non-smooth) RMFC-T.

We introduce several novel techniques to obtain our results. One key technique to obtain a PTAS for the smooth version is a novel LP-guided guessing approach, which substantially improves upon a previous LP-guided approach of Adjiashvili, Baggio, and Zenklusen [1]. We will provide a more thorough overview of our techniques in Section 2 and Section 3.

Moreover, we show that our new approach, with several novel technical ingredients, can carry over to NUkC. This leads to an approximation algorithm that, by the hardness result of [7], settles the approximability of the problem with respect to $\alpha$, i.e., the increase in the number of balls (under the very mild assumption that the second criterion $\beta$ is bounded by $2^{\text{poly}(n)}$).

---

[4]Indeed, their reduction is from RMFC-T, where distinguishing between a budget of $B = 1$ and $B \geq 2$ is NP-hard, and the budget corresponds to the number of centers in NUkC. Moreover, $\beta \leq 2^{\mathrm{poly}(n)}$ follows because the reduction from RMFC-T to NUkC leads to NUkC instances that are invariant under scaling the radii by a factor of $2^{\mathrm{poly}(n)}$.

**Theorem 1.3.** *For any $\varepsilon > 0$, there is a $(2, 15 + \varepsilon)$-approximation algorithm for NUkC.*

We remark that the approximation guarantee we get is even stronger in the following sense. For any $\varepsilon > 0$, we are able to efficiently compute a solution with factor $15 + \varepsilon$ within the optimum dilation, so that for any $\ell \in \mathbb{Z}_{>0}$, we use the largest $\ell$ radii (when counting multiplicities) at most $\lfloor (1 + \varepsilon)\ell \rfloor$ times. So, we only lose a constant factor on the optimum dilation while increasing the number of centers by an arbitrarily small factor, even for any prefix of the largest $\ell$ radii. (See Section 6 for the formal statement.)

## 1.2 Further related work

Another way to interpret RMFC, is that it seeks to find a "temporal vertex cut" that cuts off the root $r$ from all the leaves. We call the vertex cut *temporal* because of the temporal aspect of when vertices can be fireproofed. Also non-temporal cut problems in the context of spreading/vaccination settings have been explored, where one can think of vaccinating vertices before the spreading starts. For further discussion on this topic, we refer the interested reader to the work of [17] and references therein.

Also the Firefighter problem is NP-hard even on trees of maximum degree three [12]. Contrary to RMFC, it is easy to obtain $O(1)$-approximations for it, even through a simple greedy algorithm. A line of research on the Firefighter problem has essentially settled its approximability on trees. Namely, after the development of several constant-factor approximations [6, 16, 19],[5] a PTAS was presented in [1].

## 1.3 Organization of the paper

We start by presenting our ideas in the context of RMFC-T, and later explain how they can be extended to NUkC. More precisely, in Section 2, we outline the high-level structure of our approach. In particular, we introduce a smooth version of RMFC-T, for which we later present a PTAS. This will imply both Theorem 1.1 and Theorem 1.2. Moreover, we expand on how we compress RMFC instances, and on crucial aspects of the natural LP, which has a high integrality gap, and how we want to get around it through an LP-guided guessing approach. Our LP-guided guessing procedure is then presented in Section 3, where we introduce it step by step to simplify the presentation. Section 4 and Section 5 contain the proof of correctness and the proof of the running time of our PTAS for the smooth version of RMFC-T, respectively. Section 6 explains how to adapt and extend our approach to obtain Theorem 1.3. Finally, Section 7, Section 8, and Section 9 contain missing proofs of some technical aspects of our approach.

# 2 High-level outline of our approach

We start with a high-level outline of our approach, on which we then expand in the subsequent sections.

To simplify the presentation, note that we can make the following standard assumption without loss of generality. Observe that in an RMFC-T instance on a tree $G = (V, E)$ (with root $r \in V$), we can always assume that the task is to protect all leaves $\Gamma \subseteq V$ of the tree. Here, we do not consider the root as a leaf, even if it has degree one, i.e., $r \notin \Gamma$. Indeed, if an arbitrary subset $S \subseteq V$ of vertices needs protection, we get an equivalent instance by (i) removing all vertices from $S$ that have an ancestor (a vertex on its path to the root) that is also in $S$, because such vertices are protected for free when protecting an ancestor, and then (ii) removing all vertices from $V \setminus S$ that are not on a path from the root to some vertex in $S$. Hence, from now, when referring to RMFC-T, we always assume that the task is to protect all leaves.

---

[5] Another constant-factor approximation follows from the observation in [2] that the Firefighter problem on trees is a special case of submodular function maximization subject to a matroid constraint, and approximation algorithms for constrained submodular maximization [11, 24].

For $\ell \in \mathbb{Z}_{\geq 1}$, we denote by $V_\ell \subseteq V$ the vertices at distance exactly $\ell$ from the root $r$, and call these the vertices on *level* $\ell$. For brevity, we also write $V_{\leq \ell} := \bigcup_{i=1}^{\ell} V_i$, and analogously use $V_{<\ell}$, $V_{\geq \ell}$, and $V_{>\ell}$. Thus, a solution to an RMFC-T instance on a tree $G = (V, E)$ with budget $B$ is a set $R \subseteq V \setminus \{r\}$ of vertices to fireproof such that $|R \cap V_\ell| \leq B$ for each $\ell \in \mathbb{Z}_{\geq 1}$, and every leaf $t \in \Gamma$ has an ancestor in $R$.

Contrary to the common representation of trees in Computer Science as growing top-down, we think of our trees as growing bottom-up (like real trees), i.e., with the root at the bottom and the leaves at the top. This has the advantage that larger indices correspond to higher levels in the tree, and makes talking about levels above and below a threshold $h \in \mathbb{Z}_{\geq 1}$ more natural. In particular, levels below $h$ have smaller indices than $h$ and are closer to the root, whereas those above $h$ have indices larger than $h$ and are father away from the root. In short, bottom levels are closer to the root, and top ones farther away from the root.

## 2.1 Smooth RMFC on trees

We start by introducing a smooth variant of RMFC-T, which has several advantages. In particular, it simplifies and makes more natural multiple steps of our approach. Moreover, contrary to (classical) RMFC-T, it admits a PTAS, and this PTAS will imply both Theorem 1.1 and Theorem 1.2. Finally, we believe that it is a natural and interesting problem in its own right.

Let $L \in \mathbb{Z}_{>0}$ be the *height* of the tree $G$, i.e., the maximum distance of any vertex from the root $r$. (Thus, $L$ is the largest integer with $V_L \neq \emptyset$.) To get to the smooth variant, we start with a first generalization of RMFC-T, which has already been used in [1]. Namely, we allow for non-uniform budgets $B_1, ..., B_L \in \mathbb{R}_{\geq 0}$ for the different levels of the tree. Hence, the task is to find the smallest $\alpha \in \mathbb{R}_{>0}$ together with a vertex set $R \subseteq V \setminus \{r\}$ that protects all leaves and satisfies $|R \cap V_\ell| \leq \alpha \cdot B_\ell$ for all $\ell \in [L]$. We call $\alpha$ the *stretch* of the solution. Hence, the task is to find a solution with minimum stretch. Note that the budget condition can be equivalently written as

$$|R \cap V_{\leq \ell}| \leq \sum_{j=1}^{\ell} \lfloor \alpha \cdot B_j \rfloor \qquad \forall \ell \in [L]. \tag{1}$$

The only difference when using condition (1) is that the budget on level $\ell$ can also be used at a higher level. This does not change the problem, because budgets on lower levels are more powerful than budgets on higher ones. Indeed, instead of fireproofing a certain vertex, one can always fireproof any ancestor of it, i.e., a vertex on its path to the root, and protect at least the same leaves. In *smooth RMFC on trees* (SRMFC-T), we change condition (1) by removing the floor in the right-hand side sum, i.e.,

$$|R \cap V_{\leq \ell}| \leq \alpha \sum_{j=1}^{\ell} B_j \qquad \forall \ell \in [L]. \tag{2}$$

The main difference is that fractional budget units can be accumulated over several levels and used as soon as their sum reaches one. This can be interpreted as allowing different speeds of fire propagation and protection. For example, we can model a fire that spreads 5 times faster than we can fireproof vertices by setting $B_j = 1$ for all $j \in [L]$ and $\alpha = \frac{1}{5}$. Then, in the smooth RMFC variant, one can fireproof a vertex on level 5, whereas in the classical one, no fireproofing of any vertex is possible for stretch $\alpha = \frac{1}{5}$. This allows for a more fine-grained and smoother way to adjust budgets through the stretch $\alpha$. We call an SRMFC-T instance $\alpha$-*feasible* if it admits a solution with stretch at most $\alpha$. The following statement shows that approximations for the smooth variant imply approximations for the classical variant. (Its proof is deferred to Section 10.)

**Theorem 2.1.** *Given an $\alpha$-approximation algorithm for SRMFC-T, we can efficiently compute a solution to an RMFC-T instance of budget at most $\lceil \alpha B_{\text{OPT}} \rceil$, where $B_{\text{OPT}}$ is the optimal budget.*

We will prove the following, which, together with Theorem 2.1, implies both Theorem 1.1 and Theorem 1.2 as direct corollaries.

**Theorem 2.2.** *There is a PTAS for SRMFC-T.*

## 2.2 Compression

As originally observed in [1], RMFC-T can first be preprocessed to only have logarithmically in $|V|$ many levels while approximately preserving the optimal budget. Similar to [1, 23], we also first compress our instance to reduce the height of the tree. Our compression result follows ideas of [1], with a new splitting operation. It leads to instances that we call $\varepsilon$-*compressed*, and which are defined as follows.

**Definition 2.3.** *For $\varepsilon > 0$, an SRMFC-T instance $(G, r, B)$ with $L$ levels is called $\varepsilon$-compressed if*
- $L \leq \lceil \log_{1+\varepsilon}(n) \rceil + 1$, *where $n$ is the number of vertices in $G$, and*
- *the budget up to level $\ell$ is given by $B_{\leq \ell} = (1+\varepsilon)^{\ell-1}$ for $\ell \in [L]$.*

Note that in this case $B_\ell = \varepsilon(1+\varepsilon)^{\ell-2}$ for $\ell \in [L] \setminus \{1\}$ and $B_1 = 1$. Also observe that since we consider the smooth variant, we do not have to round the budgets to integers, which will simplify the presentation.

The statement below formalizes the key implication of our compression result. It shows that a PTAS for 1-feasible $\varepsilon$-compressed SRMFC-T implies a PTAS for SRMFC-T. The fact that we can focus on 1-feasible instances follows because the optimal stretch of an instance can be guessed and the budgets can then be rescaled to obtain 1-feasibility. (See Section 7 for the proof of Theorem 2.4.)

**Theorem 2.4.** *Let $\varepsilon > 0$. An $\alpha$-approximation for 1-feasible $\varepsilon$-compressed SRMFC-T implies a $(1+\varepsilon)\alpha$-approximation for SRMFC-T.*

Working with compressed instances has several advantages. In particular, as we will discuss next, certain properties of the natural linear programming relaxation depend on the number of levels, and become stronger with fewer levels. Also, under certain circumstances, we can use dynamic programming approaches that are efficient on few levels to optimally solve subproblems.

## 2.3 Linear programming considerations and top-bottom leaf separation

A key ingredient in our approach is a natural linear programming (LP) relaxation for SRMFC-T; or, more precisely, of the polyhedron that describes all fractional solutions with respect to a given stretch $\alpha$. To this end, for an instance $(G, r, B)$ of SRMFC-T, we define the following polyhedron for $\alpha \in \mathbb{R}_{>0}$ and any subset of the leaves $\Gamma' \subseteq \Gamma$, which describes fractional solutions for stretch $\alpha$ that protect all leaves in $\Gamma'$:

$$Q_\alpha(\Gamma') := \left\{ x \in \mathbb{R}_{\geq 0}^{V \setminus \{r\}} : x(V_{\leq \ell}) \leq \alpha \sum_{i=1}^\ell B_i \ \forall \ell \in [L] \ \text{ and } \ x(P_t) \geq 1 \ \forall t \in \Gamma' \right\},$$

where, for any $v \in V$, the vertex set $P_v \subseteq V \setminus \{r\}$ denotes all vertices on the unique path from the root $r$ to $v$ (excluding $r$). Moreover, $x(U) := \sum_{u \in U} x_u$ for any $U \subseteq V \setminus \{r\}$.

Hence, if an SRMFC-T instance is feasible for stretch $\alpha$, then, in particular, $Q_\alpha(\Gamma) \neq \emptyset$. The natural LP relaxation for SRMFC-T is $\min\{\alpha : x \in Q_\alpha(\Gamma)\}$, and has, as shown for the natural LP relaxation for RMFC-T in [8], a super-constant integrality gap of $\Theta(\log^* |V|)$, as we discuss in Section 9.3. In line with [1, 23], we use the LP relaxation in a more subtle way, namely as a guide for a guessing procedure that identifies certain structures of an optimal solution. The following lemma, which follows from an LP sparsity result of [1], shows that only instances with a low budget on the first level can have a large integrality gap (see Section 9 for the proof).

**Lemma 2.5.** *Let $\varepsilon > 0$ and $\alpha \in \mathbb{R}_{>0}$. Then, for any SRMFC-T instance with $Q_\alpha(\Gamma) \neq \emptyset$ and $B_1 \geq \frac{L}{\varepsilon}$, where $L$ is the number of levels of the instance, we can efficiently compute an $(\alpha + \varepsilon)$-feasible solution.*

Lemma 2.5 has some profound consequences. In particular, given an $\varepsilon$-compressed SRMFC-T instance, then, if we knew what vertices $R_{\text{bot}} \subseteq V$ an optimal solution fireproofs in the bottom $h = O_\varepsilon(\log L) = O_\varepsilon(\log \log |V|)$[6] levels (we recall that these are the $h$ levels closest to the root), then we can use Lemma 2.5 to protect the leaves not protected by $R_{\text{bot}}$, by fireproofing an appropriate set of vertices in the top $L - h$ levels of the tree. Indeed, given $R_{\text{bot}}$, let $\Gamma' \subseteq \Gamma$ be the leaves not protected by $R_{\text{bot}}$. We can then define a new SRMFC-T instance on the tree $G'$ obtained from $G$ by contracting $V_{\leq h}$ into the root and removing all leaves protected by $R_{\text{bot}}$. This instance has a large budget on the first level, and we can apply Lemma 2.5 to it. Hence, the hard part, from an LP perspective, are only the very bottom levels of the tree, more precisely the bottom $O_\varepsilon(\log \log |V|)$ levels.

Interestingly, if we had an SRMFC-T instance with few levels and low budgets, then a straightforward dynamic programming (DP) approach would solve it optimally, as highlighted by the following result. (For completeness, we provide a proof in Section 8.)

**Theorem 2.6.** *SRMFC-T can be solved in time* $O\big(n\big(1 + \max_{\ell \in [L]} B_\ell\big)^{2L}\big)$, *where $n$ is the number of vertices and $L$ is the number of levels.*

Note that Theorem 2.6 leads to an efficient algorithm for $\varepsilon$-compressed SRMFC-T instances if $L = O\big(\sqrt{\log |V|}\big)$. (We recall that $B_\ell = \varepsilon(1 + \varepsilon)^{\ell - 2}$ for $\ell \in [L]$ in $\varepsilon$-compressed instances.)

Hence, for a general SRMFC-T instance, it seems natural to try to combine an LP-based approach for the top $L - O_\varepsilon(\log \log |V|)$ levels of the tree with a DP-based approach for the bottom $O_\varepsilon(\log \log |V|)$ levels. However, it is highly unclear how to combine these approaches.

Nevertheless, we aim at exploiting the strength of both approaches by trying to find a good way to split the SRMFC-T problem into a problem on the bottom levels and one on the top levels. We do this by computing a partition $(\Gamma_{\text{bot}}, \Gamma_{\text{top}})$ of the leaves $\Gamma$ into *bottom-protected* leaves $\Gamma_{\text{bot}}$ and *top-protected* leaves $\Gamma_{\text{top}}$. To this end, we define the threshold height $\hat{h}$ as the smallest $h \in \mathbb{Z}_{>0}$ satisfying $B_{h+1} \geq \frac{L}{\varepsilon}$, i.e.,

$$\hat{h} = \left\lceil \log_{1+\varepsilon}\left(\frac{L}{\varepsilon^2}\right) \right\rceil + 1,$$

and interpret a partition $(\Gamma_{\text{bot}}, \Gamma_{\text{top}})$ as follows. $\Gamma_{\text{bot}}$ contains all leaves $t \in \Gamma$ to be protected by fireproofing a vertex in $P_t \cap V_{\leq \hat{h}}$, whereas $\Gamma_{\text{top}}$ are the leaves $t \in \Gamma$ to be protected by fireproofing a vertex in $P_t \cap V_{>\hat{h}}$.[7]

Protecting the leaves in $\Gamma_{\text{bot}}$ or $\Gamma_{\text{top}}$ by fireproofing on the bottom $\hat{h}$ levels or top $L - \hat{h}$ levels of the tree, respectively, are itself SRMFC-T instances. More precisely, the instance to protect $\Gamma_{\text{bot}}$ consists of the subtree of $G$ over $\big(\bigcup_{t \in \Gamma_{\text{bot}}} P_t\big) \cap V_{\leq \hat{h}}$ and the root, with budgets $B_1, \ldots, B_{\hat{h}}$. We call this the $\Gamma_{\text{bot}}$-*bottom instance* or simply *bottom instance*. Analogously, the instance to protect $\Gamma_{\text{top}}$ is obtained by first contracting the first $\hat{h}$ levels of $G$ into the root—this can be achieved by contracting all tree edges with both endpoints in $V_{\leq \hat{h}}$—and then restricting to $\big(\bigcup_{t \in \Gamma_{\text{top}}} P_t\big) \cap V_{>\hat{h}}$ and the root. Also this top instance carries over the budgets $B_{\hat{h}+1}, \ldots, B_L$, and we call it the $\Gamma_{\text{top}}$-*top instance* or simply *top instance*.

For the value of $\hat{h}$ chosen above, one can easily verify that, given any partition $(\Gamma_{\text{bot}}, \Gamma_{\text{top}})$ of the leaves $\Gamma$, the bottom instance can be solved optimally and efficiently using Theorem 2.6, and the top instance can be solved near-optimally and efficiently using Lemma 2.5. Ideally, if we could compute a partition $(\Gamma_{\text{bot}}, \Gamma_{\text{top}})$ of $\Gamma$ that corresponds to an optimal solution—i.e., $\Gamma_{\text{bot}}$ contains exactly the leaves that an optimal solution protects by fireproofing vertices on the bottom levels—then we would be done.

---

[6]When subscripting the big-O notation with $\varepsilon$, i.e., $O_\varepsilon(\cdot)$, the hidden constant may depend on $\varepsilon$.

[7]We can assume without loss of generality that the solution we are looking for is minimal, implying that no leaf is protected by both fireproofing a vertex below level $\hat{h}$ and one above level $\hat{h}$.

We will not be able to achieve this. However, our main contribution is that we can efficiently compute polynomially many partitions such that at least one of them is close to optimal, as stated below.

**Theorem 2.7.** *Let $\varepsilon \in \left(0, \frac{1}{7}\right]$ and consider a 1-feasible $\varepsilon$-compressed SRMFC-T instance with $L$ levels. Then there is an efficient procedure that computes polynomially many partitions of the leaves $\Gamma$ containing one partition $\left(\Gamma_{\text{bot}}, \Gamma_{\text{top}}\right)$ so that both the bottom instance and the top instance are $(1 + 7\varepsilon)$-feasible.*

Proving Theorem 2.7 is the main technical contribution of this work, and is achieved through our novel LP-guided guessing approach, which we describe in Section 3. Note that Theorem 2.7 implies our PTAS for SRMFC-T, i.e., Theorem 2.2, due to the above discussion (see Section 10 for a formal proof).

## 3 LP-guided guessing

We now present our LP-guided guessing procedure used to prove Theorem 2.7. This is the main technical part of our approach, and we introduce it step by step to simplify the presentation. We begin with a basic version of our procedure that highlights several main ideas and facilitates both drawing parallels and understanding differences compared to prior work.

Consider an $\varepsilon$-compressed and 1-feasible SRMFC-T instance $(G = (V, E), r, B)$. Our LP-guided guessing procedure iteratively computes a point $x \in Q_1(\Gamma)$ and guesses properties based on vertices in its support. However, to limit the support of $x$ (and therefore the number of guesses), we first sparsify $x$. Apart from reducing the support, our sparsification also leads to high values on support vertices, which is useful for accounting purposes later on. We note that our sparsification lemma, shown below, is not simply a statement about vertex solutions of $Q_1(\Gamma)$, as is common in many LP-based procedures. In contrast, it modifies $x$ to get a point $y$ in the slightly larger polytope $Q_{1+3\varepsilon}(\Gamma)$ to obtain the desired properties.

Due to the super-constant integrality gap of $Q_1(\Gamma)$, we cannot obtain constant lower bounds on the values of $y$ on all its support vertices. However, we can obtain constant lower bounds above a threshold height $\check{h} = \Theta_\varepsilon\left(\log^{(2)} L\right)$, which we set to

$$\check{h} := \left\lceil \log_{1+\varepsilon}\left(\frac{\hat{h}}{\varepsilon^2}\right) \right\rceil + 1.$$

**Lemma 3.1.** *Let $\varepsilon \in \left(0, \frac{1}{7}\right]$, and consider an $\varepsilon$-compressed SRMFC-T instance. Let $x \in Q_1(\Gamma)$. Then we can efficiently compute a point $y \in Q_{1+3\varepsilon}(\Gamma)$ with*
- $\text{supp}(y) \subseteq \text{supp}(x)$,
- $y_v \geq \frac{\varepsilon}{h}$ *for each $v \in \text{supp}(y) \cap V_{\leq \check{h}}$, and*
- $y_v \geq \varepsilon$ *for each $v \in \text{supp}(y) \cap V_{>\check{h}} \cap V_{\leq \hat{h}}$.*

The proof of Lemma 3.1 is postponed to Section 9. Throughout this section, $\varepsilon \leq \frac{1}{7}$ denotes the fixed compression parameter of our $\varepsilon$-compressed instance.

### 3.1 A basic LP-guided guessing procedure

We now present the basic outline of our LP-guided guessing procedure. This first version can be thought of as a refinement of prior work. It allows for understanding the interactions between different ingredients introduced so far, builds a basis for improvements presented later, and already improves prior results.

We work with the basic relaxation $Q_1(\Gamma)$, and iteratively improve it by maintaining a set $D$ of vertices that will not be fireproofed. Formally, for $D \subseteq V \setminus \{r\}$, $\alpha \in \mathbb{R}_{>0}$, and $\Gamma' \subseteq \Gamma$, we define

$$Q_\alpha^D(\Gamma') := \left\{ x \in \mathbb{R}_{\geq 0}^{V \setminus \{r\}} : x(V_{\leq \ell}) \leq \alpha \sum_{i=1}^\ell B_i \ \forall \ell \in [L] \ \text{ and } \ x(P_t) \geq 1 \ \forall t \in \Gamma' \ \text{ and } \ x(D) = 0 \right\},$$

to be all points in $Q_\alpha(\Gamma')$ that are zero for every vertex in $D$.[8] (Indeed, this follows from $x(D) = 0$ together with non-negativity of $x$.)

Throughout the procedure, we enumerate different sets $D$. These enumerations aim at guessing part of the structure of an optimal solution OPT. (If there are multiple optimal solutions, we fix an arbitrary one; moreover, as usual, we choose OPT so that it does not contain vertices with an ancestry relationship.) Hence, we are interested in sets $D$ with $\text{OPT} \cap D = \emptyset$, which we call OPT-*compatible*. Of course, we want to guess an OPT-compatible $D$ that is as small as necessary to obtain a good partition $(\Gamma_{\text{bot}}, \Gamma_{\text{top}})$ of the leaves $\Gamma$. To this end, for a generic threshold $h \in [L]$, we define the following set $V_{\text{core}}^h \subseteq V_{\leq h} \setminus \text{OPT}$ of what we call *core vertices*:

$$V_{\text{core}}^h := \bigcup_{v \in \text{OPT} \cap V_{\leq h}} (P_v \setminus \{v\}).$$

We use a generic threshold $h$ to draw the line between the *bottom levels* $V_{\leq h}$ and *top levels* $V_{>h}$. Our final procedure will set $h$ to $\hat{h} = \lceil \log_{1+\varepsilon}(\frac{L}{\varepsilon^2}) \rceil + 1$. Thus, we abbreviate $V_{\text{core}} := V_{\text{core}}^{\hat{h}}$. However, a generic threshold $h \in [L]$ is helpful to better understand the simpler guessing procedures we introduce first and what they imply.

$V_{\text{core}}$ is designed so that a point in $Q_1^D(\Gamma)$ without support in $V_{\text{core}}$ allows for deriving a small number of partitions of $\Gamma$ into $(\Gamma_{\text{bot}}, \Gamma_{\text{top}})$, one of which will admit both a 1-feasible integral solution for the bottom instance and a strong fractional solution for the top instance. We recall that, as discussed previously, the integrality gap of the natural LP is due to the bottom levels. This allows us to derive a strong integral solution, as we discuss after the statement below.

**Lemma 3.2.** *Let $h \in [L]$, $\alpha > 0$, and let $y \in Q_\alpha^D(\Gamma)$. In time $O(|V| \cdot 2^{|\text{supp}(y) \cap V_{\leq h}|})$, we can compute a set of at most $2^{|\text{supp}(y) \cap V_{\leq h}|}$ partitions $(\Gamma_{\text{bot}}, \Gamma_{\text{top}})$ of $\Gamma$ so that if $\text{supp}(y) \cap V_{\text{core}}^h = \emptyset$, then for one of these partitions*
- *the bottom instance is 1-feasible, and*
- *the top instance admits a fractional solution for stretch $1 + \alpha$.*

Note that for $h \leq \hat{h}$, the number of computed partitions is polynomial in the number of vertices of $G$. For brevity, we say that $y \in \mathbb{R}^{V \setminus \{r\}}$ is $V_{\text{core}}^h$-*free* if $\text{supp}(y) \cap V_{\text{core}}^h = \emptyset$.

*Proof.* Assume $y$ is $V_{\text{core}}^h$-free. Set $\Gamma_{\text{bot}} = \emptyset$. For each $v \in \text{supp}(y) \cap V_{\leq h}$, we guess whether $P_v \cap \text{OPT} \neq \emptyset$. If so, we add all leaves in $T_v$ to $\Gamma_{\text{bot}}$. We claim that $(\Gamma_{\text{bot}}, \Gamma \setminus \Gamma_{\text{bot}})$ satisfies the desired properties. By enumerating all possible outcomes of our guesses, we then have the desired set of partitions. A key property of $V_{\text{core}}^h$ is that for each $v \notin V_{\text{core}}^h$, either $P_v \cap \text{OPT} \neq \emptyset$, or all leaves in $T_v$ are protected by $\text{OPT} \cap V_{>h}$. Therefore, as $\text{supp}(y) \cap V_{\text{core}}^h = \emptyset$, the bottom vs. top partition of leaves we guess is compatible with OPT for the leaves above any vertex in $\text{supp}(y) \cap V_{\leq h}$.

By construction, OPT restricted to the bottom instance (i.e., the vertices $V_{\leq h}$) is 1-feasible. We claim that $y + \chi^{\text{OPT}}$ restricted to the top instance protects all leaves $\Gamma \setminus \Gamma_{\text{bot}}$ of the top instance, where $\chi^{\text{OPT}}$ denotes the characteristic vector of OPT. Indeed, for any $t \in \Gamma \setminus \Gamma_{\text{bot}}$, either $P_t \cap \text{supp}(y) \cap V_{\leq h} = \emptyset$, in which case $y(P_t \cap V_{>h}) = 1$. Or there is some $v \in P_t \cap \text{supp}(y) \cap V_{\leq h}$, which implies together with $t \in \Gamma \setminus \Gamma_{\text{bot}}$ that all leaves in $T_v$ (including $t$) are protected by OPT-vertices in $V_{>h}$. Because $y \in Q_\alpha(\Gamma)$, the point $y + \chi^{\text{OPT}}$ (restricted to the top instance) has stretch at most $1 + \alpha$. □

If we get a partition $(\Gamma_{\text{bot}}, \Gamma_{\text{top}})$ with the guarantees described in Lemma 3.2, then we transform it into an integral solution as follows. Theorem 2.6 leads to a 1-feasible integral solution to the bottom instance and is efficient for $h = O\bigl(\sqrt{\log|V|}\bigr)$. To convert a $(1+\alpha)$-feasible fractional solution to the top instance into an integral one, we will lose a factor corresponding to the integrality gap of the relaxation, which depends on the choice of $h$. The following lemma, which follows from Lemma 2.5, formalizes this for the thresholds $h \in \{\hat{h}, \check{h}\}$, which are relevant for us. Section 9 contains a formal proof of a generalization.

---
[8] We note that prior LP-guided guessing approaches in this context [1, 7, 23] additionally maintained a subset of vertices to be fireproofed. From a technical perspective, requiring a vertex to be fireproofed can also be achieved by requiring its ancestors and descendants not to be fireproofed. Hence, this can be simulated through $D$. However, our reason for just using $D$ is that it simplifies our presentation because we do not explicitly require certain vertices to be fireproofed.

**Lemma 3.3.** *Let $h \in \{\hat{h}, \check{h}\}$ and $\alpha \in \mathbb{R}_{>0}$. Given an $\varepsilon$-compressed SRMFC-T instance with $L$ levels, and $y \in Q_\alpha(\Gamma)$ with $\mathrm{supp}(y) \cap V_{\leq h} = \emptyset$, we can efficiently compute a solution not fireproofing any vertex in $V_{\leq h}$ and with stretch*
- *$2\alpha + \varepsilon$ if $h = \check{h}$, and*
- *$\alpha + \varepsilon$ if $h = \hat{h}$.*

In our first basic guessing procedure, we set $h = \check{h}$, and show how to efficiently obtain a sparse $V^h_{\mathrm{core}}$-free point $y \in Q^D_{1+3\varepsilon}(\Gamma)$. By Lemma 3.2 and Lemma 3.3, this immediately leads to a $(4 + O(\varepsilon))$-feasible solution to SRMFC-T. Combined with Theorem 2.1, we thus get an efficient algorithm for RMFC-T returning a solution of budget at most $\lceil (4+\varepsilon) B_{\mathrm{OPT}} \rceil$, where $B_{\mathrm{OPT}}$ is the budget of an optimal solution. Even though this already improves upon the previously best solution guarantee of $\lceil (5+\varepsilon) B_{\mathrm{OPT}} \rceil$ [23], its main purpose is to serve as a basis for further improvements. We therefore omit a formal proof of the budget guarantee, as it is weaker than guarantees obtained (and proved) through later improvements.

Algorithm 1 describes how a basic LP-guided guessing procedure allows for efficiently obtaining a sparsified $V^h_{\mathrm{core}}$-free point $y \in Q_{1+3\varepsilon}(\Gamma)$. (Algorithm 2 to its right will be used later.) It iteratively first computes a sparsification $y$ of a point $x \in Q^D_1(\Gamma)$, and guesses the set $A_{\mathrm{core}} := \mathrm{supp}(y) \cap V^h_{\mathrm{core}}$, by exploring all options for $A_{\mathrm{core}} \subseteq \mathrm{supp}(y) \cap V_{\leq h}$, and then adds $A_{\mathrm{core}}$ to $D$.

**Algorithm 1: BASIC-EXPLORATION($D, \gamma$)**

**if** $\gamma > 0$ and $Q^D_1(\Gamma) \neq \emptyset$ **then**
    Determine a point $x \in Q^D_1(\Gamma)$.
    $y \leftarrow$ Sparsification of $x$ through Lemma 3.1.
    **for** each $A_{\mathrm{core}} \subseteq \mathrm{supp}(y) \cap V_{\leq h}$ **do**
        $D \leftarrow D \cup A_{\mathrm{core}}$.
        **if** $A_{\mathrm{core}} = \emptyset$ **then**
            Add to output partitions $(\Gamma_{\mathrm{bot}}, \Gamma_{\mathrm{top}})$ generated by Lemma 3.2 applied to $y$.
        **else**
            BASIC-EXPLORATION($D, \gamma - 1$).
        **end**
    **end**
**end**

**Algorithm 2: EXPLORATION-WITH-MIXING($D, \gamma, Y$)**

**if** $\gamma > 0$ and $Q^D_1(\Gamma) \neq \emptyset$ **then**
    Determine a point $x \in Q^D_1(\Gamma)$.
    $y \leftarrow$ Sparsification of $x$ through Lemma 3.1.
    **for** any disjoint $A_{\mathrm{core}}, A_{\mathrm{top}} \subseteq \mathrm{supp}(y) \cap V_{\leq h}$ **do**
        $D \leftarrow D \cup A_{\mathrm{core}} \cup D^h_{\mathrm{top}}(A_{\mathrm{top}})$.
        **if** $A_{\mathrm{core}} = \emptyset$ **then** add $y$ to $Y$.
        **if** $|Y| = N$ **then**
            Add $\bigl(\Gamma \setminus \Gamma^h_{\mathrm{top}}(\overline{Y}), \Gamma^h_{\mathrm{top}}(\overline{Y})\bigr)$ to output.
        **else**
            EXPLORATION-WITH-MIXING($D, \gamma - 1, Y$).
        **end**
    **end**
**end**

In each execution path, each vertex of $V^h_{\mathrm{core}}$ will be included at most once in a set $A_{\mathrm{core}}$ because we add this set to $D$. We therefore set the parameter $\gamma$, which controls the recursion depth (number of guesses), to a value $\gamma \geq |V^h_{\mathrm{core}}| + 1$. This guarantees, by the pigeonhole principle, that we encounter a sparsified $V^h_{\mathrm{core}}$-free point $y$ as desired, when following the execution path that always correctly guesses $A_{\mathrm{core}} = \mathrm{supp}(y) \cap V^h_{\mathrm{core}}$. For an explicit upper bound on $|V^h_{\mathrm{core}}|$, we use the relation $|V^h_{\mathrm{core}}| \leq h B_{\leq h}$, which follows because $|\mathrm{OPT} \cap V_{\leq h}| \leq B_{\leq h}$, and each vertex in $\mathrm{OPT} \cap V_{\leq h}$ contributes at most $h$ vertices to $V^h_{\mathrm{core}}$. Hence, we can set $\gamma = h B_{\leq h} + 1$.

To analyze the runtime of Algorithm 1, note that by choosing $h = \check{h}$, we have $B_{\leq h} = (1+\varepsilon)^{h-1} = O_\varepsilon(\log L)$. This, in turn, implies

$$\gamma = \check{h} B_{\leq \check{h}} + 1 = O_\varepsilon(\log L \cdot \log \log L).$$

Moreover, for any sparsification $y$ of a point in $Q^D_1(\Gamma)$, we have by Lemma 3.1 that

$$|\text{supp}(y) \cap V_{\leq \check{h}}| \leq B_{\leq \check{h}} \cdot \left(\frac{\varepsilon}{\check{h}}\right)^{-1} = O_\varepsilon(\log L \cdot \log \log L).$$

The number of subsets of $\text{supp}(y) \cap V_{\leq \check{h}}$ is thus bounded by $2^{O_\varepsilon(\log L \cdot \log \log L)} = L^{O_\varepsilon(\log \log L)}$, which leads to a total number of execution paths of

$$\left(L^{O_\varepsilon(\log \log L)}\right)^\gamma = L^{O_\varepsilon(\log L \log^2 \log L)},$$

which is polynomial in $|V|$ as $L = O_\varepsilon(\log |V|)$, thus leading to a polynomial-time algorithm for $h = \check{h}$.

**Obstacles to obtain a PTAS**

In order to obtain a $(1 + O(\varepsilon))$-feasible solution to the top instance, we need to overcome two main obstacles. Currently, the top instance
- is only guaranteed to have a fractional solution with stretch about 2 (see Lemma 3.2), and
- is only guaranteed to have an integrality gap of roughly 2 (see Lemma 3.3).

To overcome the second hurdle, we would like to choose $h = \hat{h}$ in Algorithm 1, which would reduce the integrality gap of the top instance to close to one (Lemma 3.3). However, this increases the running time, and, in its current version, the runtime of Algorithm 1 would become quasi-polynomial for $h = \hat{h}$. We solve this issue by thinning $V_{\text{core}}$ as described in Section 3.3.1. The resulting bottom instance then might not be 1-feasible anymore, but we show the existence of a $(1 + O(\varepsilon))$-feasible solution by accounting against a well-constructed fractional point.

How to tackle the first obstacle will be discussed next.

## 3.2 Mixing fractional solutions

The key goal behind Algorithm 1 is to obtain a sparsified $V_{\text{core}}^h$-free point $y$, as highlighted by Lemma 3.2. Lemma 3.2 then leads to a partition of $\Gamma$ into $\Gamma_{\text{bot}}$ and $\Gamma_{\text{top}}$ so that the top instance can be fractionally solved with a stretch of approximately 2. We start by a simple yet useful observation that this stretch can be improved if we have the following further property on $y$, where we define (again, for a generic threshold $h \in [L]$)

$$\Gamma_{\text{top}}^h(y) := \{t \in \Gamma : y(P_t \cap V_{>h}) \geq 1 - \varepsilon\}.$$

As we choose $h = \hat{h}$ in our final algorithm, we abbreviate $\Gamma_{\text{top}}(y) := \Gamma_{\text{top}}^{\hat{h}}(y)$.

**Observation 3.4.** *Let $h \in [L]$, $\alpha > 0$, and $y \in Q_\alpha(\Gamma)$ with $\text{supp}(y) \cap V_{\text{core}}^h = \emptyset$ and $y(P_t \cap V_{>h}) \geq 1 - \varepsilon$ for each leaf $t \in \Gamma$ protected by $\text{OPT} \cap V_{>h}$. Then, for the partition $\left(\Gamma \setminus \Gamma_{\text{top}}^h(y), \Gamma_{\text{top}}^h(y)\right)$ of $\Gamma$,*
- *$\text{OPT} \cap V_{\leq h}$ is 1-feasible for the bottom instance, and*
- *$(1 + 2\varepsilon)y$ restricted to $V_{>h}$ is a fractional solution of stretch $(1 + 2\varepsilon)\alpha$ to the top instance.*

*Proof.* $(1 + 2\varepsilon)y$ is clearly feasible for the top instance, because each leaf $t \in \Gamma_{\text{bot}}(y)$ satisfies $y(P_t \cap V_{>h}) \geq 1 - \varepsilon$. (Here, we use $\varepsilon \leq \frac{1}{2}$.) Moreover, the bottom instance only contains leaves protected by $\text{OPT} \cap V_{\leq h}$, and therefore $\text{OPT} \cap V_{\leq h}$ is 1-feasible for the bottom instance. □

We show how to obtain a point $y \in Q_{1+3\varepsilon}^D(\Gamma)$ fulfilling the conditions of Observation 3.4 (for $\alpha = 1 + 3\varepsilon$). To this end, we modify Algorithm 1 as follows to obtain Algorithm 2. A key difference is that Algorithm 2 maintains a collection $Y$ (allowing multiplicities) of sparsified points $y$ computed in iterations with $A_{\text{core}} = \emptyset$ (i.e., iterations where $y$ is $V_{\text{core}}^h$-free, if the guess is correct). The desired point is then obtained by taking the average

$$\overline{Y} := \frac{1}{|Y|} \sum_{y \in Y} y$$

of the points in $Y$, which is why Algorithm 2 returns partitions of the form $(\Gamma \setminus \Gamma_{\text{top}}^h(\overline{Y}), \Gamma_{\text{top}}^h(\overline{Y}))$. Our goal in Algorithm 2 is to return the average $\overline{Y}$ of $N = \lceil \varepsilon^{-1} \rceil$ many $V_{\text{core}}^h$-free points.

To make sure that such an average $\overline{Y}$ fulfills the conditions of Observation 3.4, we extend the guessing step. For each vertex $v \in \text{supp}(y) \cap V_{\leq h}$, we do not only guess whether $v \in V_{\text{core}}^h$ (in which case we put $v$ in $A_{\text{core}}$ as before), but also whether all leaves above $v$ are protected by $\text{OPT} \cap V_{>h}$, in which case we put $v$ in another set $A_{\text{top}}$. As before, as we do not know OPT (and therefore also not $V_{\text{core}}^h$), we enumerate all disjoint sets $A_{\text{core}}, A_{\text{top}} \subseteq \text{supp}(y) \cap V_{\leq h}$. Moreover, when updating $D$, on top of adding $A_{\text{core}}$ to it, we also add

$$D_{\text{top}}^h(A_{\text{top}}) := \bigcup_{v \in A_{\text{top}}} (P_v \cup (T_v \cap V_{\leq h})) \tag{3}$$

to $D$. Again, we abbreviate $D_{\text{top}}(A_{\text{top}}) := D_{\text{top}}^{\hat{h}}(A_{\text{top}})$. Note that $D_{\text{top}}^h(A_{\text{top}})$ contains all vertices that cannot be in OPT assuming the guess encoded in $A_{\text{top}}$ is correct. In other words, because for each $v \in A_{\text{top}}$, all leaves above $v$ are protected by $\text{OPT} \cap V_{>h}$, no vertex in $P_v$ or $T_v \cap V_{\leq h}$ can be in OPT.

The statement below together with its proof shows that, for $h = \check{h}$, this extended guessing step allows us to efficiently obtain a point $\overline{Y}$ fulfilling the conditions of Observation 3.4 (for $\alpha = 1 + 3\varepsilon$). Hence, the partitions of $\Gamma$ returned by Algorithm 2 contain a partition that does not suffer from the issue we had with Algorithm 1 that the top instance needs stretch approximately 2.

**Theorem 3.5.** *Let $\varepsilon \in (0, \frac{1}{2}]$ and let $(G, r, B)$ be an $\varepsilon$-compressed 1-feasible instance of SRMFC-T. Then Algorithm 2 run with $D = \emptyset$, $Y = \emptyset$, $h = \check{h}$, $N = \lceil \frac{1}{\varepsilon} \rceil$, and $\gamma = h(1 + \varepsilon)^h + N$ runs in polynomial time and the output contains a partition $(\Gamma_{\text{bot}}, \Gamma_{\text{top}})$ of the leaves $\Gamma$, such that*
- *the bottom instance is 1-feasible, and*
- *the top instance admits a fractional solution for stretch $1 + 5\varepsilon + 6\varepsilon^2 \leq 1 + 8\varepsilon$.*

*Proof.* Note that the recursion depth suffices to fully explore the execution path that guesses correctly until $N$ iterations are encountered with a $V_{\text{core}}^h$-free point $y$, because $\gamma \geq |V_{\text{core}}^h| + N$. Let $y_1, ..., y_N$ be the sparsified $V_{\text{core}}^h$-free points computed in these $N$ iterations, i.e., $Y = (y_1, ..., y_N)$, with the numbering reflecting the order in which they have been encountered. We show that $\overline{Y}$ fulfills the conditions of Observation 3.4 (for $\alpha = 1 + 3\varepsilon$), which implies the statement, except for the running time.

Let $t \in \Gamma$ be a leaf that is top-protected by OPT, i.e., $P_t \cap \text{OPT} \cap V_{>h} \neq \emptyset$. If there is an index $i \in [N]$ and a vertex $u \in P_t \cap V_{\leq h}$ with $y_i(u) > 0$, then we must have $u \in A_{\text{top}}$ in the iteration where $y_i$ was computed, when following the execution path that guesses correctly. Indeed, because $y_i$ is $V_{\text{core}}^h$-free, we have $u \notin V_{\text{core}}^h$, which implies that all leaves above $u$ are protected by $\text{OPT} \cap V_{>h}$ if one is. However, this implies that $t$ will be entirely top-protected by $y_j$ for $j \in \{i+1, ..., N\}$, i.e., $y_j(P_t \cap V_{>h}) = 1$. Thus, at most one index $i \in [N]$ can (partially) bottom-protect $t$. Because $N = \lceil \frac{1}{\varepsilon} \rceil$, we have $\overline{Y}(P_t \cap V_{>h}) \geq \frac{N-1}{N} \geq 1 - \varepsilon$, which shows the condition of Observation 3.4, as desired.

It remains to analyze the runtime of the algorithm. Note that we can compute a point in $Q_1^D(\Gamma)$ or decide that no such point exists in polynomial time. Hence, because each execution path has length at most $\gamma$, which is polynomial in $|V|$, it suffices to bound the number of execution paths we create. For each of the up to $\gamma$ levels of recursion, we branch into at most $3^{|\text{supp}(y) \cap V_{\leq h}|}$ many subproblems. Using the sparsity guarantee of Lemma 3.1, we can thus bound the total number of execution paths by $3^{\frac{\gamma \check{h}}{\varepsilon} B_{\leq \check{h}}}$, which is polynomial in $|V|$ by our choice of $\gamma$. $\square$

Putting the pieces together, for the partition $(\Gamma_{\text{bot}}, \Gamma_{\text{top}})$ in Theorem 3.5 returned by Algorithm 2, we can efficiently compute a 1-feasible integral solution for the bottom instance using Theorem 2.6. Moreover, for the top instance, we can first compute a $1 + 8\varepsilon$-feasible fractional solution by Theorem 3.5, and then round it to a $(2 + 17\varepsilon)$-feasible integral solution using Lemma 3.3. Hence, we get a $(2 + 17\varepsilon)$-feasible solution for any $\varepsilon$-compressed 1-feasible SRMFC-T instance. Together with Theorem 2.4 and Theorem 2.1, this already yields a 3-approximation for (classical) RMFC-T, significantly improving upon prior work.

Choosing $h = \hat{h}$ would now yield the $(1 + O(\varepsilon))$-approximation for $\varepsilon$-compressed 1-feasible SRMFC-T. However, for this choice of $h$, Algorithm 2 has quasi-polynomial runtime; hence, this is not a PTAS yet.[9] We will address this issue in the remainder of this section.

## 3.3 Improving efficiency to get a PTAS

As mentioned above, choosing $h = \hat{h} = \Theta_\varepsilon(\log L)$ in Algorithm 2 yields a quasi-polynomial PTAS (a QPTAS) for SRMFC-T. To guarantee polynomial running time, we cannot afford to explore the whole core for this choice of $h$, as it can contain up to $\hat{h} B_{\leq \hat{h}} = \Omega_\varepsilon(L \log L)$ many vertices. Thus, already enumerating for $|V_{\text{core}}|$ many vertices whether they belong to the core or not would lead to a quasi-polynomial running time of $2^{\Omega_\varepsilon(L \log L)}$. We will show in Section 3.3.1 how to thin the core to a set $V_{\text{thin}} \subseteq V_{\text{core}}$ of size $O_\varepsilon(L)$.

Without additional adaptions, the running time of the resulting algorithm would still be quasi-polynomial in the number of vertices, as it can take too many iterations until we encounter a sparsified point $y$ with $\text{supp}(y) \cap V_{\text{thin}} = \emptyset$. This happens if many iterations have a fractional solution whose support has small overlap with $V_{\text{thin}}$. We discuss how to deal with this case in Section 3.3.2.

### 3.3.1 Thinning the set of critical vertices

From now on, we fix the threshold $h$ to $\hat{h} = \left\lceil \log_{1+\varepsilon}\left(\frac{L}{\varepsilon^2}\right) \right\rceil + 1$ and the parameter $N$ to $N := \left\lceil \frac{4}{\varepsilon} \right\rceil$.

Recall that $V_{\text{core}}$ is designed so that Lemma 3.2 holds. In particular, once we encounter a fractional solution $y$ with $\text{supp}(y) \cap V_{\text{core}} = \emptyset$, we obtain a partition of the leaves so that the bottom instance is 1-feasible. However, to obtain a PTAS for SRMFC-T, a $(1 + O(\varepsilon))$-feasible bottom instance suffices. We will use this slack to thin $V_{\text{core}}$ to a set $V_{\text{thin}} \subseteq V_{\text{core}}$ of size $O_\varepsilon(L)$ by fireproofing additional vertices in $V_{\leq \hat{h}}$.

To do so, we use the following observation, which states that if a vertex set $A \subseteq V$ is $\alpha$-feasible, then we can fireproof for each $v \in A$ a vertex $\kappa \in \mathbb{Z}_{\geq 1}$ levels higher up for a small fraction of the budget.

**Observation 3.6.** *Consider an $\varepsilon$-compressed SRMFC-T instance, $\kappa, h \in \mathbb{Z}_{\geq 1}$, and $\alpha \in \mathbb{R}_{>0}$. Assume $A \subseteq V_{\leq h}$ fulfills $|A \cap V_{\leq \ell}| \leq \alpha B_{\leq \ell}$ $\forall \ell \in [h]$. For every $v \in A$, let $w_v$ be a vertex at least $\kappa$ levels above $v$. Then,*

$$|\{w_v : v \in A\} \cap V_{\leq \ell}| \leq |A \cap V_{\leq \ell - \kappa}| \leq \alpha B_{\leq \ell - \kappa} = (1+\varepsilon)^{-\kappa} \alpha B_{\leq \ell} \quad \forall \ell \in [h+\kappa].$$

Consider a sparsified point $y \in Q_{1+3\varepsilon}(\Gamma)$ output by Lemma 3.1. By the lower bounds provided there, we have $y_v \geq \varepsilon$ for each $v \in \text{supp}(y) \cap V_\ell$ with $\ell \in \{\check{h}+1, ..., \hat{h}\}$. Thus, in this range of levels, we have $|\text{supp}(y) \cap V_{>\check{h}} \cap V_{\leq \ell}| \leq \frac{1+3\varepsilon}{\varepsilon} B_{\leq \ell}$. For $\kappa \in \mathbb{Z}_{\geq 1}$, we apply Observation 3.6 with $A = \text{supp}(y) \cap V_{>\check{h}} \cap V_{\leq \hat{h}}$ and $\alpha = \frac{1+3\varepsilon}{\varepsilon}$. This allows us to fireproof one vertex on level $\ell + \kappa$ for every vertex in $V_\ell \cap \text{supp}(y)$ with $\ell \in \{\check{h}+1, ..., \hat{h}\}$, while increasing the budget utilization up to level $\ell + \kappa$ by at most $(1+\varepsilon)^{-\kappa} \frac{1+3\varepsilon}{\varepsilon} B_{\leq \ell + \kappa}$. During the course of the algorithm, we will have $N$ many iterations in which we apply this argument. Recall that $N = \left\lceil \frac{4}{\varepsilon} \right\rceil$. Thus, choosing $\kappa$ large enough (depending on $\varepsilon$), we can in all of these iterations fireproof one vertex on level $\ell + \kappa$ for every vertex in $V_\ell \cap \text{supp}(y)$ without increasing the violation up to level $\ell + \kappa$ by more than $\varepsilon B_{\leq \ell + \kappa}$. We will discuss next how to choose these vertices, which then enables us to thin out the core.

For $\kappa \in \mathbb{Z}_{\geq 1}$ and $v \in V_\ell$ with $\ell \in [\hat{h}]$, we define $T_{v,\kappa} := T_v \cap V_{\leq \ell + \kappa}$ to be the descendants of $v$ up to level $\ell + \kappa$, including $v$ itself. Now, if for $v \in V_{\text{core}} \cap \text{supp}(y) \cap V_\ell$ with $\ell \in \{\check{h}+1, ..., \hat{h}\}$, all leaves in $T_v$ protected by $\text{OPT} \cap V_{\leq \hat{h}}$ are contained in a single sub-tree $T_u$ for some leaf $u$ of $T_{v,\kappa}$, then fireproofing $u$ (which lies $\kappa$ levels above $v$ as desired) protects all these leaves. In particular, all remaining leaves in $T_v$ are protected by $\text{OPT} \cap V_{>\hat{h}}$. Thus, we thin out the core by removing such vertices $v$. Note that these are precisely the core vertices $v \in V_{\text{core}} \cap V_\ell$ with $\ell \in \{\check{h}+1, ..., \hat{h}\}$ for which $T_{v,\kappa}$ neither contains a vertex in $\text{OPT} \cap V_{\leq \hat{h}}$ nor contains a vertex with degree at least three in the subtree defined by $V_{\text{core}}$.

---

[9]To be precise, for polynomial runtime one can choose $h$ so that $\hat{h} - h$ is in $\Omega(\log \log L)$. Unfortunately, our upper bound on the integrality gap of the bottom instance remains approximately 2.

Unfortunately, we cannot apply the same argument to vertices $v \in V_{\text{core}} \cap \text{supp}(y) \cap V_{\leq \check{h}}$. Due to the weaker lower bound on $y_v$ on these levels, we could end up fireproofing too many additional vertices. Nonetheless, $V_{\text{core}} \cap V_{\leq \check{h}}$ can contain up to $\check{h} B_{\leq \hat{h}} = \Omega_\varepsilon(L \log \log L)$ many vertices, as vertices of OPT on levels between $\check{h}$ and $\hat{h}$ can lead to many core vertices on levels below $\check{h}$. Thus, enumerating for $|V_{\text{core}} \cap V_{\leq \check{h}}|$ many vertices separately whether they belong to the core or not would still lead to a quasi-polynomial running time. However, we observe that we can prune our enumeration significantly using the following observation. Consider a vertex $v \in V_{\text{core}} \cap V_{\leq \check{h}}$ and assume that all leaves in $T_v$ are protected by $\text{OPT} \cap V_{>\check{h}}$. In this case, we can safely update $D$ by adding all vertices in $P_v \cup (T_v \cap V_{\leq \check{h}})$. Thus, we remove all such support vertices from the core. The resulting thinned core in $V_{\leq \check{h}}$ are the vertices in $\bigcup_{w \in \text{OPT} \cap V_{\leq \check{h}}} P_w$, which is precisely $V_{\text{core}}^{\check{h}}$.

To summarize, we define the thinned core $V_{\text{thin}} \subseteq V_{\text{core}}$ as follows.

**Definition 3.7.** *Let $\mathcal{B} \subseteq V_{\text{core}}$ be the set of vertices with degree at least three in the subtree defined by $V_{\text{core}}$ and set $\kappa := \lceil \log_{1+\varepsilon}\left(\frac{(1+3\varepsilon)N}{\varepsilon^2}\right) \rceil$. We define the* thinned core *as*

$$V_{\text{thin}} := \left(\left\{v \in V_{>\check{h}} : T_{v,\kappa} \cap \left((\text{OPT} \cap V_{\leq \hat{h}}) \cup \mathcal{B}\right) \neq \emptyset\right\} \cup V_{\text{core}}^{\check{h}}\right) \setminus \text{OPT},$$

*where $T_{v,\kappa} := T_v \cap V_{\leq \ell+\kappa}$ for $v \in V_\ell$.*

A visualization of $V_{\text{thin}}$ compared to $V_{\text{core}}$ is given in Figure 1. We call vertices in $V_{\text{dropped}} := V_{\text{core}} \setminus V_{\text{thin}}$ *dropped vertices*. Note that $|V_{\text{thin}}| = O_\varepsilon(L)$ as desired. In particular, we can efficiently enumerate for $|V_{\text{thin}}|$ many vertices whether they belong to the thinned core or not.

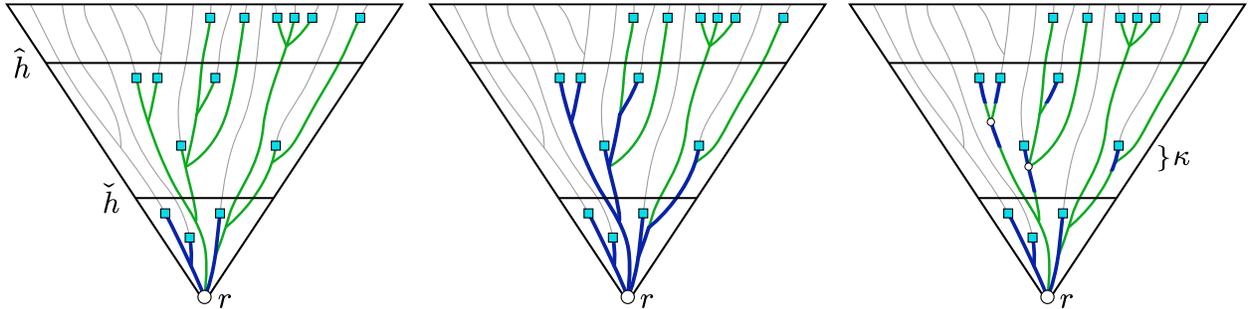

Figure 1: Illustration of the different sets we guess on. The turquoise squares indicate vertices in OPT. In the left figure, we mark the set $V_{\text{core}}^h$ for $h = \check{h}$ in blue. For this choice of $h$, we can efficiently explore $V_{\text{core}}^h$, leading to a $(2+\varepsilon)$-approximation algorithm for SRMFC-T when using Algorithm 2. To obtain a PTAS for SRMFC-T, we would like to set $h = \hat{h}$ in Algorithm 2. Extending $V_{\text{core}}^h$ to height $h = \hat{h}$ leads to the blue marked vertices in the middle figure. However, this set can have size $\Omega_\varepsilon(L \log L)$ which is too large for efficient exploration. For this reason, we thin out the core to the set $V_{\text{thin}}$ indicated in blue in the right figure. The small white cycles indicate the vertices on levels $\{\check{h}+1, ..., \hat{h}\}$ with degree at least three in the subtree defined by $V_{\text{core}}$, i.e., the set $\mathcal{B} \cap V_{>\check{h}}$.

We adapt Algorithm 2 to work with the thinned core $V_{\text{thin}}$ instead of $V_{\text{core}}$. To this end, we need to define how to deal with dropped vertices $v \in V_{\text{dropped}}$. For a vertex $v \in V_{\text{dropped}}$, we define

$$D_{\text{dropped}}(v) := \begin{cases} P_v \cup (T_v \cap V_{\leq \check{h}}) & \text{if } v \in V_{\leq \check{h}}, \\ P_v \cup (T_{v,\kappa} \cap V_{\leq \hat{h}}) & \text{if } v \in V_{>\check{h}}. \end{cases}$$

According to the two structural cases discussed above (distinguishing between $v \in V_{\leq \check{h}}$ and $v \in V_{>\check{h}}$), we can safely add $D_{\text{dropped}}(v)$ to $D$. This results in the following refinement of Algorithm 2.

**Algorithm 3:** THINNED-EXPLORATION-WITH-MIXING$(D, \gamma, Y)$

---

**if** $\gamma > 0$ and $Q_1^D(\Gamma) \neq \emptyset$ **then**
    Determine a point $x \in Q_1^D(\Gamma)$.
    $y \leftarrow$ Sparsification of $x$ through Lemma 3.1.
    **for** any disjoint $A_{\text{thin}}, A_{\text{top}}, A_{\text{dropped}} \subseteq \text{supp}(y) \cap V_{\leq \hat{h}}$ **do**
        $D \leftarrow D \cup A_{\text{thin}} \cup D_{\text{top}}(A_{\text{top}}) \cup D_{\text{dropped}}(A_{\text{dropped}})$.
        **if** $A_{\text{thin}} = \emptyset$ **then** add $y$ to $Y$.
        **if** $|Y| = N$ **then**
            Add $\left(\Gamma \setminus \Gamma_{\text{top}}(\overline{Y}), \Gamma_{\text{top}}(\overline{Y})\right)$ to output.
        **else**
            THINNED-EXPLORATION-WITH-MIXING$(D, \gamma - 1, Y)$.
        **end**
    **end**
**end**

---

Akin to before, we call an iteration $V_{\text{thin}}$-*free* if $A_{\text{thin}} = \emptyset$ and $V_{\text{thin}}$-*overlapping* otherwise. Let us give a bit of intuition of why Algorithm 3 finds a partition of $\Gamma$ satisfying the desired properties. We run the algorithm until we encountered $N = \lceil \frac{4}{\varepsilon} \rceil$ many $V_{\text{thin}}$-free iterations. Let $Y$ be the collection of sparsified points computed in the $N$ many $V_{\text{thin}}$-free iterations when following the execution path where all guesses are compatible with OPT, leading to the mixed solution $\overline{Y} = \frac{1}{N} \sum_{y \in Y} y$. Following the reasoning in Section 3.2, $(1 + 2\varepsilon)\overline{Y}$ restricted to $V_{>\hat{h}}$ is a fractional solution to the top instance for stretch $(1 + O(\varepsilon))$. For $t \in \Gamma \setminus \Gamma_{\text{top}}(\overline{Y})$, there must be at least four $V_{\text{thin}}$-free iterations in which $y(P_t \cap V_{>\hat{h}}) < 1$, as $\overline{Y}$ is the average of $N = \lceil \frac{4}{\varepsilon} \rceil$ many LP solutions. Note that in each of these iterations, there is a vertex in $\text{supp}(y) \cap P_t \cap V_{\leq \hat{h}}$. By the way we update $D$, we can show that $t$ is protected by OPT $\cap V_{\leq \hat{h}}$ or by one of the few additional vertices we fireproof using Observation 3.6. Thus, we can show that the bottom instance is $(1 + O(\varepsilon))$-feasible. We will make this argument formal in Section 4 when proving that our final algorithm stated in Section 3.3.2 computes a partition of the leaves with the desired properties.

Although the running time is already significantly smaller than the running time of Algorithm 2 for $h = \hat{h}$, it is still quasi-polynomial in the number of vertices. The reason why this algorithm is not yet efficient is that we might have only few vertices of $V_{\text{thin}}$ in the support of $y$ in each $V_{\text{thin}}$-overlapping iteration. This could lead to $|V_{\text{thin}}| = \Omega_\varepsilon(L)$ many $V_{\text{thin}}$-overlapping iterations, each of which could generate $4^{|\text{supp}(y) \cap V_{\leq \hat{h}}|} = 4^{\Omega_\varepsilon(L)}$ many subproblems. We address this issue in the following subsection, which finally leads to a PTAS for SRMFC-T.

### 3.3.2 Efficient LP-guided guessing through bulk guessing

The question arises, how can we deal with the case when the support of $y$ contains only few vertices of $V_{\text{thin}}$? Recall that we have to guarantee the existence of a $(1 + O(\varepsilon))$-feasible solution to the bottom instance. Consider an iteration in which $|\text{supp}(y) \cap V_{\text{thin}}|$ is small. The key observation is that in this case, fireproofing all vertices of $\text{supp}(y) \cap V_{\text{thin}}$ only fireproofs a small number of vertices, while allowing us to proceed as in $V_{\text{thin}}$-free iterations.

However, one has to be a bit more careful. The intersection of the support with $V_{\text{thin}}$ may be small overall, but it may still be too large on some levels. In order to stay within a $(1 + O(\varepsilon))$-violation, we analyze the size of $\text{supp}(y) \cap V_{\text{thin}}$ level by level. For this reason, we guess for each level $\ell \in [h]$ whether many support vertices on this level are in $V_{\text{thin}}$. If we guess that this is the case, we perform the enumeration only for support vertices on this level. We call this *bulk guessing* on level $\ell$ and call the corresponding iteration of the algorithm $V_{\text{thin}}$-*heavy*. We choose the following threshold: We perform bulk guessing on level $\ell$, if $|\text{supp}(y) \cap V_{\text{thin}} \cap V_\ell| \geq \frac{\varepsilon}{N} B_\ell$. If there is no level on which we can perform bulk guessing, we continue as in $V_{\text{thin}}$-free iterations before, and observe in the analysis that fireproofing the vertices of $\text{supp}(y) \cap V_{\text{thin}}$ can be done cheaply. We call such an iteration $V_{\text{thin}}$-*light*. Note that in these iterations, $|\text{supp}(y) \cap V_{\text{thin}} \cap V_\ell | < \frac{\varepsilon}{N} B_\ell$ for every $\ell \in [\hat{h}]$. As there are $N$ many $V_{\text{thin}}$-light iterations, we fireproof at most $\varepsilon B_{\leq \ell}$ many additional vertices up to level $\ell$ this way.

We remark that we might need many $V_{\text{thin}}$-heavy iterations until we encounter $N$ fractional solutions with small level-wise overlap with $V_{\text{thin}}$. However, we can bound the number of guesses summed over all $V_{\text{thin}}$-heavy iterations in the execution path compatible with OPT. Thus, we introduce a budget $\zeta$ for the number of guesses we perform in total during $V_{\text{thin}}$-overlapping iterations. If we perform a $V_{\text{thin}}$-overlapping iteration for level $\ell$, we decrease $\zeta$ by $|\text{supp}(y) \cap V_\ell|$. We have to choose $\zeta$ large enough so that the algorithm explores the execution path compatible with OPT, yet small enough to guarantee polynomial running time. To do so, we can exploit that $|V_{\text{thin}}| = O_\varepsilon(L)$. We combine this with the observation that we can relate the number of sets we enumerate on the execution path compatible with OPT and the number of vertices in $V_{\text{thin}}$ we add to $D$, using the lower bounds on $y_v$ for $v \in \text{supp}(y) \cap V_{\leq \hat{h}}$ given by Lemma 3.1. Our final algorithm, leading to a PTAS for SRMFC-T, is shown in Algorithm 4.

---

**Algorithm 4:** EFFICIENT-THINNED-EXPLORATION$(D, \zeta, Y)$

1: **if** $Q_1^D(\Gamma) \neq \emptyset$ **then**
2:     Determine a point $x \in Q_1^D(\Gamma)$.
3:     $y \leftarrow$ Sparsification of $x$ through Lemma 3.1.
4:     **for** each $\ell \in [\hat{h}]$ with $|\text{supp}(y) \cap V_\ell| \leq \zeta$ **do**
5:        **for** each $A_{\text{thin}} \subseteq \text{supp}(y) \cap V_\ell$ with $|A_{\text{thin}}| \geq \frac{\varepsilon}{N} B_\ell$ **do**
6:           $D \leftarrow D \cup A_{\text{thin}}$.
7:           EFFICIENT-THINNED-EXPLORATION$(D, \zeta - |\text{supp}(y) \cap V_\ell|, Y)$.
8:        **end**
9:     **end**
10:    Add $y$ to $Y$.
11:    **if** $|Y| = N$ **then**
12:        Add $\left(\Gamma \setminus \Gamma_{\text{top}}(\overline{Y}), \Gamma_{\text{top}}(\overline{Y})\right)$ to output.
13:    **else**
14:        **for** any disjoint $A_{\text{top}}, A_{\text{dropped}} \subseteq \text{supp}(y) \cap V_{\leq \hat{h}}$ **do**
15:           $D \leftarrow D \cup D_{\text{top}}(A_{\text{top}}) \cup D_{\text{dropped}}(A_{\text{dropped}})$.
16:           EFFICIENT-THINNED-EXPLORATION$(D, \zeta, Y)$.
17:        **end**
18:    **end**
19: **end**

---

For clarity and convenience, we repeat the definitions of $D_{\text{top}}, D_{\text{dropped}},$ and $\Gamma_{\text{top}}$ here. Recall that for $A \subseteq V_{\leq \hat{h}}$, we defined

$$D_{\text{top}}(A) := \bigcup_{v \in A} \left(P_v \cup \left(T_v \cap V_{\leq \hat{h}}\right)\right)$$

and

$$D_{\text{dropped}}(A) := \left(\bigcup_{v \in A \cap V_{\leq \check{h}}} P_v \cup \left(T_v \cap V_{\leq \check{h}}\right)\right) \cup \bigcup_{v \in A \cap V_{> \check{h}}} \left(P_v \cup \left(T_{v, \kappa} \cap V_{\leq \hat{h}}\right)\right).$$

Furthermore, for $\alpha \in \mathbb{R}_{>0}$ and $y \in Q_\alpha(\Gamma)$, we defined

$$\Gamma_{\text{top}}(y) := \left\{ t \in \Gamma : y(P_t \cap V_{>\hat{h}}) \geq 1 - \varepsilon \right\}.$$

We have now assembled all ingredients to prove correctness and efficiency of our algorithm. Let

$$\overline{\zeta} := \frac{2N}{\varepsilon^3}\left(\check{h}^2 B_{\leq \check{h}} + 2\kappa B_{\leq \hat{h}}\right).$$

**Theorem 3.8.** *Let $\varepsilon \in \left(0, \frac{1}{7}\right]$ and let $(G, r, B)$ be an $\varepsilon$-compressed 1-feasible SRMFC-T instance. Then, for one partition $\left(\Gamma_{\mathrm{bot}}, \Gamma_{\mathrm{top}}\right)$ of $\Gamma$ in the output of Algorithm 4, run with $D = \emptyset$, $Y = \emptyset$, $\zeta = \overline{\zeta}$, and $N = \left\lceil \frac{4}{\varepsilon} \right\rceil$,*
- *the bottom instance is $(1 + 2\varepsilon)$-feasible, and*
- *the top instance is $(1 + 7\varepsilon)$-feasible.*

The formal proof of Theorem 3.8 is provided in Section 4.

Finally, we also achieve polynomial running time. Note that it suffices to bound the number of (recursive) calls to Algorithm 4 over the whole run of the algorithm. Indeed, we can efficiently compute a point in the polytope $Q_1^D(\Gamma)$ or decide that it is empty. Moreover, Lemma 3.1 guarantees polynomial running time for the sparsification step. We give the proof of the following theorem in Section 5.

**Theorem 3.9.** *The number of calls to Efficient-Thinned-Exploration over the whole run of Efficient-Thinned-Exploration$(\emptyset, \zeta, \emptyset)$ for $\zeta \in \mathbb{Z}_{\geq 1}$ is bounded by $2^{O_\varepsilon(N(\zeta + L))}$.*

Note that Theorem 3.8 and Theorem 3.9 together imply our main result, Theorem 2.7, as $N = O_\varepsilon(1)$ and $\overline{\zeta} = O_\varepsilon(L)$. Hence the runtime is polynomial in $|V|$ for every fixed $\varepsilon$.

# 4 Proof of correctness (Theorem 3.8)

In this section, we prove Theorem 3.8. We follow the motivation given in Section 3 and closely analyze one specific execution path of Algorithm 4. We will show that for the partition $\Gamma = \Gamma_{\mathrm{bot}} \cup \Gamma_{\mathrm{top}}$ resulting from following this execution path, both the bottom and the top instance are $(1 + O(\varepsilon))$-feasible, as desired. To do so, let OPT denote a fixed 1-feasible solution to the instance. We recall that, for this OPT, we define $\mathcal{B}$ to be the vertices of degree at least 3 in the subtree induced by $V_{\mathrm{core}}$ and

$$V_{\mathrm{thin}} := \left(\left\{v \in V_{>\check{h}} : T_{v,\kappa} \cap \left(\left(\mathrm{OPT} \cap V_{\leq \hat{h}}\right) \cup \mathcal{B}\right) \neq \emptyset\right\} \cup V_{\mathrm{core}}^{\check{h}}\right) \setminus \mathrm{OPT}.$$

Additionally, we will build up a set $R_{\mathrm{add}}$, which we will add to $\mathrm{OPT} \cap V_{\leq \hat{h}}$ in the analysis to show that the bottom instance is $(1 + 2\varepsilon)$-feasible. We will show that for the execution path we investigate, the set $D$ satisfies $D \cap \mathrm{OPT} = \emptyset$ at all times. In this case we say that $D$ is OPT-*compatible*.

Now we describe the execution path we consider, called the OPT-*execution path*, in detail. Consider a run of Efficient-Thinned-Exploration. Note that by our assumption, $Q_1^D(\Gamma)$ is nonempty in each iteration, as $\chi^{\mathrm{OPT}}$ remains feasible. Hence, we never close the execution path in line 1 of Efficient-Thinned-Exploration for any OPT-compatible set $D$.

Let $y$ be the sparsified point in the polytope $Q_{1+3\varepsilon}^D(\Gamma)$ computed in line 3 of Algorithm 4. Let $S := \mathrm{supp}(y) \cap V_{\leq \hat{h}}$ denote the support of $y$ on the bottom $\hat{h}$ levels, and for $\ell \in [L]$, let $S_\ell := S \cap V_\ell$ denote the level-$\ell$-vertices in the support of $y$. If $|S_\ell \cap V_{\mathrm{thin}}| \geq \frac{\varepsilon}{N} B_\ell$ for some $\ell \in \left[\check{h}\right]$, consider an arbitrary such $\ell$. Then we follow the execution path calling Efficient-Thinned-Exploration in line 7 of the algorithm. We will show later that this execution path indeed exists, i.e., $|S_\ell| \leq \zeta$ at this point. For the specific $\ell$ we chose the subset $A_{\mathrm{thin}} := S_\ell \cap V_{\mathrm{thin}}$. Then $D$ stays OPT-compatible after updating $D = D \cup A_{\mathrm{thin}}$.

If we have $|S_\ell \cap V_{\mathrm{thin}}| < \frac{\varepsilon}{N} B_\ell$ for every $\ell \in \left[\check{h}\right]$ and $|Y| < N$, then we follow the call of Efficient-Thinned-Exploration in line 16 of the algorithm, performing a $V_{\mathrm{thin}}$-light iteration with the following subsets of $S$:

$$A_{\mathrm{top}} = \left\{v \in S : (P_v \cup T_v) \cap \mathrm{OPT} \subseteq V_{>\hat{h}}\right\}, \quad \text{and} \quad A_{\mathrm{dropped}} = S \cap V_{\mathrm{dropped}}.$$

In other words, $A_\text{top}$ contains the vertices $v \in S$ for which OPT protects the leaves in $T_v$ using only vertices of level greater than $\hat{h}$. The vertices of $A_\text{dropped}$ are the vertices $v \in S \cap V_{\leq \check{h}}$ for which OPT protects all leaves in $T_v$ using vertices of level greater than $\check{h}$, together with the vertices $v \in S \cap V_{>\check{h}}$ for which there is a unique vertex $w \in T_{v,\kappa} \cap V_{\ell+\kappa}$, where $\ell$ is the level on which $v$ lies, with $T_w \cap \text{OPT} \cap V_{\leq \hat{h}} \neq \emptyset$.

Recall that we update $D$ by adding $D_\text{top}(A_\text{top}) \cup D_\text{dropped}(A_\text{dropped})$ to it. In particular, $D$ remains OPT-compatible after this update, by definition of $A_\text{top}$ and $A_\text{dropped}$ as well as the definition of $D_\text{top}$ and $D_\text{dropped}$. We update $R_\text{add}$ as follows: Add $V_\text{thin} \cap S$ to $R_\text{add}$. Furthermore, for $v \in A_\text{dropped} \cap V_{>\check{h}}$, add to $R_\text{add}$ the unique vertex $w \in T_{v,\kappa} \cap V_{\ell+\kappa}$ with $T_w \cap \text{OPT} \cap V_{\leq \hat{h}} \neq \emptyset$, where $\ell$ is the level on which $v$ lies, i.e., $v \in V_\ell$. Note that such a unique vertex $w$ exists by definition of $V_\text{thin}$. Also note that by definition $v \in V_{\leq \hat{h}-\kappa}$, and thus $w \in V_{\leq \hat{h}}$.

This finishes the description of the execution path that we will now analyze. We denote by $\left(\Gamma_\text{bot} := \Gamma \setminus \Gamma_\text{top}(\overline{Y}), \Gamma_\text{top} := \Gamma_\text{top}(\overline{Y})\right)$ the partition of the leaves $\Gamma$ that we obtain at the end of the OPT-execution path.

**Lemma 4.1.** *The* OPT-*execution path is an execution path of* EFFICIENT-THINNED-EXPLORATION$\left(\emptyset, \overline{\zeta}, \emptyset\right)$.

*Proof.* We only have to check that in the OPT-execution path, every time we call EFFICIENT-THINNED-EXPLORATION in line 7 of Algorithm 4, we have $|\text{supp}(y) \cap V_\ell| \leq \zeta$. Let $\mathcal{U}$ be the family of all such sets $\text{supp}(y) \cap V_\ell$ encountered during the OPT-execution path. We have to show $\sum_{S \in \mathcal{U}} |S| \leq \overline{\zeta}$. Note that

$$|V_\text{thin} \cap V_{\leq \check{h}}| = |V_\text{core}^{\check{h}}| \leq \check{h} B_{\leq \check{h}}$$

and

$$|V_\text{thin} \cap V_{>\check{h}}| \leq 2\kappa B_{\leq \hat{h}} \ .$$

Let $S \in \mathcal{U}$ and $\ell \in [L]$ with $S = \text{supp}(y) \cap V_\ell$, where $y$ denotes the sparsified fractional solution in the corresponding iteration of the algorithm. By Lemma 3.1 we have

$$|S| \leq \begin{cases} \frac{\check{h}}{\varepsilon} y(V_\ell) \leq \frac{\check{h}}{\varepsilon}(1+3\varepsilon)B_{\leq \ell} & \text{if } \ell \in \left[\check{h}\right], \\ \frac{1}{\varepsilon} y(V_\ell) \leq \frac{1}{\varepsilon}(1+3\varepsilon)B_{\leq \ell} & \text{else.} \end{cases}$$

Note that $B_{\leq \ell} = (1+\varepsilon)^{\ell-1} \leq \frac{1+\varepsilon}{\varepsilon} B_\ell$ for each $\ell \in \left[\hat{h}\right]$, with equality for $\ell > 1$. By definition of $\mathcal{U}$ we have $|S \cap V_\text{thin}| \geq \frac{\varepsilon}{N} B_\ell$. Hence, using $\varepsilon \leq \frac{1}{7}$,

$$\frac{|S|}{|S \cap V_\text{thin}|} \leq \begin{cases} \frac{(1+\varepsilon)(1+3\varepsilon)N\check{h}}{\varepsilon^3} \leq \frac{2N\check{h}}{\varepsilon^3} & \text{if } S \subseteq V_{\leq \check{h}}, \\ \frac{(1+\varepsilon)(1+3\varepsilon)N}{\varepsilon^3} \leq \frac{2N}{\varepsilon^3} & \text{else.} \end{cases}$$

Altogether, this yields

$$\sum_{S \in \mathcal{U}} |S| \leq \frac{2N\check{h}}{\varepsilon^3} |V_\text{thin} \cap V_{\leq \check{h}}| + \frac{2N}{\varepsilon^3} |V_\text{thin} \cap V_{>\check{h}}|$$

$$\leq \frac{2N\check{h}}{\varepsilon^3}\left(\check{h} B_{\leq \check{h}}\right) + \frac{2N}{\varepsilon^3}\left(2\kappa B_{\leq \hat{h}}\right)$$

$$= \overline{\zeta} \ .$$

□

**Lemma 4.2.** *At the end of the* OPT-*execution path, we have $|R_\text{add} \cap V_{\leq \ell}| \leq 2\varepsilon B_{\leq \ell}$ for each $\ell \in \left[\hat{h}\right]$.*

*Proof.* There are two places where we add vertices to $R_{\mathrm{add}}$.

First, in any $V_{\mathrm{thin}}$-light iteration, for each $\ell \in [\check{h}]$, we add $|S_\ell \cap V_{\mathrm{thin}}| < \frac{\varepsilon}{N} B_\ell$ many vertices of level $\ell$ to $R_{\mathrm{add}}$. Summing over all $V_{\mathrm{thin}}$-light iterations yields at most $\varepsilon B_\ell$ vertices of level $\ell$ that we add to $R_{\mathrm{add}}$ this way.

Second, in each $V_{\mathrm{thin}}$-light iteration, for each $\ell \in \{\check{h}+1, ..., \hat{h}\}$ and each vertex $v \in A_{\mathrm{dropped}} \cap V_\ell$, we add at most one vertex of level $\ell + \kappa$ to $R_{\mathrm{add}}$. To charge this addition, we use Observation 3.6. Let $y \in Q^D_{1+3\varepsilon}(\Gamma)$ be the sparsified point in this iteration. Note that by Lemma 3.1, we have $|A_{\mathrm{dropped}} \cap V_{\leq \ell} \cap V_{>\check{h}}| \leq \frac{y(V_{\leq \ell})}{\varepsilon} \leq \frac{1+3\varepsilon}{\varepsilon} B_{\leq \ell}$ for each $\ell \in [\hat{h} - \kappa]$. Hence, by Observation 3.6 applied to $A_{\mathrm{dropped}} \cap V_{>\check{h}}$, $|R_{\mathrm{add}} \cap V_{\leq \ell+\kappa}|$ increases by at most $(1+\varepsilon)^{-\kappa} \frac{1+3\varepsilon}{\varepsilon} B_{\leq \ell+\kappa}$ in this iteration, for each $\ell \in [\hat{h} - \kappa]$. By choice of $\kappa$, this is at most $\frac{\varepsilon}{N} B_{\leq \ell+\kappa}$. Hence, summing over all $V_{\mathrm{thin}}$-light iterations, we add at most $\varepsilon B_{\leq \ell+\kappa}$ vertices of $V_{\leq \ell+\kappa}$ to $R_{\mathrm{add}}$ this way. □

**Lemma 4.3.** $\left(\mathrm{OPT} \cap V_{\leq \hat{h}}\right) \cup R_{\mathrm{add}}$ *protects all leaves in* $\Gamma_{\mathrm{bot}}$.

*Proof.* Let $t \in \Gamma_{\mathrm{bot}}$ with $P_t \cap \mathrm{OPT} \cap V_{\leq \hat{h}} = \emptyset$. We show that $t$ is protected by $R_{\mathrm{add}}$.

Let $y^1, ..., y^N$ be the sparsified fractional solutions in the $V_{\mathrm{thin}}$-light iterations of the OPT-execution path, encountered in this order. By definition of $\Gamma_{\mathrm{bot}}$, we have $\frac{1}{N}\sum_{i=1}^N y^i(P_t \cap V_{>\hat{h}}) \leq 1 - \varepsilon$. In particular, as $N = \lceil \frac{4}{\varepsilon} \rceil$, we must have $y^i(P_t \cap V_{>\hat{h}}) < 1$ for at least four different $i_1 < ... < i_4 \in [N]$. Hence, for $j = 1, ..., 4$, there is a vertex $v_j \in P_t \cap \mathrm{supp}(y^{i_j})$. If one of these four vertices is in $V_{\mathrm{thin}}$, we are done, as we would have added it to $R_{\mathrm{add}}$ in the corresponding $V$-light iteration.

Hence, assume that each of these four vertices is part of $A_{\mathrm{top}}$ or $A_{\mathrm{dropped}}$ in its corresponding $V$-light iteration. (If neither is the case, then $P_t \cap \mathrm{OPT} \cap V_{\leq \hat{h}} \neq \emptyset$, a contradiction.) Note that $v_1, v_2, v_3 \in A_{\mathrm{dropped}}$ in their corresponding $V$-light iteration. Indeed, if one of $v_1, v_2$, or $v_3$ had been part of $A_{\mathrm{top}}$, then $P_t \cap V_{\leq \hat{h}}$ would have been added to $D$, contradicting $v_4 \in \mathrm{supp}(y^{i_4})$.

Note that this implies
- $T_v \cap \mathrm{OPT} \cap V_{\leq \hat{h}} \neq \emptyset$ for $v \in \{v_1, v_2, v_3\}$, as otherwise one of $v_1, v_2$, and $v_3$ would have been part of $A_{\mathrm{top}}$, and
- $v_2, v_3 \in V_{>\check{h}}$ by our update of $D$ and definition of $D_{\mathrm{dropped}}$.

But this implies that $v_3 \in T_w$, where $w \in T_{v_2,\kappa} \cap V_{\ell+\kappa}$, with $\ell$ being the level on which $v_2$ lies, is the vertex with $T_w \cap \mathrm{OPT} \cap V_{\leq \hat{h}} \neq \emptyset$ that we added to $R_{\mathrm{add}}$. Hence, $t$ is protected by $w$. □

Now, we are ready to prove Theorem 3.8.

*Proof of Theorem 3.8.* By Lemma 4.1, the OPT-execution path is one of the execution paths of EFFICIENT-THINNED-EXPLORATION$(\emptyset, \overline{\zeta}, \emptyset)$. We will show that for the partition $\Gamma = \Gamma_{\mathrm{bot}} \cup \Gamma_{\mathrm{top}}$ generated by the OPT-execution path the bottom instance is $(1 + 2\varepsilon)$-feasible and the top instance is $(1 + 7\varepsilon)$-feasible.

The bottom instance is $(1 + 2\varepsilon)$-feasible by Lemma 4.2 and Lemma 4.3. Hence, it remains to show that the top instance is $(1 + 7\varepsilon)$-feasible. Let $\overline{Y} = \frac{1}{N}\sum_{i=1}^N y^i \in Q_{1+3\varepsilon}(\Gamma)$ be the average of the sparsified fractional solutions in the $V_{\mathrm{thin}}$-light iterations of the OPT-execution path. Let $y_{\mathrm{top}}$ denote its restriction to $V_{>\hat{h}}$. By definition of $\Gamma_{\mathrm{top}}$, we have $\overline{Y}(P_t \cap V_{>\hat{h}}) \geq 1 - \varepsilon$ for each $t \in \Gamma_{\mathrm{top}}$. Hence, $(1 + 2\varepsilon)y_{\mathrm{top}} \in Q_{1+6\varepsilon}(\Gamma_{\mathrm{top}})$, where we use $(1+2\varepsilon) \cdot (1+3\varepsilon) \leq 1 + 6\varepsilon$, which holds because we require $\varepsilon \leq \frac{1}{7}$ in Theorem 3.8 (this inequality actually only needs $\varepsilon \leq \frac{1}{6}$). Now, apply Lemma 2.5 to get a $(1+7\varepsilon)$-feasible solution to the top instance. □

# 5 Proof of runtime (Theorem 3.9)

We will now prove Theorem 3.9. We remark that we opted for a simpler presentation of the algorithm and the analysis rather than optimizing the dependence on $\varepsilon$ in the exponent of the running time. We will first analyze the size of $\mathrm{supp}(y)$, the support of the sparsified fractional solution $y$ computed in line 3 of EFFICIENT-THINNED-EXPLORATION. To be precise, we observe that its size is linear in $L$ (up to a constant depending on $\varepsilon$).

**Observation 5.1.** *By Lemma 3.1, we have* $|\mathrm{supp}(y) \cap V_{\leq \ell}| \leq \check{h}\frac{1+3\varepsilon}{\varepsilon} B_{\leq \ell}$ *for each* $\ell \leq \check{h}$, *and* $|\mathrm{supp}(y) \cap V_{\leq \ell} \cap V_{>\check{h}}| \leq \frac{1+3\varepsilon}{\varepsilon} B_{\leq \ell}$ *for each* $\check{h} < \ell \leq \hat{h}$. *In particular, this implies*

$$|\mathrm{supp}(y) \cap V_{\leq \hat{h}}| \leq \frac{1+3\varepsilon}{\varepsilon}\bigl(\check{h} B_{\leq \check{h}} + B_{\leq \hat{h}}\bigr) \leq \gamma_\varepsilon L,$$

*for a constant* $\gamma_\varepsilon$ *depending only on* $\varepsilon$. *Here we used that* $\hat{h} = O_\varepsilon(\log L)$, $\check{h} = O_\varepsilon(\log \log L)$, *and* $B_{\leq \ell} = (1+\varepsilon)^{\ell-1}$ *for* $\ell \in [L]$.

*Proof of Theorem 3.9.* For $\zeta, s \in \mathbb{Z}_{\geq 0}$ with $s \leq N$, let $\mathcal{P}(\zeta, s)$ denote the maximum number of calls to EFFICIENT-THINNED-EXPLORATION over the whole run of the algorithm when calling EFFICIENT-THINNED-EXPLORATION$(D, \zeta, Y)$, for any choice of $D$ and for any $Y$ with $|Y| = N - s$, including the initial call of EFFICIENT-THINNED-EXPLORATION$(D, \zeta, Y)$. We claim that

$$\mathcal{P}(\zeta, s) \leq C_\varepsilon^\zeta \cdot D_\varepsilon^s$$

for $C_\varepsilon = 2^{\frac{N}{\varepsilon^3}+1}$ and $D_\varepsilon = 3^{\gamma_\varepsilon L + 1}$. Note that this implies the statement of Theorem 3.9.

We prove this claim by induction on $\zeta + s$. For the base case $\zeta = s = 0$, we do not have a recursive call to EFFICIENT-THINNED-EXPLORATION, and thus $\mathcal{P}(\zeta, s) = 1$, which implies the claimed bound. For the induction step, we analyze all recursive calls to EFFICIENT-THINNED-EXPLORATION. Recall that we defined $S_\ell = \mathrm{supp}(y) \cap V_\ell$ for each $\ell \in [L]$ and $S = \mathrm{supp}(y) \cap V_{\leq \hat{h}}$. Let $I \subseteq [L]$ denote the set of indices $\ell \in [L]$ with $|S_\ell| \geq \frac{\varepsilon}{N} B_\ell$ and $|S_\ell| \leq \zeta$. First, for each level $\ell \in I$, we perform a $V_{\mathrm{thin}}$-heavy iteration on level $\ell$ and reduce $\zeta$ by $|S_\ell|$. Thus, we can bound the number of recursive calls to EFFICIENT-THINNED-EXPLORATION in line 7 of Algorithm 4, over all choices of $\ell$, by

$$\sum_{\ell \in I} 2^{|S_\ell|} \mathcal{P}(\zeta - |S_\ell|, s).$$

Applying the induction hypothesis, we can upper bound this by

$$\sum_{\ell \in I} 2^{|S_\ell|} C_\varepsilon^{\zeta - |S_\ell|} D_\varepsilon^s = C_\varepsilon^\zeta D_\varepsilon^s \sum_{\ell \in I} \left(\frac{2}{C_\varepsilon}\right)^{|S_\ell|}. \qquad (4)$$

Note that for $\ell \in I$, we have

$$|S_\ell| \geq \frac{\varepsilon}{N} B_\ell = \frac{\varepsilon}{N}(B_{\leq \ell} - B_{\leq \ell-1}) \geq \frac{\varepsilon}{N}\bigl((1+\varepsilon)^{\ell-1} - (1+\varepsilon)^{\ell-2}\bigr) \geq \frac{\varepsilon^2 (1+\varepsilon)^{\ell-2}}{N}, \qquad (5)$$

where the second inequality above holds with equality if $\ell > 1$ (because $B_{\leq \ell} = (1+\varepsilon)^{\ell-1}$) and is a strict inequality for $\ell = 1$ (because $B_{\leq 0} = 0$). This allows us to upper bound the sum on the right-hand side of (4) by

$$\sum_{\ell \in I} \left(\frac{2}{C_\varepsilon}\right)^{|S_\ell|} \leq \sum_{\ell=1}^{\hat{h}} \left(\frac{2}{C_\varepsilon}\right)^{\frac{\varepsilon^2(1+\varepsilon)^{\ell-2}}{N}} = \sum_{\ell=1}^{\hat{h}} \left(\frac{1}{2}\right)^{\frac{(1+\varepsilon)^{\ell-2}}{\varepsilon}} \leq \sum_{\ell=1}^{\hat{h}} \left(\frac{1}{2}\right)^{\frac{1+(\ell-2)\varepsilon}{\varepsilon}} = \left(\frac{1}{2}\right)^{\frac{1}{\varepsilon}-2} \sum_{\ell=1}^{\hat{h}} \left(\frac{1}{2}\right)^\ell$$

$$\leq \left(\frac{1}{2}\right)^{\frac{1}{\varepsilon}-2} \leq \frac{1}{32},$$

where the first inequality uses (5) and $C_\varepsilon \geq 2$, the first equality follows from $C_\varepsilon = 2^{\frac{N}{\varepsilon^3}+1}$, and the last inequality holds because $\varepsilon \leq \frac{1}{7}$. Thus, together with (4), we can bound the number of recursive calls during the $V_{\text{thin}}$-heavy iterations by

$$\frac{1}{32} C_\varepsilon^\zeta D_\varepsilon^s.$$

Finally, if $s > 0$, we apply a $V_{\text{thin}}$-light iteration on $S$ and reduce $s$ by one. For each choice of two disjoint subsets of $S$ we perform at most $\mathcal{P}(\zeta, s-1)$ recursive calls to Efficient-Thinned-Exploration in line 16 of Algorithm 4. Hence, we can bound the total number of recursive calls in this $V_{\text{thin}}$-light iteration by $3^{|S|}\mathcal{P}(\zeta, s-1)$. By Observation 5.1, we have $|S| \leq \gamma_\varepsilon L$. Combining this with the induction hypothesis, we get

$$3^{|S|}\mathcal{P}(\zeta, s-1) \leq 3^{|S|} \cdot C_\varepsilon^\zeta D_\varepsilon^{s-1} \leq 3^{\gamma_\varepsilon L} \cdot C_\varepsilon^\zeta D_\varepsilon^{s-1} = \frac{C_\varepsilon^\zeta D_\varepsilon^s}{3},$$

by choice of $D_\varepsilon$.

Combining both cases, and adding the initial call to Efficient-Thinned-Exploration, we have shown that

$$\mathcal{P}(\zeta, s) \leq \frac{1}{32} C_\varepsilon^\zeta D_\varepsilon^s + \frac{1}{3} C_\varepsilon^\zeta D_\varepsilon^s + 1 \leq C_\varepsilon^\zeta D_\varepsilon^s$$

as desired, where the second inequality holds because $C_\varepsilon \geq 2$, $D_\varepsilon \geq 2$, and $\zeta + s \geq 1$. □

## 6 Adaption to Non-Uniform $k$-Center

As already mentioned in the introduction, there is a strong link between RMFC-T and the Non-Uniform $k$-Center problem (NUkC). In the following we will introduce NUkC and adapt the approach presented for RMFC-T to NUkC to prove Theorem 1.3. Similar to RMFC-T, we also consider a smooth version of NUkC, which allows to use the budget for a given radius also for smaller radii. For this smooth version, we will show that we can find a solution that violates the budgets by a factor of $1 + \varepsilon$ only and approximates the optimal dilation of the radii by a constant factor. A key technique is an LP-aware reduction from NUkC to RMFC-T introduced by [7]. This reduction allows to extend our sparsity results shown for SRMFC-T, cf. Lemma 3.1, to the smooth NUkC variant. The main additional challenge compared to RMFC-T is that we can no longer exploit the underlying tree structure. New techniques, using the fact that dilating the radii by a constant factor yields similar structural properties, are introduced to overcome this issue. This is then combined with an enumeration approach similar to the one used for RMFC-T.

### 6.1 Problem definition and main result

We begin by formally introducing the Non-Uniform $k$-Center problem (NUkC). An instance $(V, d, k, r)$ of NUkC consists of a metric space $(V, d)$ and a set of radii $r \in \mathbb{R}_{\geq 0}^L$ with budgets $k \in \mathbb{Z}_{\geq 0}^L$ for $L \in \mathbb{Z}_{\geq 1}$. We assume that the radii are sorted in non-increasing order, i.e., $r_\ell \geq r_{\ell+1}$ for each $\ell \in [L-1]$. For brevity, for $\ell \in [L]$ and $A \subseteq V \times [L]$, we write $A_\ell := \{(v, \ell') \in A : \ell' = \ell\}$, and analogously use $A_{\leq \ell}, A_{<\ell}, A_{\geq \ell}$, and $A_{>\ell}$. For $\alpha, \beta \in \mathbb{R}_{\geq 0}$ we say that $(V, d, k, r)$ is $(\alpha, \beta)$-*feasible* if we can cover $V$ by opening at most $\alpha k_\ell$ balls of radius $\beta r_\ell$ for each $\ell \in [L]$. To be precise, seeing a pair $(v, \ell) \in V \times [L]$ as opening a ball of radius $r_\ell$ at $v$, the instance is $(\alpha, \beta)$-feasible if there is a set of pairs $C \subseteq V \times [L]$ such that
- $|C_\ell| \leq \alpha k_\ell$ for each $\ell \in [L]$, and
- $\bigcup_{(v,\ell) \in C} \text{Ball}(v, \beta r_\ell) = V$,

where for $v \in V$ and $r \in \mathbb{R}_{\geq 0}$, $\mathrm{Ball}(v, r) := \{u \in V : d(u, v) \leq r\}$ denotes the ball of radius $r$ around $v$. We also call such a set $C$ an $(\alpha, \beta)$-*feasible solution.* An $(\alpha, \beta)$-approximation algorithm for NUkC is a polynomial-time algorithm that, given an instance $(V, d, k, r)$ of NUkC, either determines that the instance is $(1, 1)$-infeasible, or computes an $(\alpha, \beta)$-feasible solution to the instance. We remark that the original definition of NUkC given in [7] is obtained by setting $\alpha = 1$ and minimizing $\beta$. The definition of $(\alpha, \beta)$-approximation is equivalent to the definition given in [7].[10]

We recall that computing an $(\alpha, \beta)$-approximation for NUkC with $\alpha < 2$ and $\beta \leq 2^{\mathrm{poly}(n)}$ is NP-hard. More precisely, if $k_\ell = 1$ for each $\ell \in [L]$, it is NP-hard to decide whether there is a $(1, \beta)$-feasible solution for any $\beta \leq 2^{\mathrm{poly}(n)}$ [7]. Akin to RMFC-T, we show that hardness for $1 < \alpha < 2$ is an artifact of the level-wise approximation guarantee. To be precise, we introduce a smooth version of NUkC, where we allow to use the budget for a given radius also for smaller radii. Similar to RMFC, this allows to accumulate fractional parts of the budgets.

Let us now formally introduce the *smooth Non-Uniform k-Center problem* (SNUkC). The instance $(V, d, k, r)$ is defined as for NUkC, except that we do not require $k$ to be integral. Moreover, we do not require a bound on the number of centers of a given radius, but rather on the number of centers with at least a given radius: For $\alpha, \beta \in \mathbb{R}_{\geq 0}$, we say that $(V, d, k, r)$ is $(\alpha, \beta)$-*feasible* if there is a set of pairs $C \subseteq V \times [L]$ such that
- $\sum_{j=1}^{\ell} |C_j| \leq \alpha \sum_{j=1}^{\ell} k_j$ for each $\ell \in [L]$, and
- $\bigcup_{(v,\ell) \in C} \mathrm{Ball}(v, \beta r_\ell) = V$.

Again, we call $C$ an $(\alpha, \beta)$-*feasible solution* to the SNUkC problem. An $(\alpha, \beta)$-approximation algorithm for SNUkC is a polynomial-time algorithm that, given an instance $(V, d, k, r)$ of SNUkC, either determines that the instance is $(1, 1)$-infeasible, or computes an $(\alpha, \beta)$-feasible solution.

We remark that for integral $\alpha$ and $k$ the feasible solutions to both problems coincide. Furthermore, if we are only interested in finding $(1, 1)$-feasible solutions, then SNUkC is equivalent to NUkC for budgets $k'_\ell = \left\lfloor \sum_{j=1}^{\ell} k_j \right\rfloor - \left\lfloor \sum_{j=1}^{\ell-1} k_j \right\rfloor$ for each $\ell \in [L]$. There is an even stronger connection between the two problems, as the following observation shows.

**Observation 6.1.** *Let $\alpha, \beta \in \mathbb{R}_{\geq 0}$. An $(\alpha, \beta)$-approximation algorithm for SNUkC implies an $(\lceil \alpha \rceil, \beta)$-approximation algorithm for NUkC.*

*Proof.* Consider an instance $(V, d, k, r)$ of NUkC as an instance of SNUkC. Let $C$ be the $(\alpha, \beta)$-feasible solution computed by the given approximation algorithm for SNUkC. We will modify $C$ such that $|C_\ell| \leq \lceil \alpha k_\ell \rceil$ for each $\ell \in [L]$. By definition of SNUkC, we have

$$\sum_{j=1}^{\ell} |C_j| \leq \sum_{j=1}^{\ell} \alpha k_j$$

for each $\ell \in [L]$. If there is an $\ell \in [L]$ such that $|C_\ell| > \lceil \alpha k_\ell \rceil$, then there must be an $\ell' < \ell$ such that $|C_{\ell'}| < \lceil \alpha k_{\ell'} \rceil$. Replace $(v, \ell)$ in $C$ by $(v, \ell')$ for some $(v, \ell) \in C_\ell$. Iterate this until $|C_\ell| \leq \lceil \alpha k_\ell \rceil$ for each $\ell \in [L]$. □

Using Observation 6.1, we directly reduce Theorem 1.3 to the following theorem for SNUkC.

**Theorem 6.2.** *For any $\varepsilon > 0$, there is a polynomial-time $(1 + \varepsilon, 15 + \varepsilon)$-approximation algorithm for SNUkC.*

---

[10]In the definition of [7], the second approximation factor is defined relative to the best possible dilation of the radii. Yet one can efficiently guess the optimal dilation, as there are at most $\frac{n(n-1)}{2}$ different distances in the metric space and at most $L$ different radii. As the optimal dilation must equal at least one distance to at least one radius, this gives at most $\frac{n(n-1)}{2}L$ different candidates.

Note that Theorem 6.2 provides a constant-factor approximation on the dilation of the radii by opening just slightly more centers than the budgets allow, while taking into account that balls with larger radius are more valuable. More precisely, for every $\ell \in \mathbb{Z}_{>0}$, the algorithm opens at most $(1+\varepsilon)k_{\leq \ell}$ many balls of the largest $\ell$ radii, where $k_{\leq \ell} := \sum_{j=1}^{\ell} k_j$. We believe that this reflects the desired approximation guarantee in many applied settings. For example, if opening centers incurs some (possibly radius-dependent) cost, a level-wise guarantee on the number of centers opened may not be necessary. Therefore, we believe that the smooth version of NUkC is a natural problem in its own right.

The remainder of this section is devoted to the proof of Theorem 6.2.

## 6.2 Ingredients of the algorithm for smooth NUkC

In the following, we give a high-level overview of the main algorithmic ingredients that we will use to prove Theorem 6.2. The details of the adapted enumeration approach follow in Section 6.3.

As for SRMFC-T, we start with compressing the instance to reduce the number of different radii significantly. However, before we do the compression, we slightly round the radii so that two different radii differ by at least a factor $1 + \varepsilon$. This will be useful later to keep the dilation of the radii small, as we can guarantee that radii $\kappa$ levels apart differ by at least a factor $(1+\varepsilon)^\kappa$. Then we essentially use the same compression as for SRMFC-T, resulting in the following definition. Note that compared to Definition 2.3, we cannot guarantee that the budget up to level $\ell$ is given by $(1+\varepsilon)^{\ell-1}$ since we might need to aggregate level with the same radius to maintain the property that two different radii differ by at least a factor $1 + \varepsilon$. We still guarantee that the budget grows at least geometrically.

**Definition 6.3.** *For $\varepsilon > 0$, an SNUkC instance $(V, d, k, r)$ with $L$ levels and $n$ vertices is $\varepsilon$-compressed if*
- $L \leq \lceil \log_{1+\varepsilon}(n) \rceil + 1$,
- $k_1 \geq 1$ and $k_{\ell+1} \geq (1+\varepsilon)k_\ell$ for each $\ell \in [L-1]$, and
- $r_\ell \geq (1+\varepsilon)r_{\ell+1}$ for each $\ell \in [L-1]$.

Akin to SRMFC-T, approximation results for $(1,1)$-feasible $\varepsilon$-compressed instances carry over to general instances. (See Section 7 for the proof of Theorem 6.4.)

**Theorem 6.4.** *Let $\varepsilon > 0$. An $(\alpha, \beta)$-approximation for $(1,1)$-feasible $\varepsilon$-compressed SNUkC implies a $((1+\varepsilon)\alpha, (1+\varepsilon)\beta)$-approximation for (general) SNUkC instances.*

Given a $(1,1)$-feasible $\varepsilon$-compressed instance, let OPT denote a fixed $(1,1)$-feasible solution.

Akin to SRMFC-T, a key ingredient to our approach is the following natural polyhedral formulation for SNUkC. We define the polytope in more generality, allowing a controlled violation of the budgets and dilation of the radii. Additionally, as we will fix parts of the vertices as being covered during the course of the algorithm, we also allow to consider only a subset of the vertices. For this purpose, let $\alpha \in \mathbb{R}_{\geq 0}$, $\beta \in \mathbb{R}_{\geq 0}^L$ and $U \subseteq V$. We define

$$Q_{\alpha,\beta}(U) := \left\{ x \in [0,1]^{V \times [L]} \,\middle|\, \begin{array}{l} \sum_{\ell=1}^{L} \sum_{u \in \text{Ball}(v, \beta_\ell r_\ell)} x_{u,\ell} \geq 1 \quad \forall v \in U \\ \sum_{v \in V} \sum_{\ell'=1}^{\ell} x_{v,\ell'} \leq \alpha k_{\leq \ell} \forall \ell \in [L] \end{array} \right\}.$$

For $\beta' \in \mathbb{R}_{\geq 0}$, we often abbreviate $Q_{\alpha,\beta'}(U) := Q_{\alpha,\beta}(U)$ with $\beta_\ell = \beta'$ for each $\ell \in [L]$. Note that integral points in $Q_{\alpha,\beta}(V)$ correspond to $(\alpha, \beta)$-feasible solutions and vice versa. In an integer solution, the variable $x_{v,\ell}$ tells us whether we open a ball of radius $\beta r_\ell$ at $v$. A very similar formulation was already used for NUkC in [7]. As for SRMFC-T, we will use points in this polytope as guidance for a guessing procedure that identifies certain parts of OPT. We call points in $Q_{\alpha,\beta}(V)$ *fractional $(\alpha, \beta)$-feasible solutions*.

The following lemma, which follows from an LP-aware reduction from NUkC to RMFC-T first used by Chakrabarty, Goyal, and Krishnaswamy [7], shows that only instances with low budget can have a large integrality gap (see Section 9 for the proof).

**Lemma 6.5.** *Let $\varepsilon > 0$ and let $(V, d, k, r)$ be an instance of SNUkC with $L$ levels and $k_\ell \geq \frac{L}{\varepsilon}$ for each $\ell \in [L]$. Let $\alpha, \beta \in \mathbb{R}_{\geq 0}$. If $Q_{\alpha,\beta}(V) \neq \emptyset$, then we can efficiently determine an $((1+\varepsilon)(\alpha + \varepsilon), 8\beta)$-feasible solution.*

An interesting application of Lemma 6.5 is the following. Here and henceforth, for a given $\varepsilon$-compressed SNUkC instance on $L$ levels, let $\hat{h} \in [L]$ be minimal such that $k_{\hat{h}+1} \geq \frac{L}{\varepsilon}$. Assume that we know the set $V_{\text{small}} \subseteq V$ covered by balls of radius smaller than $r_{\hat{h}}$ in OPT. Then we can use Lemma 6.5 to find a $((1+\varepsilon)^2, 8)$-feasible solution for $V_{\text{small}}$, even when restricting $k$ and $r$ to levels $\{\hat{h}+1, ..., L\}$.

Even if we did know $V_{\text{small}}$, the question remains how to cover $V \setminus V_{\text{small}}$ using only balls of radius at least $r_{\hat{h}}$. In contrast to SRMFC-T, we do not know how to solve this with a simple dynamic program. This is due to the lack of an underlying tree structure, which gave a laminar structure in SRMFC-T. Although we still end up with a dynamic programming approach, it will heavily rely on guesses made during our enumeration. For this reason, in contrast to our enumeration approach for SRMFC-T, we will not return a partition of the vertices, but instead move the dynamic program and the LP rounding into the enumeration approach.

To be precise, we will show the following theorem.

**Theorem 6.6.** *Let $(V, d, k, r)$ be a $(1, 1)$-feasible $\varepsilon$-compressed instance of SNUkC. Then there is an efficient algorithm that computes a $(1 + 14\varepsilon, 15 + 6\varepsilon)$-feasible solution to $(V, d, k, r)$.*

Note that our main result for SNUkC, namely Theorem 6.2, is a direct consequence of Theorem 6.4 and Theorem 6.6.

In the next subsection, we explain how to adapt the enumeration we used for SRMFC-T to prove Theorem 6.6.

## 6.3 Enumeration approach for SNUkC

Let $(V, d, k, r)$ be an $\varepsilon$-compressed instance of SNUkC, for some $\varepsilon \in (0, 1]$. Assume that the instance is $(1, 1)$-feasible, and let OPT be a corresponding solution. We will show how to compute a $(1 + O(\varepsilon), 15 + O(\varepsilon))$-feasible solution, using an enumeration approach similar to the one used for SRMFC-T. Throughout the presentation we will highlight the additional challenges that arise when adapting the algorithm to SNUkC.

Implicitly, the approach for SRMFC-T exploited in many ways that the tree guarantees a laminar structure on the leaves. To be precise, let $(G, r, B)$ be an instance of SRMFC-T. For every vertex $v$ of $G$ let $\Gamma_v$ denote the set of leaves in the subtree $T_v$ of $G$ rooted at $v$. Then the sets $(\Gamma_v)_{v \in V}$ form a laminar family. For SNUkC, we do not have such a structure. In particular, balls of radius $r_\ell$ around different vertices $v$ might intersect in complicated ways. We will compensate for this by allowing balls of dilated radius to be opened at vertices close to centers in OPT, so that vertices covered by this center in OPT are covered by the dilated ball. As we will see, this idea can be used to sparsify fractional solutions, which bounds the width of our enumeration. We can also use this idea to bound the depth of our enumeration exploiting that, instead of identifying centers in OPT, it suffices to identify nearby vertices.

Let us now describe the enumeration procedure in more detail. We will achieve a dilation of the radii by a factor of $\rho := 15 + 6\varepsilon$. We proceed in multiple iterations, maintaining sets $C, D \subseteq V \times [L]$. The set $C$ contains pairs $(v, \ell)$ for which we commit to open a ball of radius $\rho r_\ell$ at $v$ in the final solution. The set $D$ on the other hand contains pairs $(v, \ell)$ for which we guarantee that we will not open a ball on level $\ell$ at $v$ in the final solution. As such, we say that a point $x \in Q_{\alpha,\beta}(U)$ is $(C, D)$-*compatible* if
- $x(D) = 0$, and

- $\sum_{v \in V} x_{v,\ell} \leq \alpha k_\ell^C$ for each $\ell \in [L]$, where $k_\ell^C := k_\ell - |C_\ell|$ are the adjusted budgets assuming that we already opened centers for pairs in $C$.

In each iteration of the enumeration procedure, we compute a $(C, D)$-compatible point $x \in Q_{1,1}(V \setminus V(C))$, where $V(C)$ is the set of vertices covered by the centers in $C$ when dilating their radii by factor $\rho$, namely

$$V(C) := \bigcup_{(v,\ell) \in C} \mathrm{Ball}(v, \rho r_\ell).$$

Similar to SRMFC-T, we first sparsify the point $x$, relative to two thresholds. Let $\hat{h} \in [L]$ be minimal such that $k_{\hat{h}+1} \geq \frac{L}{\varepsilon}$, and additionally, let $\check{h} \in [L]$ be minimal such that $k_{\check{h}+1} \geq \frac{\hat{h}}{\varepsilon}$. Note that $\hat{h} \leq \lceil \log_{1+\varepsilon}(\frac{L}{\varepsilon^2}) \rceil + 1$ and $\check{h} \leq \lceil \log_{1+\varepsilon}(\frac{\hat{h}}{\varepsilon^2}) \rceil + 1$ by definition of $\varepsilon$-compressed instances. Then we distinguish between levels $\ell \leq \check{h}$, levels $\check{h} < \ell \leq \hat{h}$, and levels $\ell > \hat{h}$.

**Lemma 6.7.** *Let $0 < \varepsilon \leq \frac{1}{7}$ and $\lambda \in \mathbb{Z}_{>0}$. Consider an $\varepsilon$-compressed instance $(V, d, k, r)$ of SNUkC. Let $U \subseteq V$. For $x \in Q_{1,1}(U)$, we can efficiently compute a point $y \in Q_{1+7\varepsilon, \beta(\lambda)}(U)$ with*

$$\beta(\lambda)_\ell := \begin{cases} 1 + \frac{2}{1-2(1+\varepsilon)^{-\lambda}} & \text{if } h \in [\hat{h}] \\ 1 & \text{else .} \end{cases}$$

*for each $\ell \in [L]$ such that*
- $\mathrm{supp}(y) \subseteq \mathrm{supp}(x)$,
- $y_{v,\ell} \geq \frac{\varepsilon}{\lambda \check{h}}$ *for every* $(v, \ell) \in \mathrm{supp}(y)$ *with* $\ell \leq \check{h}$,
- $y_{v,\ell} \geq \frac{\varepsilon}{\lambda}$ *for every* $(v, \ell) \in \mathrm{supp}(y)$ *with* $\check{h} < \ell \leq \hat{h}$.

The proof of this lemma is postponed to Section 9. We want to bound the factor we lose on the dilation of the radii due to the sparsity guarantee. To do so, first note that $\lim_{\lambda \to \infty} 1 + \frac{2}{1-2(1+\varepsilon)^{-\lambda}} = 3$. Let $\lambda_\varepsilon$ be the smallest possible choice of $\lambda$ so that $1 + \frac{2}{1-2(1+\varepsilon)^{-\lambda}} \leq 3 + \frac{\varepsilon}{2}$. Set $\sigma := 3 + \varepsilon$.

For this choice of $\lambda$, the algorithm repeatedly computes a sparse $(C, D)$-compatible point $y \in Q_{1+\varepsilon, \beta(\lambda)}(V \setminus V(C))$ by using Lemma 6.7. Consider a pair $(v, \ell) \in \mathrm{supp}(y)$ with $\ell \leq \hat{h}$. We would like to guess for each vertex $u \in \mathrm{Ball}(v, \beta_\ell r_\ell)$ whether $u$ is covered by some ball of radius at least $r_{\hat{h}}$ in OPT. If we could afford to do so, we could compute a partition of the vertices into two sets $V_{\mathrm{big}}$ and $V_{\mathrm{small}}$, so that $V_{\mathrm{big}}$ admits a $(1, 1)$-feasible solution when only using radii of size at least $r_{\hat{h}}$, and $V_{\mathrm{small}}$ admits a good fractional solution when only using radii smaller than $r_{\hat{h}}$, akin to Lemma 3.2 for SRMFC-T. However, guessing this for each vertex separately results in exponentially many outcomes. Hence, we guess how OPT covers the vertices of $\mathrm{Ball}(v, \sigma r_\ell)$ by identifying structural cases. Note that $\mathrm{Ball}(v, \sigma r_\ell)$ contains all vertices that get coverage by $(v, \ell) \in \mathrm{supp}(y)$, plus some small buffer zone that will be used later to compensate for the non-laminar structure of balls on different levels.

We define a partition of $\mathrm{supp}(y)_{\leq \hat{h}}$ into four sets $A_{\mathrm{big}}$, $A_{\mathrm{small}}$, $A_{\mathrm{sep}}$, and $A_{\mathrm{non\text{-}sep}}$, depending on how a pair $(v, \ell)$ lies relative to centers in OPT. While we cannot compute this partition directly, as we do not know OPT, we will later enumerate over all possible partitions. We define the parts of the partition as follows.

(i) $A_{\mathrm{big}}$ contains the pairs $(v, \ell) \in \mathrm{supp}(y)$ such that there is some $(w, \ell') \in \mathrm{OPT}$ with $w \in \mathrm{Ball}(v, \sigma(r_\ell + r_{\ell'}))$ and $\ell' \leq \ell$.

(ii) $A_{\mathrm{small}}$ contains the pairs $(v, \ell) \in \mathrm{supp}(y) \setminus A_{\mathrm{big}}$ such that there is no $(w, \ell') \in \mathrm{OPT}$ with $w \in \mathrm{Ball}(v, \sigma(r_\ell + r_{\ell'}))$ and $\ell' \leq \hat{h}$.

(iii) $A_{\mathrm{sep}}$ contains the pairs $(v, \ell) \in \mathrm{supp}(y) \setminus (A_{\mathrm{big}} \cup A_{\mathrm{small}})$ with
- $\ell \leq \check{h}$, and there is no $(w, \ell') \in \mathrm{OPT}$ with $w \in \mathrm{Ball}(v, \sigma(r_\ell + r_{\ell'}))$ and $\ell' \leq \check{h}$, or

- $\ell > \check{h}$, and there is no $(w, \ell') \in \text{OPT}$ with $w \in \text{Ball}(v, \sigma(r_\ell + r_{\ell'}))$ and $\ell' \leq \ell + \kappa$, and additionally, for each pair of centers $(w_1, \ell_1), (w_2, \ell_2) \in \text{OPT}$ with $w_i \in \text{Ball}\big(v, \sigma\big(r_\ell + r_{\ell_i}\big)\big)$ and $\ell_i \leq \hat{h}$ for $i \in \{1, 2\}$, we have $d(w_1, w_2) \leq \mu r_{\ell+\kappa}$. We choose $\mu := 2$ and

$$\kappa := \left\lceil \max\left\{ \log_{1+\varepsilon}\left(\frac{4\sigma}{\varepsilon}\right), 2\log_{1+\varepsilon}(4\sigma + 2) \right\} \right\rceil.$$

Set $A_{\text{non-sep}} := \text{supp}(y)_{\leq \hat{h}} \setminus \big(A_{\text{big}} \cup A_{\text{small}} \cup A_{\text{sep}}\big)$. We say that pairs in $A_{\text{non-sep}}$ are *non-separable*. First, we investigate the structure of non-separable pairs.

**Observation 6.8.** *Let $(v, \ell) \in \text{supp}(y)$ be non-separable. Then one of the following properties holds:*
  (i) $\ell \leq \check{h}$, *and there is some* $(w, \ell') \in \text{OPT}$ *with* $w \in \text{Ball}(v, \sigma(r_\ell + r_{\ell'}))$ *and* $\ell < \ell' \leq \check{h}$.
  (ii) $\ell > \check{h}$, *and there is some* $(w, \ell') \in \text{OPT}$ *with* $w \in \text{Ball}(v, \sigma(r_\ell + r_{\ell'}))$ *and* $\ell < \ell' \leq \ell + \kappa$.
  (iii) $\ell > \check{h}$, *and there are centers* $(w_1, \ell_1), (w_2, \ell_2) \in \text{OPT}$ *with* $w_i \in \text{Ball}\big(v, \sigma\big(r_\ell + r_{\ell_i}\big)\big)$ *and* $\ell + \kappa < \ell_i \leq \hat{h}$ *for* $i \in \{1, 2\}$, *such that* $d(w_1, w_2) > \mu r_{\ell+\kappa}$.

The pairs in $A_{\text{sep}}$ correspond to vertices in $A_{\text{dropped}}$ for SRMFC-T and the vertices in $A_{\text{non-sep}}$ correspond to $V_{\text{thin}}$. Yet, there is a crucial difference to SRMFC-T. Due to the tree structure, we have been able to show that the number of vertices in $V_{\text{thin}}$ is linear in the number of levels in the case of SRMFC-T. This is not possible for $A_{\text{non-sep}}$, as there could be many vertices in $A_{\text{non-sep}}$ that are all close to the same center in OPT. For this reason, we have to further sparsify $A_{\text{non-sep}}$.

Let $A_{\text{thin}} \subseteq A_{\text{non-sep}}$ be maximal such that for any two pairs $(v, \ell), (v', \ell) \in A_{\text{thin}}$, we have $d(v, v') > 4\sigma r_\ell$. Note that $A_{\text{thin}}$ depends on the sparsified fractional solution $y$. We observe the following.

**Observation 6.9.**
  (i) *For any $(v, \ell) \in A_{\text{non-sep}}$, there exists a pair $(v', \ell) \in A_{\text{thin}}$ such that $d(v, v') \leq 4\sigma r_\ell$.*
  (ii) *For any $\ell \in [\hat{h}]$ and any pair $(w, \ell') \in \text{OPT}_{\leq \hat{h}}$ with $\ell' \geq \ell$ there is at most one pair $(v, \ell) \in A_{\text{thin}}$ such that $d(v, w) \leq \sigma(r_\ell + r_{\ell'})$.*

The first property will be useful to cover $A_{\text{non-sep}}$ by opening a small number of centers in $A_{\text{thin}}$. The second property allows to associate, for each $\ell \in [L]$, each pair in $\text{OPT}_{\leq \hat{h}}$ with at most one pair $(v, \ell) \in A_{\text{thin}}$. This will be crucial to bound the running time of our algorithm, as it limits the number of pairs in $A_{\text{thin}}$ that we have to consider over the course of the algorithm.

Akin to the algorithm for SRMFC-T, there will be two kinds of iterations in our enumeration procedure. If there is a level $\ell \in [\hat{h}]$ such that $A_{\text{thin}}$ contains many vertices on this level, we call the iteration $A_{\text{thin}}$-*heavy*, otherwise we call it $A_{\text{thin}}$-*light*. We will make these definitions precise later. In the following, we describe how to handle both types of iterations.

### $A_{\text{thin}}$-heavy iterations

We start with the case that there is some level $\ell \in [\hat{h}]$ such that $|(A_{\text{thin}})_\ell|$ is large. Let $(v, \ell) \in (A_{\text{thin}})_\ell$. Unlike for SRMFC-T, we cannot simply add $(v, \ell)$ to $D$, as then we could encounter too many vertices in $A_{\text{thin}}$ in future iterations.

Instead, we want to add $(u, \ell)$ to $D$ for all nearby vertices $u$. The idea is that if $(v, \ell)$ is non-separable "because of" a close center $(w, \ell') \in \text{OPT}$ with $\ell' \geq \ell$, then we want to guarantee that in future iterations, no other support vertex $(v', \ell)$ on this level can be non-separable "because of $(w, \ell')$". Yet this faces a problem, as for one of these nearby vertices $u$ we could have $(u, \ell) \in \text{OPT}$. Hence, in order to maintain the existence of a $(C, D)$-compatible solution, we guess for every $(v, \ell) \in (A_{\text{thin}})_\ell$ whether there is a close center $(w, \ell) \in \text{OPT}$. To be precise, we guess for every $(v, \ell) \in (A_{\text{thin}})_\ell$ whether there is a $(w, \ell) \in \text{OPT}$ with $w \in \text{Ball}(v, 4\sigma r_\ell)$. If this is the case, add $(v, \ell)$ to $C$. Note that then $\text{Ball}(w, r_\ell) \subseteq V(C)$.

If we guess that there is no such pair $(w, \ell)$, we add $(u, \ell)$ to $D$ for all vertices $u \in \mathrm{Ball}(v, 4\sigma r_\ell)$. This guarantees the following: For $(w, \ell') \in \mathrm{OPT}$ with $w \in \mathrm{Ball}(v, \sigma(r_\ell + r_{\ell'}))$ and $\ell' \geq \ell$, we have $w \notin \mathrm{Ball}(v', \sigma(r_\ell + r_{\ell'}))$ for any other support vertex $(v', \ell) \in A_\mathrm{thin}$ by the second property of Observation 6.9, as well as $w \notin \mathrm{Ball}(v', \sigma(r_\ell + r_{\ell'}))$ for any support vertex $(v', \ell)$ in future iterations of the algorithm. The latter holds, as, if $w \in \mathrm{Ball}(v', \sigma(r_\ell + r_{\ell'}))$, we added $(v', \ell)$ to $D$ since $d(v, v') \leq 4\sigma r_\ell$ by the triangle inequality. In particular, a support vertex $(v', \ell)$ can no longer be non-separable "because of $(w, \ell')$". We will make this more precise later. This reasoning will be essential to bound the runtime of our algorithm. Summing up, we define for $A \subseteq A_\mathrm{thin}$

$$D_\mathrm{thin}(A) := \bigcup_{(v,\ell) \in A} \{(u, \ell) : u \in \mathrm{Ball}(v, 4\sigma r_\ell)\}$$

and update $D$ by adding $D_\mathrm{thin}(A)$.

### $A_\mathrm{thin}$-light iterations

In an $A_\mathrm{thin}$-light iteration, we want to make progress in identifying whether vertices are covered by balls of small or of large radius in OPT. To do so, we will use the partition $(A_\mathrm{big}, A_\mathrm{small}, A_\mathrm{sep}, A_\mathrm{non\text{-}sep})$ defined to update the sets $C$ and $D$. For the sets $A_\mathrm{small}$ and $A_\mathrm{sep}$, we define the following sets of pairs to be added to $D$. They precisely correspond to the pairs that cannot be in OPT, by definition of $A_\mathrm{small}$ and $A_\mathrm{sep}$.

$$D_\mathrm{small}(A_\mathrm{small}) := \bigcup_{(v,\ell) \in A_\mathrm{small}} \{(u, \ell') : u \in \mathrm{Ball}(v, \sigma(r_\ell + r_{\ell'})), \ell' \leq \hat{h}\},$$

$$D_\mathrm{sep}(A_\mathrm{sep}) := \left( \bigcup_{(v,\ell) \in A_\mathrm{sep}: \ell \leq \check{h}} \{(u, \ell') : u \in \mathrm{Ball}(v, \sigma(r_\ell + r_{\ell'})), \ell' \leq \check{h}\} \right)$$
$$\cup \bigcup_{(v,\ell) \in A_\mathrm{sep}: \ell > \check{h}} \{(u, \ell') : u \in \mathrm{Ball}(v, \sigma(r_\ell + r_{\ell'})), \ell' \leq \min(\hat{h}, \ell + \kappa)\}.$$

As mentioned above, we do not know OPT. Hence, we cannot directly compute the partition $(A_\mathrm{big}, A_\mathrm{small}, A_\mathrm{sep}, A_\mathrm{non\text{-}sep})$. Instead, we will enumerate over all possible partitions. As the support of $y$ is sparse, using Lemma 6.7, this enumeration can be done efficiently. For every possible partition, we update $D$ by adding $D_\mathrm{small}(A_\mathrm{small})$ and $D_\mathrm{sep}(A_\mathrm{sep})$ to $D$. If the guessed partition is correct, then we maintain existence of a $(C, D)$-compatible solution $x \in Q_{1,1}(V \setminus V(C))$. Moreover, the update of $D$ guarantees that vertices, which get coverage from $(v, \ell) \in \mathrm{supp}(y)$, will not get coverage from centers with certain radii in future iterations. In particular, note that vertices that get coverage from a pair in $A_\mathrm{small}$ will be fully covered by balls of small radius in future iterations of the algorithm since we add $D_\mathrm{small}(A_\mathrm{small})$ to $D$.

For pairs in $A_\mathrm{big}$, we add another level of guessing to keep the dilation of the radii small. Let $(v, \ell) \in A_\mathrm{big}$. By definition, there is a close center $(w, \ell') \in \mathrm{OPT}$ with $\ell' \leq \ell$. We guess whether $\ell - \ell' \leq \kappa_2$ for $\kappa_2 = \lceil \log_{1+\varepsilon}(\frac{2\sigma}{\varepsilon}) \rceil$. To be precise, for $j \in [\kappa_2]$, let $A_\mathrm{big}^j$ be the set of pairs $(v, \ell) \in A_\mathrm{big}$ for which there is some $(w, \ell') \in \mathrm{OPT}$ with $w \in \mathrm{Ball}(v, \sigma(r_\ell + r_{\ell'}))$ and $\ell' + j = \ell$. Furthermore, let $A_\mathrm{big}^\mathrm{close} = \bigcup_{j \in [\kappa_2]} A_\mathrm{big}^j$. We will add $(v, \ell - j)$ to $C$ for $(v, \ell) \in A_\mathrm{big}^j$ being part of a well-chosen subset of $A_\mathrm{big}^\mathrm{close}$ such that
- on one hand, all vertices covered by pairs in $A_\mathrm{big}^\mathrm{close}$ are covered, and
- on the other hand, we do not add too many center to $C$.

To be precise, for each $(w, \ell') \in \mathrm{OPT}$ such that there is a $(v, \ell) \in A_\mathrm{big}^\mathrm{close}$ with $w \in \mathrm{Ball}(v, \sigma(r_\ell + r_{\ell'}))$ and $\ell \leq \ell' + \kappa_2$ we add $(v^w, \ell')$ to $C$ for an arbitrary pair $(v^w, \ell) \in A_\mathrm{big}^\mathrm{close}$ satisfying this property. Then $\mathrm{Ball}(w, r_{\ell'}) \subseteq V(C)$. We also have $\bigcup_{(v,\ell) \in A_\mathrm{big}^\mathrm{close}} \mathrm{Ball}(v, \sigma r_\ell) \subseteq V(C)$, as for each $(v, \ell) \in A_\mathrm{big}^\mathrm{close}$ there is some $(w, \ell') \in \mathrm{OPT}$ with $w \in \mathrm{Ball}(v, \sigma(r_\ell + r_{\ell'}))$ and $\ell \leq \ell' + \kappa_2$. Then $\mathrm{Ball}(v, \sigma r_\ell) \subseteq \mathrm{Ball}(v^w, \rho r_{\ell'}) \subseteq V(C)$, as $d(v, v^w) \leq d(v, w) + d(w, v^w) \leq \sigma(r_\ell + r_{\ell'}) + \sigma(r_\ell + r_{\ell'}) \leq 4\sigma r_{\ell'} < \rho r_{\ell'}$.

The pairs in $A_{\text{big}} \setminus A_{\text{big}}^{\text{close}}$ that are not covered by $C$ will be covered by the solution to the dynamic program that we run on the bottom instance in the very end, using that for $(v_1, \ell_1), (v_2, \ell_2) \in A_{\text{big}}$ for which there is some $(w, \ell') \in \text{OPT}$ with $w \in \text{Ball}(v_i, \sigma(r_{\ell_i} + r_{\ell'}))$ and $\ell' \leq \ell_i$ for $i \in \{1, 2\}$, we can cover both $\text{Ball}(v_1, \sigma r_{\ell_1})$ and $\text{Ball}(v_2, \sigma r_{\ell_2})$ with $\text{Ball}(v_1, \sigma r_{\ell_1} + 2\sigma r_{\ell'} + 2\sigma r_{\ell_2}) \subseteq \text{Ball}(v_1, \rho r_{\ell'})$. We remark that, as we will show later, if we have added for $(w, \ell')$ some center to $C$, then both $\text{Ball}(v_1, \sigma r_{\ell_1})$ and $\text{Ball}(v_2, \sigma r_{\ell_2})$ are part of $V(C)$.

The pairs in $A_{\text{sep}}$ correspond to vertices in $A_{\text{dropped}}$ for SRMFC-T. We do not know how to partition the vertices that get coverage by pairs in $A_{\text{sep}}$ into those that are covered by $\text{OPT}_{\leq \hat{h}}$, and those that are covered by $\text{OPT}_{>\hat{h}}$. Still, akin to SRMFC-T, we will see that these vertices can be covered by adding a small number of extra centers, if they get coverage by pairs in $A_{\text{sep}}$ in constantly many iterations of our algorithm.

It remains to discuss how to handle pairs in $A_{\text{non-sep}}$. These will be covered by adding the pairs in $A_{\text{thin}}$ to $C$. By assumption on the iteration being $A_{\text{thin}}$-light, there are only few such pairs on each level, hence we do not violate the budget too much. On the other hand, by definition of $A_{\text{thin}}$, opening balls of radius $\rho r_\ell$ at these pairs covers all vertices that get coverage from pairs in $A_{\text{non-sep}}$.

## 6.4 The NUkC algorithm

We still need to define how we obtain a partition of $V$ into vertices covered by balls of large radius and vertices covered by balls of small radius. For a fractional solution $y \in Q_{\alpha,\beta}(U)$, we define

$$V_{\text{small}}(y) := \left\{ v \in V : \sum_{\ell=\hat{h}+1}^{L} \sum_{u \in \text{Ball}(v, \beta_\ell r_\ell)} y_{u,\ell} \geq 1 - \varepsilon \right\}.$$

Akin to SRMFC-T, we store the sparsified fractional solutions computed in $A_{\text{thin}}$-light iterations in a set $Y$. We then consider their average

$$\overline{Y} := \frac{1}{|Y|} \sum_{y \in Y} y \ .$$

Note that $\overline{Y} \in Q_{1+7\varepsilon, \beta(\lambda)}(V \setminus V(C))$. By restricting $\overline{Y}$ to levels $\hat{h}+1, ..., L$ and scaling it by $(1-\varepsilon)^{-1}$ we obtain a point in $Q_{1+10\varepsilon, 1}(V_{\text{small}}(\overline{Y}))$, using $\varepsilon \leq \frac{1}{7}$.

Now we are almost ready to state our enumeration algorithm. As mentioned earlier, we use dynamic programming to compute a solution on $V_{\text{big}} := V \setminus (V(C) \cup V_{\text{small}}(\overline{Y}))$. To do so, we exploit the guesses made during the enumeration algorithm. More precisely, while applying dynamic programming to all vertices in $V_{\text{big}}$ faces runtime issues, we can restrict the dynamic program to a subset $\mathcal{S}$ of the support of the sparsified solutions in $A_{\text{thin}}$-light iterations. On the correct execution path, we can show that every vertex of $V_{\text{big}}$ is close to a vertex in $\mathcal{S}$, and hence the dynamic program will return a $(1 + O(\varepsilon), 15 + O(\varepsilon))$-feasible solution.

The following theorem formalizes the guarantees of our dynamic program. We postpone its proof to Section 8.

**Theorem 6.10.** *Let $(V, d, k, r)$ be an instance of SNUkC with $L$ levels and $n$ vertices. Let $\beta \in \mathbb{R}_{\geq 0}$. For $\mathcal{S} \subseteq V \times [L]$, we can in time $\mathcal{O}(L4^n)$ compute a set $C \subseteq V \times [L]$ such that*
- *$C$ satisfies the budget constraints, i.e., for every $\ell \in [L]$, we have $|C_{\leq \ell}| \leq k_{\leq \ell}$,*
- *for every $(v, \ell) \in \mathcal{S}$, there is some $(v', \ell') \in C$ with $\ell' \leq \ell$ and $d(v, v') \leq \beta r_{\ell'}$,*

*or decide that no such set exists.*

Finally, we introduce the two subinstances that we will use in our algorithm. For $h \in [L]$, let $(V, d, k, r)_{\leq h} = (V, d, k', r')$ be the SNUkC instance with the same vertices, where $k$ and $r$ are restricted to the first $h$ levels, i.e., $(V, d, k', r')$ has $h$ levels and $k'_\ell = k_\ell$ and $r'_\ell = r_\ell$ for $1 \leq \ell \leq h$. Analogously, we denote by $(V, d, k, r)_{>h} = (V, d, k', r')$ the SNUkC instance with the same vertices, where we omit the first $h$ levels, i.e., $(V, d, k', r')$ has $L - h$ levels and $k'_\ell = k_{\ell+h}$ and $r'_\ell = r_{\ell+h}$ for $1 \leq \ell \leq L - h$. Now, given an $\varepsilon$-compressed instance $(V, d, k, r)$ of SNUkC, we run the following enumeration approach.

---

**Algorithm 5:** EFFICIENT-THINNED-EXPLORATION-SNUkC$(C, D, \zeta, Y, A_{\mathrm{DP}}^{\mathrm{total}})$

1: **if** there is a $(C, D)$-compatible point $x$ in $Q_{1,1}(V \setminus V(C))$ **then**
2:    $y \leftarrow$ Sparsification of $x$ through Lemma 6.7 with $\lambda = \lambda_\varepsilon$ so that $\mathrm{Ball}(v, \sigma r_\ell) \setminus V(C) \neq \emptyset$ for each $(v, \ell) \in \mathrm{supp}(y)$.
3:    **for** each $\ell \in [\hat{h}]$ with $|\mathrm{supp}(y)_\ell| \leq \zeta$ **do**
4:      **for** each $A_{\mathrm{thin}}^\ell \subseteq \mathrm{supp}(y)_\ell$ with $|A_{\mathrm{thin}}^\ell| \geq \frac{\varepsilon}{N} k_\ell$ and each $A_{\mathrm{thin}}^C \subseteq A_{\mathrm{thin}}^\ell$ **do**
5:         $D \leftarrow D \cup D_{\mathrm{thin}}(A_{\mathrm{thin}}^\ell \setminus A_{\mathrm{thin}}^C)$
6:         EFFICIENT-THINNED-EXPLORATION-SNUkC$(C \cup A_{\mathrm{thin}}^C, D, \zeta - |\mathrm{supp}(y)_\ell|, Y, A_{\mathrm{DP}}^{\mathrm{total}})$.
7:      **end**
8:    **end**
9:    Add $y$ to $Y$.
10:   **if** $|Y| = N$ **then**
11:      $V_{\mathrm{DP}} \leftarrow \{v \in V : (v, \ell) \in A_{\mathrm{DP}}^{\mathrm{total}} \text{ for some } \ell \in [L]\}$
12:      $A_{\mathrm{DP}}^{\mathrm{final}} \leftarrow \{(v, \ell) \in A_{\mathrm{DP}}^{\mathrm{total}} : \mathrm{Ball}(v, \sigma r_\ell) \setminus V(C) \neq \emptyset\}$
13:      **if** Theorem 6.10 applied to $(V_{\mathrm{DP}}, d, k^C + 2\varepsilon k, r)_{\leq \hat{h}}$ with $\mathcal{S} = A_{\mathrm{DP}}^{\mathrm{final}}$ and $\beta = 4\sigma$ returns a set $C_{\mathrm{big}}$ **then**
14:         $C_{\mathrm{small}} \leftarrow$ Output of Lemma 6.5 applied to $(V_{\mathrm{small}}(\overline{Y}), d, k, r)_{>\hat{h}}$.
15:         Output $C \cup C_{\mathrm{big}} \cup C_{\mathrm{small}}$.
16:     **end**
17:   **else**
18:      **for** each partition $(A_{\mathrm{big}}^{\mathrm{close}}, A_{\mathrm{small}}, A_{\mathrm{sep}}^D, A_{\mathrm{DP}})$ of $\mathrm{supp}(y)_{\leq \hat{h}}$ **do**
19:        **for** each subset $A_{\mathrm{big}}^C$ of $A_{\mathrm{big}}^{\mathrm{close}}$ and each $f : A_{\mathrm{big}}^C \to [\kappa_2]$ **do**
20:           $C \leftarrow C \cup \{(v, \ell - f(v, \ell)) : (v, \ell) \in A_{\mathrm{big}}^C\}$.
21:           $D \leftarrow D \cup D_{\mathrm{small}}(A_{\mathrm{small}}) \cup D_{\mathrm{sep}}(A_{\mathrm{sep}}^D)$.
22:           EFFICIENT-THINNED-EXPLORATION-SNUkC$(C, D, \zeta, Y, A_{\mathrm{DP}}^{\mathrm{total}} \cup A_{\mathrm{DP}})$.
23:       **end**
24:      **end**
25:   **end**
26: **end**

---

We remark that the property that $\mathrm{Ball}(v, \sigma r_\ell) \setminus V(C) \neq \emptyset$ for each $(v, \ell) \in \mathrm{supp}(y)$ in step 2 of the algorithm can be easily obtained by setting $y_{v,\ell} = 0$ for each each $(v, \ell)$ with $\mathrm{Ball}(v, \sigma r_\ell) \setminus V(C) = \emptyset$. Note that this modification maintains feasibility.

Akin to SRMFC-T, we call the instance in step 14 of the algorithm the *top instance* and $(V_{\mathrm{big}}, d, k^C + 2\varepsilon k, r)_{\leq \hat{h}}$ the *bottom instance*, where

$$V_{\mathrm{big}} := V \setminus \left(V(C) \cup V_{\mathrm{small}}(\overline{Y})\right).$$

We will later show that if the dynamic program returns a set $C_{\mathrm{big}}$ in step 13 of the algorithm, then $C_{\mathrm{big}}$ is $(1, 5\sigma)$-feasible for the top instance.

**Theorem 6.11.** *Let $0 < \varepsilon \leq \frac{1}{7}$. Let $N = \lceil \frac{3}{\varepsilon} \rceil$ and $\overline{\zeta} = \frac{4N\lambda_\varepsilon}{\varepsilon^5}\left(\check{h}^2\hat{h} + 3\kappa L\right)$. Given a $(1,1)$-feasible $\varepsilon$-compressed instance of SNUkC, EFFICIENT-THINNED-EXPLORATION-SNUkC$\left(\emptyset, \emptyset, \overline{\zeta}, \emptyset, \emptyset\right)$ computes a $(1 + 14\varepsilon, 15 + 6\varepsilon)$-feasible solution.*

We prove Theorem 6.11 in the following subsection.

Finally, we show that for the choice of $N$ and $\overline{\zeta}$ in Theorem 6.11, Algorithm 5 has polynomial running time.

**Theorem 6.12.** *For $N = \lceil \frac{3}{\varepsilon} \rceil$ and $\overline{\zeta} = \frac{4N\lambda_\varepsilon}{\varepsilon^5}\left(\check{h}^2\hat{h} + 3\kappa L\right)$, the running time of EFFICIENT-THINNED-EXPLORATION-SNUkC$\left(\emptyset, \emptyset, \overline{\zeta}, \emptyset, \emptyset\right)$ is polynomial in the number of vertices of the given $\varepsilon$-compressed instance of SNUkC.*

The proof of Theorem 6.12 is postponed to Section 6.6.

Note that Theorem 6.11 and Theorem 6.12 together imply our main result for SNUkC, Theorem 6.6.

## 6.5 Correctness (Proof of Theorem 6.11)

The proof in this section follows the high-level overview given in Section 6.3.

### 6.5.1 Description of the execution path we analyze

Let OPT denote a fixed $(1,1)$-feasible solution to the $\varepsilon$-compressed SNUkC instance. We will closely analyze one specific execution path of Algorithm 5, which we call the OPT-*execution path*. We will show that this execution path leads to a $(1 + O(\varepsilon), 15 + O(\varepsilon))$-feasible solution.

During the run of the algorithm, we add vertices to the sets $C$ and $D$. In order to control the size of $C$ and to guarantee feasibility, we maintain a set $\text{OPT}^C \subseteq \text{OPT}_{\leq \hat{h}}$ together with a surjective assignment $\varphi : \text{OPT}^C \to C$. We guarantee the following properties throughout the execution path:
- For every $(w, \ell) \in \text{OPT}^C$, we have that $d(v, w) \leq 4\sigma r_\ell$ for $(v, \ell) = \varphi((w, \ell))$. In particular, $\text{Ball}(w, \sigma r_\ell) \subseteq \text{Ball}(v, \rho r_\ell) \subseteq V(C)$.
- For every $(w, \ell) \in \text{OPT}_{\leq \hat{h}} \setminus \text{OPT}^C$, we have $(w, \ell) \notin D$.

These properties ensure that the characteristic vector $\chi^{\text{OPT} \setminus \text{OPT}^C}$ of $\text{OPT} \setminus \text{OPT}^C$ is a $(C, D)$-compatible solution covering $V \setminus V(C)$. Indeed, the first property guarantees that the budgets are not violated and that all vertices of $V \setminus V(C)$ are covered. Additionally, the second property guarantees that no center in $\text{OPT} \setminus \text{OPT}^C$ is blocked by $D$.

Furthermore, we will construct a set $C_{\text{add}}$ of pairs, which we add to $\text{OPT} \setminus \text{OPT}^C$ at the end of the algorithm to obtain a feasible solution for the bottom instance. The pairs in $C_{\text{add}}$ will help us to show that there is a solution to the bottom instance, which we then compute via dynamic programming. In order not to violate the budgets to much, we will ensure that $|(C_{\text{add}})_{\leq \ell}| \leq 2\varepsilon k_{\leq \ell}$ for every $\ell \in \left[\hat{h}\right]$.

Consider a run of EFFICIENT-THINNED-EXPLORATION-SNUkC. Note that by our invariant there is a $(C, D)$-compatible point $x$ in $Q_{1,1}(V \setminus V(C))$. Let $y$ be the sparsification of $x$ obtained by applying Lemma 6.7 with $\lambda = \lambda_\varepsilon$ and $\text{Ball}(v, \sigma r_\ell) \setminus V(C) \neq \emptyset$ for each $(v, \ell) \in \text{supp}(y)$. Let $\left(A_{\text{big}}, A_{\text{small}}, A_{\text{sep}}, A_{\text{non-sep}}\right)$ be the partition of $\text{supp}(y)_{\leq \hat{h}}$ and $A_{\text{thin}} \subseteq A_{\text{non-sep}}$ as described in Section 6.3. Moreover, let the sets $A_{\text{big}}^j \subseteq A_{\text{big}}$ for $j \in [\kappa_2]$ be as described in Section 6.3. First, we have to decide when to perform an $A_{\text{thin}}$-heavy iteration and when to perform an $A_{\text{thin}}$-light iteration.

#### $A_{\text{thin}}$-heavy iterations

If there is a level $\ell$ such that $|(A_{\text{thin}})_\ell| \geq \frac{\varepsilon}{N} k_\ell$, then we call this iteration $A_{\text{thin}}$-heavy, and call it $A_{\text{thin}}$-light otherwise. If it exists, we consider an arbitrary such $\ell$. For this $\ell$, we follow the execution path calling EFFICIENT-THINNED-EXPLORATION-SNUkC in step 6 of Algorithm 5 with the following choices. Let $\text{OPT}_\ell^C$ be the set of pairs $(w, \ell) \in \text{OPT} \setminus \text{OPT}^C$ such that there is some $(v, \ell) \in (A_{\text{thin}})_\ell$ with $d(v, w) \leq$ $4\sigma r_\ell$. For each $(w, \ell) \in \mathrm{OPT}_\ell^C$, we pick an arbitrary pair $(v^w, \ell) \in (A_{\mathrm{thin}})_\ell$ with $d(v^w, w) \leq 4\sigma r_\ell$. Add $(w, \ell)$ to $\mathrm{OPT}_\ell^C$ and $(v^w, \ell)$ to $C$. Furthermore, set $\varphi((w, \ell)) = (v^w, \ell)$. Let $A_{\mathrm{thin}}^C$ be the set of these chosen pairs $(v^w, \ell)$. We follow the execution path choosing this $A_{\mathrm{thin}}^C$ in step 6 of Algorithm 5. We set $A_{\mathrm{thin}}^\ell := (A_{\mathrm{thin}})_\ell$. Note that we update $D$ by adding $D_{\mathrm{thin}}\big((A_{\mathrm{thin}})_\ell \setminus A_{\mathrm{thin}}^C\big)$. We will show later that this execution path indeed exists, i.e., $|\mathrm{supp}(y)_\ell| \leq \zeta$.

Note that our invariant is still satisfied. Indeed, for each $(w, \ell) \in \mathrm{OPT}_\ell^C$, we have that $\varphi((w, \ell)) = (v^w, \ell)$ with $d(v^w, w) \leq 4\sigma r_\ell$ by construction. Moreover, consider a pair $(w, \ell) \in \mathrm{OPT}_{\leq \hat{h}} \setminus \mathrm{OPT}^C$. Then, by definition $d(v, w) > 4\sigma r_\ell$ for all $(v, \ell) \in (A_{\mathrm{thin}})_\ell$. Hence, since we added $D_{\mathrm{thin}}\big((A_{\mathrm{thin}})_\ell \setminus A_{\mathrm{thin}}^C\big)$ to $D$, we have $(w, \ell) \notin D$.

### $A_{\mathrm{thin}}$-light iterations

If there is no such $\ell$, then we perform an $A_{\mathrm{thin}}$-light iteration. In the following, we describe how we build up the sets $A_{\mathrm{sep}}^D$ and $A_{\mathrm{DP}}$, as well as which vertices to add to $C_{\mathrm{add}}$. Let $(v, \ell) \in A_{\mathrm{sep}}$. We call $(v, \ell)$ a *successor* of a pair $(v', \ell')$, if $(v', \ell')$ was part of $A_{\mathrm{sep}}$ in a previous $A_{\mathrm{thin}}$-light iteration, $\check{h} < \ell' < \ell - \kappa$, and $d(v, v') \leq \sigma(r_{\ell'} - r_\ell)$. We distinguish the following two cases:
- If $(v, \ell)$ is a successor of some pair $(v', \ell')$, then we add $(v, \ell)$ to $A_{\mathrm{DP}}$.
- Otherwise, we add $(v, \ell)$ to $A_{\mathrm{sep}}^D$.

Additionally, if this is the first iteration a pair $(v', \ell')$ has a successor, then we add $(v, \ell' + \frac{\kappa}{2})$ to $C_{\mathrm{add}}$, where $(v, \ell)$ is an arbitrary successor of $(v', \ell')$ we encounter in this iteration.

Moreover, we add $A_{\mathrm{big}}$ and $A_{\mathrm{non\text{-}sep}}$ to $A_{\mathrm{DP}}$, and $A_{\mathrm{thin}}$ to $C_{\mathrm{add}}$. Finally, we have to define $A_{\mathrm{big}}^{\mathrm{close}}$ and $A_{\mathrm{big}}^C$. For $j \in [\kappa_2]$, let $A_{\mathrm{big}}^j$ be the set of pairs $(v, \ell) \in A_{\mathrm{big}}$ for which there is some $(w, \ell') \in \mathrm{OPT} \setminus \mathrm{OPT}^C$ with $w \in \mathrm{Ball}(v, \sigma(r_\ell + r_{\ell'}))$ and $\ell' + j = \ell$. Furthermore, let $A_{\mathrm{big}}^{\mathrm{close}} = \bigcup_{j \in [\kappa_2]} A_{\mathrm{big}}^j$. For every $(w, \ell') \in \mathrm{OPT} \setminus \mathrm{OPT}^C$ such that there is a $(v, \ell) \in A_{\mathrm{big}}^{\mathrm{close}}$ with $w \in \mathrm{Ball}(v, \sigma(r_\ell + r_{\ell'}))$ and $\ell' < \ell \leq \ell' + \kappa_2$ we add $(v^w, \ell)$ to $A_{\mathrm{big}}^C$ with $f((v^w, \ell)) = \ell - \ell'$ for an arbitrary pair $(v^w, \ell) \in A_{\mathrm{big}}^{\mathrm{close}}$ satisfying this property. Add $(w, \ell')$ to $\mathrm{OPT}^C$ and set $\varphi((w, \ell')) = (v^w, \ell)$. Note that our update of $C$ adds $(v^w, \ell')$ to $C$.

We again have to verify that our invariant is still satisfied. First, our construction of $\varphi$ ensures that for every $(w, \ell) \in \mathrm{OPT}^C$, we have that $d(v, w) \leq 2\sigma r_\ell$ for $(v, \ell) = \varphi((w, \ell))$. Hence, the first property of our invariant is satisfied. To verify the second property, we have to check our updates of $D$. Note that we add $D_{\mathrm{small}}(A_{\mathrm{small}})$ and $D_{\mathrm{sep}}(A_{\mathrm{sep}}^D)$ to $D$. By the way we have chosen the partition $(A_{\mathrm{big}}, A_{\mathrm{small}}, A_{\mathrm{sep}}, A_{\mathrm{non\text{-}sep}})$, we have that for each $(w, \ell) \in \mathrm{OPT} \setminus \mathrm{OPT}^C$, $(w, \ell) \notin D_{\mathrm{small}}(A_{\mathrm{small}})$ and $(w, \ell) \notin D_{\mathrm{sep}}(A_{\mathrm{sep}}^D)$.

This finishes the description of the OPT-execution path. In the following subsections, we analyze this execution path and show that it leads to a $(1 + O(\varepsilon), 15 + O(\varepsilon))$-feasible solution. To do so, we proceed in three steps.
- First, we show that the execution path is indeed one of the execution paths explored by EFFICIENT-THINNED-EXPLORATION-SNUkC$(\emptyset, \emptyset, \overline{\zeta}, \emptyset, \emptyset)$. To do so, we have to bound the size of the sets $A_{\mathrm{thin}}$ encountered throughout the execution path and show that our choice of $\overline{\zeta}$ suffices.
- Second, we show that on the OPT-execution path, the dynamic program in step 14 of Algorithm 5 returns a set $C_{\mathrm{big}}$, and that this set is a $(1 + 2\varepsilon, 5\sigma)$-feasible solution for the bottom instance.
- Finally, we conclude with a proof of Theorem 6.11.

### 6.5.2 Bounding the number of $A_{\mathrm{thin}}$-heavy iterations

We begin by showing that for $\overline{\zeta} = \frac{4N\lambda_\varepsilon}{\varepsilon^5}\big(\check{h}^2 \hat{h} + 3\kappa L\big)$, the OPT-execution path is one of the execution paths of EFFICIENT-THINNED-EXPLORATION-SNUkC$(\emptyset, \emptyset, \overline{\zeta}, \emptyset, \emptyset)$.

To prove this, let $A_{\mathrm{thin}}^{\mathrm{total}}$ denote the union of the sets $A_{\mathrm{thin}}^\ell$ in the $A_{\mathrm{thin}}$-heavy iterations of the OPT-execution path. The main ingredient is that we can bound the size of $A_{\mathrm{thin}}^{\mathrm{total}}$ as follows.

**Lemma 6.13.** *On the* OPT*-execution path, we have*
- $|(A_{\text{thin}}^{\text{total}})_{\leq \check{h}}| \leq \frac{1+\varepsilon}{\varepsilon^2}\check{h}\hat{h}$ *and*
- $|(A_{\text{thin}}^{\text{total}})_{>\check{h}}| \leq \frac{3\kappa(1+\varepsilon)}{\varepsilon^2}L.$

We will prove Lemma 6.13 in the remainder of this subsection, but first we show how it implies that the OPT-execution path is one of the execution paths of Algorithm 5.

**Lemma 6.14.** *The* OPT*-execution path is one of the execution paths of* Efficient-Thinned-Exploration-SNUkC$(\emptyset, \emptyset, \overline{\zeta}, \emptyset, \emptyset)$.

*Proof.* We only have to check that on the OPT-execution path, we have $|\text{supp}(y)_\ell| \leq \zeta$ every time we call Efficient-Thinned-Exploration-SNUkC in line 6 of Algorithm 5. Let $\mathcal{U}$ be the family of all sets $\text{supp}(y)_\ell$ in $A_{\text{thin}}$-heavy iterations of the OPT-execution path. We have to show $\sum_{S \in \mathcal{U}} |S| \leq \overline{\zeta}$.

Let $S \in \mathcal{U}$ and $\ell \in [L]$ with $S = \text{supp}(y)_\ell$, where $y$ denotes the sparsified fractional solution in the corresponding iteration of the algorithm. By Lemma 6.7 we have

$$|S| \leq \begin{cases} \frac{\lambda_\varepsilon \check{h}}{\varepsilon} y(V_\ell) \leq \frac{\lambda_\varepsilon \check{h}}{\varepsilon}(1+7\varepsilon)k_{\leq \ell} & \text{if } \ell \in [\check{h}], \\ \frac{\lambda_\varepsilon}{\varepsilon} y(V_\ell) \leq \frac{\lambda_\varepsilon}{\varepsilon}(1+7\varepsilon)k_{\leq \ell} & \text{else.} \end{cases}$$

Note that $k_{\leq \ell} \leq \frac{1+\varepsilon}{\varepsilon} k_\ell$ for each $\ell \in [\check{h}]$ by definition of $\varepsilon$-compressed instances. By definition of $\mathcal{U}$ we have $|A_{\text{thin}}^\ell| \geq \frac{\varepsilon}{N} k_\ell$. Hence, using $\varepsilon \leq \frac{1}{7}$,

$$\frac{|S|}{|A_{\text{thin}}^\ell|} \leq \begin{cases} \frac{(1+\varepsilon)(1+7\varepsilon)N\lambda_\varepsilon \check{h}}{\varepsilon^3} \leq \frac{3N\lambda_\varepsilon \check{h}}{\varepsilon^3} & \text{if } S \subseteq V_{\leq \check{h}}, \\ \frac{(1+\varepsilon)(1+7\varepsilon)N\lambda_\varepsilon}{\varepsilon^3} \leq \frac{3N\lambda_\varepsilon}{\varepsilon^3} & \text{else.} \end{cases}$$

Together with Lemma 6.13, and again using $\varepsilon \leq \frac{1}{7}$, this yields

$$\sum_{S \in \mathcal{U}} |S| \leq \frac{3N\lambda_\varepsilon \check{h}}{\varepsilon^3} |(A_{\text{thin}}^{\text{total}})_{\leq \check{h}}| + \frac{3N\lambda_\varepsilon}{\varepsilon^3} |(A_{\text{thin}}^{\text{total}})_{>\check{h}}|$$

$$\leq \frac{4N\lambda_\varepsilon}{\varepsilon^5}\left(\check{h}^2\hat{h} + 3\kappa L\right)$$

$$= \overline{\zeta}.$$

$\square$

In the remainder of this subsection we will prove Lemma 6.13. The proof is unfortunately more involved than for SRMFC-T, as we can no longer define a set $V_{\text{thin}}$ upfront.

*Proof of Lemma 6.13.* First, observe that our update of $C$ and $D$ in $A_{\text{thin}}$-heavy iterations gives the following crucial property.

**Observation 6.15.** *Let* $\ell \in [\hat{h}]$. *For* $(v, \ell), (v', \ell) \in A_{\text{thin}}^{\text{total}}$, *we have* $d(v, v') > 4\sigma r_\ell$. *In particular, for every* $(w, \ell') \in \text{OPT}_{\leq \hat{h}}$ *with* $\ell' > \ell$, *there is at most one pair* $(\hat{v}, \ell)$ *in* $A_{\text{thin}}^{\text{total}}$ *with* $w \in \text{Ball}(\hat{v}, \sigma(r_\ell + r_{\ell'}))$.

*Proof.* If both pairs are added to $A_{\text{thin}}^{\text{total}}$ in the same $A_{\text{thin}}$-heavy iteration, this directly follows by definition of $A_{\text{thin}}$. Otherwise, let $(v, \ell)$ be the pair first added to $A_{\text{thin}}^{\text{total}}$. We either add $(v, \ell)$ to $C$, or we add $(v', \ell)$ to $D$ for all $v' \in \text{Ball}(v, 4\sigma r_\ell)$. In both cases, in no further iteration of the algorithm, we can have a pair $(v', \ell)$ in the support of $y$ with $d(v, v') \leq 4\sigma r_\ell$. Indeed, if we add $(v, \ell)$ to $C$, this directly follows by $\text{Ball}(v, \sigma r_\ell) \setminus V(C) \neq \emptyset$ for each $(v, \ell) \in \text{supp}(y)$ as guaranteed in line 2 of Algorithm 5. □

We now distinguish the three cases of a non-separable $(v, \ell) \in \text{supp}(y)$, summarized in Observation 6.8. We quickly recall them here for convenience. Let $(v, \ell) \in \text{supp}(y)$ be non-separable. Then one of the following holds:
(i) $\ell \leq \check{h}$, and there is some $(w, \ell') \in \text{OPT}$ with $w \in \text{Ball}(v, \sigma(r_\ell + r_{\ell'}))$ and $\ell < \ell' \leq \check{h}$.
(ii) $\ell > \check{h}$, and there is some $(w, \ell') \in \text{OPT}$ with $w \in \text{Ball}(v, \sigma(r_\ell + r_{\ell'}))$ and $\ell < \ell' \leq \ell + \kappa$.
(iii) $\ell > \check{h}$, and there are centers $(w_1, \ell_1), (w_2, \ell_2) \in \text{OPT}$ with $w_i \in \text{Ball}(v, \sigma(r_\ell + r_{\ell_i}))$ and $\ell + \kappa < \ell_i \leq \hat{h}$ for $i \in \{1, 2\}$, such that $d(w_1, w_2) > \mu r_{\ell+\kappa}$.

Note that to bound $\left(A_{\text{thin}}^{\text{total}}\right)_{\leq \check{h}}$, we only have to consider the first case of Observation 6.8, while to bound $\left(A_{\text{thin}}^{\text{total}}\right)_{> \check{h}}$, we have to consider the second and third case of Observation 6.8. So let us first bound how many times the first case of Observation 6.8 can occur in $A_{\text{thin}}^{\text{total}}$.

By Observation 6.15, for every $(w, \ell) \in \text{OPT}_{\leq \check{h}}$ there are at most $\check{h}$ pairs $(v, \ell') \in A_{\text{thin}}^{\text{total}}$ with $\ell' < \ell$ so that $w \in \text{Ball}(v, \sigma(r_{\ell'} + r_\ell))$. Hence, we have

$$\left|\left(A_{\text{thin}}^{\text{total}}\right)_{\leq \check{h}}\right| \leq \check{h} k_{\leq \check{h}} \leq \check{h} \frac{1+\varepsilon}{\varepsilon} k_{\check{h}} \leq \check{h} \frac{1+\varepsilon}{\varepsilon^2} \hat{h} \ .$$

Next, we bound the pairs in $A_{\text{thin}}^{\text{total}}$ satisfying the second property in Observation 6.8. For every $(w, \ell) \in \text{OPT}_{\leq \hat{h}} \setminus \text{OPT}_{\leq \check{h}}$, there are at most $\kappa$ pairs $(v, \ell') \in A_{\text{thin}}^{\text{total}}$ with $\ell' < \ell \leq \ell' + \kappa$ so that $w \in \text{Ball}(v, \sigma(r_{\ell'} + r_\ell))$, again using Observation 6.15. Hence, in total, there are at most

$$\kappa k_{\leq \hat{h}} \leq \kappa \frac{1+\varepsilon}{\varepsilon} k_{\hat{h}} \leq \kappa \frac{1+\varepsilon}{\varepsilon^2} L$$

many non-separable pairs in $A_{\text{thin}}^{\text{total}}$ satisfying the second property in Observation 6.8.

Finally, we need to bound the pairs in $A_{\text{thin}}^{\text{total}}$ satisfying the third property in Observation 6.8. Here the argumentation is more involved. We build a set $\mathcal{C}$ of *critical* triples $(w_1, w_2, \ell)$ with $(w_1, \ell_1), (w_2, \ell_2) \in \text{OPT}_{\leq \hat{h}}$ and $\ell_1, \ell_2 > \ell + \kappa$. Let $(v, \ell) \in A_{\text{thin}}^{\text{total}}$ satisfy the third property in Observation 6.8. For this $(v, \ell)$ let $(w_1, \ell_1), (w_2, \ell_2) \in \text{OPT}_{\leq \hat{h}}$ be an arbitrary pair of OPT-pairs with $\ell_1, \ell_2 > \ell + \kappa$, $d(w_1, w_2) > \mu r_{\ell+\kappa}$, and $w_i \in \text{Ball}\left(v, \sigma\left(r_\ell + r_{\ell_i}\right)\right)$ for $i \in \{1, 2\}$. Add $(w_1, w_2, \ell)$ to $\mathcal{C}$. Note that Observation 6.15 guarantees that we do not add the same triple to $\mathcal{C}$ twice. By construction, the number of pairs in $A_{\text{thin}}^{\text{total}}$ satisfying the third property in Observation 6.8 is bounded by the number of triples added to $\mathcal{C}$ over the course of the OPT-execution path. We will now bound the size of $\mathcal{C}$. First, we analyze the distances between the centers of critical triples.

**Claim 6.16.** *Let $(w_1, w_2, \ell), (w_3, w_4, \ell) \in \mathcal{C}$ be critical for the same level $\ell$. Then $d(w_1, w_3) > r_\ell$.*

*Proof.* Let $(w_1, w_2, \ell)$ be added to $\mathcal{C}$ for $(v, \ell) \in A_{\text{thin}}^{\text{total}}$ and $(w_3, w_4, \ell)$ be added to $\mathcal{C}$ for $(v', \ell) \in A_{\text{thin}}^{\text{total}}$. For $(w, \ell') \in \text{OPT}$ let $r(w)$ denote the radius $r_{\ell'}$ of the ball opened at $w$ in OPT. By Observation 6.15, we have

$$\begin{aligned} d(w_1, w_3) &\geq d(v, v') - d(v, w_1) - d(v', w_3) \\ &\geq 4\sigma r_\ell - \sigma(r_\ell + r(w_1)) - \sigma(r_\ell + r(w_3)) \\ &\geq 2\sigma(r_\ell - r_{\ell+\kappa}) \\ &> r_\ell, \end{aligned} \quad (6)$$

where we used Observation 6.15 in the second inequality, and our choice of $\kappa$ in the final inequality. □

Next, observe that $(w_1, w_2, \ell)$ being critical implies that

$$\mu r_{\ell+\kappa} < d(w_1, w_2) \leq 2\sigma(r_\ell + r_{\ell+\kappa}). \tag{7}$$

Let $\tau := \kappa + \lceil \log_{1+\varepsilon}\left(\frac{8\sigma}{\mu}\right) \rceil$. In particular, for $(w_1, w_2, \ell)$ critical, we have that $(w_1, w_2, \ell')$ can only be critical if $|\ell' - \ell| < \tau$, as otherwise (w.l.o.g. $\ell' < \ell$)

$$d(w_1, w_2) \leq 2\sigma(r_\ell + r_{\ell+\kappa}) \leq 4\sigma r_\ell \leq 4\sigma r_{\ell'+\tau} \leq 4\sigma(1+\varepsilon)^{\kappa-\tau} r_{\ell'+\kappa} \leq \mu r_{\ell'+\kappa},$$

contradicting (7). For $i \in [\tau]$, we define the *slice* $\mathcal{C}_i = \{(w_1, w_2, \ell) \in \mathcal{C} : i \equiv \ell \bmod \tau\}$.

We claim that $|\mathcal{C}_i| \leq |\mathrm{OPT}_{\leq \hat{h}}| - 1$ for each $i \in [\tau]$. This implies that $|\mathcal{C}|$ is bounded by

$$\tau |\mathrm{OPT}_{\leq \hat{h}}| \leq \tau k_{\leq \hat{h}} \leq \tau \frac{1+\varepsilon}{\varepsilon} k_{\hat{h}} \leq \tau \frac{1+\varepsilon}{\varepsilon^2} L \leq 2\kappa \frac{1+\varepsilon}{\varepsilon^2} L,$$

finishing the proof of Lemma 6.13.

Let $V_{\mathrm{OPT}} := \{w \in V : (w, \ell) \in \mathrm{OPT} \text{ for some } \ell \in [\hat{h}]\}$. To show $|\mathcal{C}_i| \leq |\mathrm{OPT}_{\leq \hat{h}}| - 1$ for each $i \in [\tau]$, we define an undirected graph $H_i = (V_{\mathrm{OPT}}, E_i)$ with an edge $\{w_1, w_2\} \in E$ if there is a critical triple $(w_1, w_2, \ell)$ in $\mathcal{C}_i$. We will show that $G$ does not contain a cycle. This finishes the proof for the following reason: By the above discussion, every edge in $H_i$ corresponds to a unique critical triple in $\mathcal{C}_i$. Furthermore, every critical triple of $\mathcal{C}_i$ gives rise to an edge in $H_i$. Hence, $|\mathcal{C}_i| = |E_i| \leq |\mathrm{OPT}_{\leq \hat{h}}| - 1$ for each $i \in [\tau]$.

Before we proceed with the proof of the claim, we will investigate paths in $H_i$.

**Claim 6.17.** *Let $w_1, w_2, ..., w_k$ be a path in $H_i$ with corresponding critical triples $(w_1, w_2, \ell_1), (w_2, w_3, \ell_2), ..., (w_{k-1}, w_k, \ell_{k-1})$ in $\mathcal{C}_i$. Assume further that $\ell_1 < \ell_j$ for $j \in \{2, ..., k\}$. Then*
- $d(w_2, w_k) \leq r_{\ell_1+\kappa}$, *and*
- *there is no $w \in V_{\mathrm{OPT}}$ with $(w_k, w, \ell_1) \in \mathcal{C}$.*

*Proof.* The second property follows from the first property by applying Claim 6.16. We show the first property by backward induction on $\ell_1$. First, if $\ell_1 = \hat{h}$, then there is no smaller radius that we can consider. Hence, we have $k = 2$ and $d(w_2, w_2) = 0$.

Assume now that the first property holds for all levels larger than $\ell_1$. By applying the induction assumption, we know that there is a unique minimizer of $\ell_i$ over $i \in \{2, ..., k\}$ — the path between any two minimizers would contradict the second property. Let $j$ be the index of this minimizer. Consider the two paths $w_{j+1}, w_j, ..., w_2$ and $w_j, w_{j+1}, ..., w_k$ in $H_i$, both starting with $\ell_j$. Using the induction assumption again, we have that $d(w_j, w_2) \leq r_{\ell_j+\kappa} \leq r_{\ell_1+\tau+\kappa}$ and $d(w_{j+1}, w_k) \leq r_{\ell_j+\kappa} \leq r_{\ell_1+\tau+\kappa}$. In particular,

$$\begin{aligned}
d(w_2, w_k) &\leq d(w_2, w_j) + d(w_j, w_{j+1}) + d(w_{j+1}, w_k) \\
&\leq 2\sigma r_{\ell_1+\tau} + (2\sigma + 2) r_{\ell_1+\tau+\kappa} \\
&\leq (4\sigma + 2) r_{\ell_1+\tau} \\
&\leq (4\sigma + 2)(1+\varepsilon)^{\kappa-\tau} r_{\ell_1+\kappa} \leq r_{\ell_1+\kappa}
\end{aligned}$$

where we used (7) in the second inequality and the definition of $\tau$ in the last inequality. □

Let us conclude that $H_i$ does not contain a cycle. Assume for contradiction that there was a cycle $w_1, w_2, ..., w_k, w_1$ in $H_i$ corresponding to critical triples $(w_1, w_2, \ell_1), (w_2, w_3, \ell_2), ..., (w_k, w_1, \ell_k)$ in $\mathcal{C}_i$. Without loss of generality, we can assume that $\ell_1 \leq \ell_j$ for $j \in [k]$. By Claim 6.17, we know that there is no $j \in \{2, ..., k\}$ such that $\ell_j = \ell_1$, as the path on the cycle from $w_1$ to $w_{j-1}$ would then contradict the second part of Claim 6.17. Hence, $\ell_2, ..., \ell_k > \ell_1$ which gives $d(w_2, w_k) \leq r_{\ell_1+\kappa}$ by the first part of Claim 6.17. Now

$$d(w_1, w_k) \geq d(w_1, w_2) - d(w_2, w_k) > \mu r_{\ell_1+\kappa} - r_{\ell_1+\kappa} = r_{\ell_1+\kappa} \geq 2\sigma\left(r_{\ell_1+\tau} + r_{\ell_1+\tau+\kappa}\right)$$

by the choice of $\tau$ and $\mu$. This contradicts (7) and the fact that $(w_1, w_k, \ell_k)$ is critical, as $\ell_k \geq \ell_1 + \tau$. As discussed above, this implies that $|\mathcal{C}_i| \leq |\mathrm{OPT}_{\leq \hat{h}}| - 1$ for each $i \in [\tau]$, and hence finishes the proof of Lemma 6.13. □

### 6.5.3 Showing that the OPT-execution path computes a good solution for $V_{\mathrm{big}}$

Next, we show that the OPT-execution path computes a good solution for the bottom instance, by showing that the dynamic program in step 13 of Algorithm 5 returns a good solution. To ensure that the bottom instance is $(1 + O(\varepsilon), 15 + O(\varepsilon))$-feasible in the OPT-execution path, we first show that $V_{\mathrm{big}}$ is covered by $A_{\mathrm{DP}}^{\mathrm{final}}$. This set may not yet satisfy the budget constraints, which will be ensured by the dynamic program.

**Lemma 6.18.** *On the* OPT*-execution path, the set $A_{\mathrm{DP}}^{\mathrm{final}}$ satisfies*

$$V_{\mathrm{big}} \subseteq \bigcup_{(v,\ell) \in A_{\mathrm{DP}}^{\mathrm{final}}} \mathrm{Ball}(v, \sigma r_\ell).$$

*Proof.* Let $v \in V_{\mathrm{big}}$. Then, by definition of $V_{\mathrm{big}}$, we have $\sum_{\ell=\hat{h}+1}^{L} \sum_{u \in \mathrm{Ball}(v,\beta_\ell r_\ell)} \overline{Y}_{u,\ell} < 1 - \varepsilon$. Recall that $\beta_\ell = \left(3 + \frac{4}{\lambda_\varepsilon}\right)(\lambda_\varepsilon + 1)^{\frac{1}{\lambda_\varepsilon}} \leq \sigma$ for each $\ell \in [\hat{h}]$, cf. Lemma 6.7. Also recall that $\overline{Y} = \frac{1}{N} \sum_{y \in Y} y$ is the mean of the sparsified points in the $A_{\mathrm{thin}}$-light iterations. In particular, as $N = \lceil \frac{3}{\varepsilon} \rceil$ we must have at least four $A_{\mathrm{thin}}$-light iterations in which $\sum_{\ell=\hat{h}+1}^{L} \sum_{u \in \mathrm{Ball}(v,\beta_\ell r_\ell)} y_{u,\ell} < 1$. We investigate these iterations in the following.

Consider a $A_{\mathrm{thin}}$-light iteration in which $\sum_{\ell=1}^{\hat{h}} \sum_{u \in \mathrm{Ball}(v,\beta_\ell r_\ell)} y_{u,\ell} > 0$. In particular, there is some $(u, \ell) \in \mathrm{supp}(y)$ with $\ell \in [\hat{h}]$ and $d(u,v) \leq \beta_\ell r_\ell$. Note that this implies $(u, \ell) \notin D$. We distinguish cases, based on how $(u, \ell)$ was classified in this iteration and show that in at most three $A_{\mathrm{thin}}$-light iterations we have $(u, \ell) \notin A_{\mathrm{DP}}^{\mathrm{total}}$. This finishes the proof, as $(v, \ell)$ is then covered by $(u, \ell) \in A_{\mathrm{DP}}^{\mathrm{total}}$, which is also part of $A_{\mathrm{DP}}^{\mathrm{final}}$ since $v \in V_{\mathrm{big}}$ and thus $v \notin V(C)$.

- If $(u, \ell) \in A_{\mathrm{DP}}^{\mathrm{total}}$, we are done.
- If $(u, \ell) \in A_{\mathrm{small}}$, then we add $(w, \ell')$ to $D$ for each $\ell' \in [\hat{h}]$ and $w \in \mathrm{Ball}(v, \sigma(r_\ell + r_{\ell'}))$. Hence, we have $\sum_{\ell=1}^{\hat{h}} \sum_{u \in \mathrm{Ball}(v,\beta_\ell r_\ell)} y_{u,\ell} = 0$ in all future iterations. Thus, this case can happen at most once.
- If $(u, \ell) \in A_{\mathrm{sep}}^{D}$ and $\ell \in [\check{h}]$, then we add $(w, \ell')$ to $D$ for each $\ell' \in [\check{h}]$ and $w \in \mathrm{Ball}(v, \sigma(r_\ell + r_{\ell'}))$. Hence, we have $\sum_{\ell=1}^{\check{h}} \sum_{u \in \mathrm{Ball}(v,\beta_\ell r_\ell)} y_{u,\ell} = 0$ in all future iterations. Thus, also this case can happen at most once.
- If $(u, \ell) \in A_{\mathrm{big}}^{\mathrm{close}}$, then there is a $(w, \ell') \in \mathrm{OPT} \setminus \mathrm{OPT}^C$ with $w \in \mathrm{Ball}(u, \sigma(r_\ell + r_{\ell'}))$ and $\ell' < \ell \leq \ell' + \kappa_2$. Hence, we add $(v^w, \ell')$ to $C$. Note that $d(v, v^w) \leq d(v, u) + d(u, w) + d(w, v^w) \leq \sigma r_\ell + \sigma(r_\ell + r_{\ell'}) + \sigma(r_\ell + r_{\ell'}) \leq \rho r_{\ell'}$. Hence, $v$ is covered by $(v^w, \ell') \in C$ and hence $v \notin V_{\mathrm{big}}$, a contradiction.

It remains to consider the case $(u, \ell) \in A_{\mathrm{sep}}^{D}$ and $\ell > \check{h}$. We finish the prove by showing that also this case can happen at most once. To derive a contradiction, let $(u', \ell')$ be another such pair in some later $A_{\mathrm{thin}}$-light iteration. Note that this imples $\ell' > \ell + \kappa$ by our update of $D$. Then, we have

$$\begin{aligned}
d(u, u') &\leq d(u, v) + d(v, u') \\
&\leq \beta_\ell r_\ell + \beta_\ell r_{\ell'} \\
&\leq \sigma(r_\ell - 2r_{\ell+\kappa}) + \sigma r_{\ell'} \\
&\leq \sigma(r_\ell - r_{\ell'}),
\end{aligned}$$

making use of the small buffer zone (i.e., $\sigma \geq \beta_\ell + \frac{\varepsilon}{2}$) and our choice of $\kappa$. Thus, $(u', \ell')$ is a successor of $(u, \ell)$ by definition, and hence part of $A_\mathrm{DP}^\mathrm{total}$, a contradiction. □

Lemma 6.18 shows that $A_\mathrm{DP}^\mathrm{final}$ covers $V_\mathrm{big}$. Hence, to show that there is a good solution covering $V_\mathrm{big}$, it suffices to show that we can cover $A_\mathrm{DP}^\mathrm{final}$. Then increasing the dilation of the radii by another factor of $\sigma$ will ensure that $V_\mathrm{big}$ is covered. This is formally shown in the following lemma.

**Lemma 6.19.** *If Theorem 6.10 returns a set $C_\mathrm{big}$ in step 13 of the algorithm, then $C_\mathrm{big}$ is a $(1, 5\sigma)$-feasible solution to the bottom instance $\left(V_\mathrm{big}, d, k^C + 2\varepsilon k, r\right)_{\leq \hat{h}}$.*

*Proof.* We have to check that the budget constraints are satisfied and that every vertex of $V_\mathrm{big}$ is covered. Consider first the budget constraints. Remember that we call Theorem 6.10 with budgets $k^C + 2\varepsilon$. Hence, by the first property of Theorem 6.10, we have $|(C_\mathrm{big})_{\leq \ell}| \leq (k^C + 2\varepsilon k)_{\leq \ell}$ for each $\ell \in [\hat{h}]$.

Next, consider a vertex $u \in V_\mathrm{big}$. Then, due to Lemma 6.18, there must be a pair $(v, \ell)$ in $A_\mathrm{DP}^\mathrm{final}$ such that $d(v, u) \leq \sigma r_\ell$. By the second property of Theorem 6.10, there is a $(v', \ell') \in C_\mathrm{big}$ with $\ell' \leq \ell$ and $d(v, v') \leq 4\sigma r_{\ell'}$. Thus, we have $d(u, v') \leq 4\sigma r_{\ell'} + \sigma r_\ell \leq 5\sigma r_{\ell'}$ by the triangle inequality. □

So, it remains to show that on the OPT-execution path, the dynamic program finds a solution $C_\mathrm{big}$. To do so, we have to analyze the set $A_\mathrm{DP}^\mathrm{final}$ further. We add vertices to $A_\mathrm{DP}^\mathrm{total}$ in three places of each $A_\mathrm{thin}$-light iteration of the OPT-execution path:

- We add $A_\mathrm{big}$ to $A_\mathrm{DP}^\mathrm{total}$.
- We add $A_\mathrm{non\text{-}sep}$ to $A_\mathrm{DP}^\mathrm{total}$.
- We add vertices of $A_\mathrm{sep}$ to $A_\mathrm{DP}^\mathrm{total}$ if they are successors of some pair.

We will first show, that pairs of the second and third type are covered by $C_\mathrm{add}$, i.e., we have

$$A_\mathrm{non\text{-}sep} \cup (A_\mathrm{sep} \cap A_\mathrm{DP}) \subseteq \bigcup_{(v,\ell) \in C_\mathrm{add}} \mathrm{Ball}(v, 4\sigma r_\ell).$$

First, for pairs in $A_\mathrm{non\text{-}sep}$, this follows from the definition of $A_\mathrm{thin}$ and Observation 6.9. For every $(v, \ell) \in A_\mathrm{non\text{-}sep}$, there is some $(w, \ell) \in A_\mathrm{thin}$ with $d(v, w) \leq 4\sigma r_\ell$ that we added to $C_\mathrm{add}$. Next, we consider pairs in $A_\mathrm{sep} \cap A_\mathrm{DP}$, where we get an even stronger property.

**Lemma 6.20.** *For each $(v, \ell) \in A_\mathrm{DP} \cap A_\mathrm{sep}$, there is a pair $(v', \ell') \in C_\mathrm{add}$ with $\ell' \leq \ell$ and $d(v, v') \leq r_{\ell'}$.*

*Proof.* Since $(v, \ell) \in A_\mathrm{DP} \cap A_\mathrm{sep}$, $(v, \ell)$ is successor of some pair $(\hat{v}, \hat{\ell})$. Let $(v', \ell')$ be the successor of $(\hat{v}, \hat{\ell})$ for which we added $(v', \hat{\ell} + \frac{\kappa}{2})$ to $C_\mathrm{add}$. Note that $\min\{\ell, \ell'\} \geq \hat{\ell} + \kappa$. In particular, $\hat{\ell} + \frac{\kappa}{2} \leq \ell$. We will show that $d(v, v') \leq r_{\hat{\ell} + \frac{\kappa}{2}}$.

Let $(w_1, \ell_1)$ and $(w_2, \ell_2)$ be two pairs in $\mathrm{OPT}_{\leq \hat{h}}$ with $\ell_1 \geq \ell$ and $\ell_2 \geq \ell'$ such that $w_1 \in \mathrm{Ball}(v, \sigma(r_\ell + r_{\ell_1}))$ and $w_2 \in \mathrm{Ball}(v', \sigma(r_{\ell'} + r_{\ell_2}))$. They both exist, as $(v, \ell)$ and $(v', \ell')$ are in $A_\mathrm{sep}$. Furthermore, $w_1 \in \mathrm{Ball}(\hat{v}, \sigma(r_{\hat{\ell}} + r_{\ell_1}))$ and $w_2 \in \mathrm{Ball}(\hat{v}, \sigma(r_{\hat{\ell}} + r_{\ell_2}))$, as $\mathrm{Ball}(v, \sigma r_\ell), \mathrm{Ball}(v', \sigma r_{\ell'}) \subseteq \mathrm{Ball}(\hat{v}, \sigma r_{\hat{\ell}})$ since both $(v, \ell)$ and $(v', \ell')$ are successors of $(\hat{v}, \hat{\ell})$. As $(\hat{v}, \hat{\ell})$ was added to $A_\mathrm{sep}$ in a previous iteration, we must have $d(w_1, w_2) \leq \mu r_{\hat{\ell} + \kappa}$. Then

$$\begin{aligned} d(v, v') &\leq d(v, w_1) + d(w_1, w_2) + d(w_2, v') \\ &\leq 2\sigma r_\ell + \mu r_{\hat{\ell} + \kappa} + 2\sigma r_{\ell'} \\ &\leq (4\sigma + 2) r_{\hat{\ell} + \kappa} \\ &\leq r_{\hat{\ell} + \frac{\kappa}{2}}, \end{aligned}$$

where we used $\mu = 2$ and $\min\{\ell, \ell'\} \geq \hat{\ell} + \kappa$ in the second inequality and our choice of $\kappa$ in the final inequality. □

Next, we bound the number of vertices that $C_{\text{add}}$ contains on each level.

**Lemma 6.21.** *We have* $|(C_{\text{add}})_{\leq \ell}| \leq 2\varepsilon k_{\leq \ell}$ *for each* $\ell \in [\hat{h}]$.

*Proof.* There are two places where we add vertices to $C_{\text{add}}$. First of all, in every $A_{\text{thin}}$-light iteration, we add $A_{\text{thin}}$ to $C_{\text{add}}$. Note that for each $\ell \in [\hat{h}]$, we have $|(A_{\text{thin}})_\ell| \leq \frac{\varepsilon}{N} k_\ell$, as else we would have performed an $A_{\text{thin}}$-heavy iteration. Summing over all $A_{\text{thin}}$-light iterations yields at most $\varepsilon k_\ell$ vertices of level $\ell$ that we add to $C_{\text{add}}$ this way.

Second, for each $\ell \in \{\check{h}+1, ..., \hat{h}\}$ and each pair $(v, \ell)$ that is in $A_{\text{sep}}$ in some $A_{\text{thin}}$-light iteration, we add at most one vertex of level $\ell + \frac{\kappa}{2}$ to $C_{\text{add}}$. Let $A_{\text{sep}}^{\text{total}}$ denote the union of all sets $A_{\text{sep}}$ over all $A_{\text{thin}}$-light iterations in the OPT-execution path. We have to bound the size of $(A_{\text{sep}}^{\text{total}})_{\leq \ell} \setminus (A_{\text{sep}}^{\text{total}})_{\leq \check{h}}$ for each $\ell \in \{\check{h}+1, ..., \hat{h}\}$. To do so, consider a specific $A_{\text{thin}}$-light iteration and let $y \in Q_{1+7\varepsilon, \beta}(V \setminus V(C))$ be the sparsified $(C, D)$-compatible point used in this iteration. Note that by Lemma 6.7 we have $y_{v,\ell} \geq \frac{\varepsilon}{\lambda_\varepsilon}$ for each $\check{h} < \ell \leq \hat{h}$ and each $(v, \ell) \in \text{supp}(y)$. Thus, for $\check{h} < \ell \leq \hat{h}$, we add at most $\frac{\lambda_\varepsilon}{\varepsilon} y\big(V \times \{\check{h}+1, ..., \ell\}\big) \leq \frac{(1+7\varepsilon)\lambda_\varepsilon}{\varepsilon} k_{\leq \ell}$ vertices to $(A_{\text{sep}}^{\text{total}})_{\leq \ell} \setminus (A_{\text{sep}}^{\text{total}})_{\leq \check{h}}$. Hence, in the end

$$\left|(A_{\text{sep}}^{\text{total}})_{\leq \ell} \setminus (A_{\text{sep}}^{\text{total}})_{\leq \check{h}}\right| \leq \frac{N(1+7\varepsilon)\lambda_\varepsilon}{\varepsilon} k_{\leq \ell} \leq \varepsilon k_{\leq \ell + \frac{\kappa}{2}}$$

by choice of $\kappa$. Hence, we add at most $\varepsilon k_{\leq \ell + \frac{\kappa}{2}}$ many vertices of level at most $\ell + \frac{\kappa}{2}$ to $C_{\text{add}}$ this way. □

It remains to cover the vertices of $A_{\text{big}}$ that are part of $A_{\text{DP}}^{\text{final}}$. To do so, note that for every pair $(v, \ell) \in A_{\text{big}}$, there is some $(w, \ell') \in \text{OPT}_{\leq \hat{h}}$ with $\ell' \leq \ell - \kappa_2$ and $d(v, w) \leq \sigma(r_\ell + r_{\ell'})$. Set $\psi(v, \ell) = (w, \ell')$ for such a pair. Note that for any two $(v_1, \ell_1), (v_2, \ell_2) \in \psi^{-1}(w, \ell')$, we have $d(v_1, v_2) \leq \sigma\big(2r_{\ell'} + r_{\ell_1} + r_{\ell_2}\big)$ by the triangle inequality. We will use this in the following.

We assembled all the ingredients to show that the dynamic program in step 13 of Algorithm 5 returns a good solution for the bottom instance on the OPT-execution path.

**Lemma 6.22.** *On the* OPT*-execution path, the call to Theorem 6.10 in step 13 of the algorithm returns a set* $C_{\text{big}}$.

*Proof.* We will explicitly construct a set $C_{\text{big}}$ that satisfies the conditions of Theorem 6.10. Then, by the correctness of Theorem 6.10, the dynamic program will return such a set.

For every $(w, \ell') \in \text{OPT}_{\leq \hat{h}} \setminus \text{OPT}^C$, for which $\psi^{-1}(w, \ell')$ is non-empty, we pick an arbitrary $(v, \ell) \in \psi^{-1}(w, \ell')$ and add $(v, \ell')$ to $C_{\text{big}}$. Additionally, we add all pairs in $C_{\text{add}}$ to $C_{\text{big}}$. We have to show that this set satisfies the conditions of Theorem 6.10, i.e.,

- $C_{\text{big}}$ satisfies the budget constraints, i.e., for every $\ell \in [\hat{h}]$, we have $|(C_{\text{big}})_{\leq \ell}| \leq (k^C + 2\varepsilon k)_{\leq \ell}$.
- For every $(v, \ell) \in A_{\text{DP}}^{\text{final}}$, there is a $(v', \ell') \in C_{\text{big}}$ with $\ell' \leq \ell$ and $d(v, v') \leq 4\sigma r_{\ell'}$.

Note that then Theorem 6.10 is guaranteed to return such a set $C_{\text{big}}$.

To see the first property, note that by our invariant we have $|C_\ell| = |\text{OPT}_\ell^C|$ for each $\ell \in [\hat{h}]$. By Lemma 6.21, we have $|(C_{\text{add}})_{\leq \ell}| \leq 2\varepsilon k_{\leq \ell}$ for each $\ell \in [\hat{h}]$. Hence, $|(C_{\text{big}})_{\leq \ell}| \leq (k_{\leq \ell} - |\text{OPT}_{\leq \ell}^C|) + 2\varepsilon k_{\leq \ell} = (k^C + 2\varepsilon k)_{\leq \ell}$ for each $\ell \in [\hat{h}]$.

To show the second property, recall that $(v, \ell) \in A_{\text{DP}}^{\text{final}}$ implies that in some $A_{\text{thin}}$-light iteration, we have $(v, \ell) \in A_{\text{big}} \cup A_{\text{non-sep}} \cup (A_{\text{sep}} \setminus A_{\text{sep}}^D)$. If $(v, \ell) \in A_{\text{sep}} \setminus A_{\text{sep}}^D$, then by Lemma 6.20 there is a $(v', \ell') \in C_{\text{add}}$ with $\ell' \leq \ell$ and $d(v, v') \leq r_{\ell'} \leq 4\sigma r_{\ell'}$, finishing this case. If $(v, \ell) \in A_{\text{non-sep}}$, then by definition of $V_{\text{thin}}$ we add a $(v', \ell)$ to $C_{\text{add}}$ with $d(v, v') \leq 4\sigma r_\ell$ in this $V_{\text{thin}}$-light iteration, finishing this case as well.

Hence assume $(v, \ell) \in A_{\text{big}} \cap A_{\text{DP}}^{\text{final}}$. Then there is some $(w, \ell') \in \text{OPT}_{\leq \hat{h}}$ with $\ell' \leq \ell - \kappa_2$ and $d(v, w) \leq \sigma(r_\ell + r_{\ell'})$. First we show that $w \in \text{OPT}_{\leq \hat{h}} \setminus \text{OPT}^C$. To derive a contradiction, assume that $w \in \text{OPT}^C$. Then there is a pair $(v', \ell') = \varphi(w, \ell') \in C$ with $d(v', w) \leq 4\sigma r_{\ell'}$. Then

$$d(v, v') \leq d(v, w) + d(w, v') \leq \sigma(r_\ell + r_{\ell'}) + 4\sigma r_{\ell'} \leq 5\sigma r_{\ell'} + \sigma r_\ell$$

by the triangle inequality. By choice of $\rho$ and $\kappa_2$ (recall that $\rho = 15 + 6\varepsilon$ and $\kappa_2 = \lceil \log_{1+\varepsilon}(\frac{2\sigma}{\varepsilon}) \rceil$), we have $5\sigma r_{\ell'} + 2\sigma r_\ell \leq \rho r_{\ell'}$ and thus $\text{Ball}(v, \sigma r_\ell) \subseteq \text{Ball}(v', \rho r_{\ell'}) \subseteq V(C)$, a contradiction to $(v, \ell) \in A_{\text{DP}}^{\text{final}}$.

Thus, $w \in \text{OPT}_{\leq \hat{h}} \setminus \text{OPT}^C$, and by construction of $C_{\text{big}}$ we chose a representative $(\overline{v}, \overline{\ell}) \in \psi^{-1}(w, \ell')$ to be in $C_{\text{big}}$. Then, as $\psi(\overline{v}, \overline{\ell}) = \psi(v, \ell) = (w, \ell')$, we have $d(v, \overline{v}) \leq \sigma(2r_{\ell'} + r_\ell + r_{\overline{\ell}}) \leq 4\sigma r_{\ell'}$ by the triangle inequality. Thus, the second property holds as well. □

### 6.5.4 Proof of Theorem 6.11

Now, we are ready to prove Theorem 6.11.

*Proof of Theorem 6.11.* By Lemma 6.14, the execution path described above is one of the execution paths of EFFICIENT-THINNED-EXPLORATION-SNUkC$(\emptyset, \emptyset, \overline{\zeta}, \emptyset, \emptyset)$. We will show that the output $C \cup C_{\text{big}} \cup C_{\text{small}}$ is $(1 + 14\varepsilon, 5\sigma + \varepsilon)$-feasible.

By Lemma 6.19 and Lemma 6.22, $C \cup C_{\text{big}}$ is $(1 + 2\varepsilon, 5\sigma + \varepsilon)$-feasible for $(V \setminus V_{\text{small}}(\overline{Y}), d, k, r)_{\leq \hat{h}}$. Note that $\overline{Y}$ scaled with factor $\frac{1}{1-\varepsilon}$ induces a fractional $(\frac{1+7\varepsilon}{1-\varepsilon}, 1)$-feasible solution to the top instance. Hence, by Lemma 6.5 and using $\varepsilon \leq \frac{1}{7}$, $C_{\text{small}}$ is a $(1 + 14\varepsilon, 8)$-feasible solution to the top instance. Combining both solutions yields a $(1 + 14\varepsilon, 5\sigma + \varepsilon)$-feasible solution to the original instance. □

## 6.6 Proof of runtime (Theorem 6.12)

We will now prove Theorem 6.12. The proof is very similar as for SRMFC-T. However, we additionally need to guarantee that each call to Theorem 6.10 has polynomial running time.

We will first analyze the size of $\text{supp}(y)$, the support of the sparsified fractional solution $y$ computed in line 2 of Algorithm 5. To be precise, we observe that its size is linear in $L$ (up to a constant depending on $\varepsilon$).

**Observation 6.23.** *By Lemma 6.7, we have $|\text{supp}(y)_{\leq \ell}| \leq \lambda_\varepsilon \check{h} \frac{1+7\varepsilon}{\varepsilon} k_{\leq \ell}$ for each $\ell \leq \check{h}$, and $|\text{supp}(y)_{\leq \ell}| \leq \lambda_\varepsilon \frac{1+7\varepsilon}{\varepsilon} k_{\leq \ell}$ for each $\check{h} < \ell \leq \hat{h}$. In particular, this implies*

$$|\text{supp}(y)_{\leq \hat{h}}| \leq \lambda_\varepsilon \frac{1+7\varepsilon}{\varepsilon} \left(\check{h} k_{\leq \check{h}} + k_{\leq \hat{h}}\right)$$

$$\leq \lambda_\varepsilon \frac{(1+7\varepsilon)(1+\varepsilon)}{\varepsilon^2} \left(\check{h} k_{\check{h}} + k_{\hat{h}}\right)$$

$$\leq \lambda_\varepsilon \frac{(1+7\varepsilon)(1+\varepsilon)}{\varepsilon^3} \left(\check{h}\hat{h} + L\right)$$

$$\leq a_\varepsilon L,$$

*for a constant $a_\varepsilon$ depending only on $\varepsilon$. Here we used that $\hat{h} = O_\varepsilon(\log L)$, $\check{h} = O_\varepsilon(\log \log L)$, and $k_{\leq \ell} \leq \frac{1+\varepsilon}{\varepsilon} k_\ell$ for $\ell \in [L]$.*

Using Observation 6.23, we get that each call of Theorem 6.10 runs in time polynomial in $n$ as $|V_{\text{DP}}| \leq |A_{\text{DP}}^{\text{total}}| \leq Na_\varepsilon L$. Thus, it suffices to bound the number of calls to EFFICIENT-THINNED-EXPLORATION-

SNUkC over the whole run of the algorithm. Indeed, we can efficiently compute a point in the polytope $Q_1^D(\Gamma)$ or decide that it is empty. Moreover, Lemma 6.7 guarantees polynomial running time.

**Theorem 6.24.** *The number of calls to* Efficient-Thinned-Exploration-SNUkC *over the whole run of* Efficient-Thinned-Exploration-SNUkC$(\emptyset, \emptyset, \zeta, \emptyset, \emptyset)$ *for* $\zeta \in \mathbb{Z}_{\geq 1}$ *is bounded by* $2^{O_\varepsilon(N(\zeta+L))}$.

Choosing $\zeta = \overline{\zeta}$, the number of recursive calls is polynomial in $n$ for fixed $\varepsilon$, proving Theorem 6.12. The proof of Theorem 6.24 is very similar to the proof of Theorem 3.9 in Section 5.

*Proof.* For $\zeta, s \in \mathbb{Z}_{\geq 0}$ with $s \leq N$, let $\mathcal{P}(\zeta, s)$ denote the maximum number of calls to Efficient-Thinned-Exploration-SNUkC over the whole run of the algorithm when calling Efficient-Thinned-Exploration-SNUkC$(C, D, \zeta, Y, A_{\text{DP}}^{\text{total}})$, for any choice of $C, D$, and $A_{\text{DP}}^{\text{total}}$ and for any $Y$ with $|Y| = N - s$, including the initial call of Efficient-Thinned-Exploration-SNUkC$(C, D, \zeta, Y, A_{\text{DP}}^{\text{total}})$ itself. We claim that
$$\mathcal{P}(\zeta, s) \leq C_\varepsilon^\zeta \cdot D_\varepsilon^s$$
for $C_\varepsilon = 3^{\frac{N}{\varepsilon^3}+1}$ and $D_\varepsilon = (\kappa_2 + 4)^{a_\varepsilon L+1}$. Note that this implies the statement of Theorem 3.9.

We prove this claim by induction on $\zeta + s$. For the base case $\zeta = s = 0$, we do not have a recursive call to Efficient-Thinned-Exploration-SNUkC, and thus $\mathcal{P}(\zeta, s) = 1$, which implies the claimed bound. For the induction step, we analyze all recursive calls to Efficient-Thinned-Exploration-SNUkC. Set $S_\ell = \text{supp}(y)_\ell$ for each $\ell \in [L]$ and $S = \text{supp}(y)_{\leq \hat{h}}$. Let $I \subseteq [L]$ denote the set of indices $\ell \in [L]$ with $|S_\ell| \geq \frac{\varepsilon}{N} k_\ell$ and $|S_\ell| \leq \zeta$. First, for each level $\ell \in I$, we perform a $A_{\text{thin}}$-heavy iteration on level $\ell$ and reduce $\zeta$ by $|S_\ell|$. Thus, we can bound the number of recursive calls to Efficient-Thinned-Exploration-SNUkC in line 6 of Algorithm 5, over all choices of $\ell$, by
$$\sum_{\ell \in I} 3^{|S_\ell|} \mathcal{P}(\zeta - |S_\ell|, s),$$
using the number of options in line 4 of Algorithm 5 to choose $A_{\text{thin}}^\ell \subseteq S_\ell$ with $|A_{\text{thin}}^\ell| \geq \frac{\varepsilon}{N} k_\ell$ and $A_{\text{thin}}^C \subseteq A_{\text{thin}}^\ell$ is upper bounded by $3^{|S_\ell|}$, as each pair in $S_\ell$ is in one of the following three sets: (i) $S_\ell \setminus A_{\text{thin}}^\ell$, (ii) $A_{\text{thin}}^\ell \setminus A_{\text{thin}}^C$, or (iii) $A_{\text{thin}}^C$. Applying the induction hypothesis, we can upper bound this by
$$\sum_{\ell \in I} 3^{|S_\ell|} C_\varepsilon^{\zeta - |S_\ell|} D_\varepsilon^s = C_\varepsilon^\zeta D_\varepsilon^s \sum_{\ell \in I} \left(\frac{3}{C_\varepsilon}\right)^{|S_\ell|}. \tag{8}$$
Note that for each $\ell \in I$, we have
$$|S_\ell| \geq \frac{\varepsilon}{N} k_\ell = \frac{\varepsilon}{N}(k_{\leq \ell} - k_{\leq \ell-1}) \geq \frac{\varepsilon}{N}((1+\varepsilon)^{\ell-1} - (1+\varepsilon)^{\ell-2}) \geq \frac{\varepsilon^2(1+\varepsilon)^{\ell-2}}{N}, \tag{9}$$
where we use $k_1 \geq 1$ and $k_{\ell+1} \geq (1+\varepsilon)k_\ell$ for each $\ell \in [L-1]$ in the second inequality. This allows us to upper bound the sum on the right-hand side of (8) by
$$\sum_{\ell \in I} \left(\frac{3}{C_\varepsilon}\right)^{|S_\ell|} \leq \sum_{\ell=1}^{\hat{h}} \left(\frac{3}{C_\varepsilon}\right)^{\frac{\varepsilon^2(1+\varepsilon)^{\ell-2}}{N}} = \sum_{\ell=1}^{\hat{h}} \left(\frac{1}{3}\right)^{\frac{(1+\varepsilon)^{\ell-2}}{\varepsilon}} \leq \sum_{\ell=1}^{\hat{h}} \left(\frac{1}{3}\right)^{\frac{1+(\ell-2)\varepsilon}{\varepsilon}} = \left(\frac{1}{3}\right)^{\frac{1}{\varepsilon}-2} \sum_{\ell=1}^{\hat{h}} \left(\frac{1}{3}\right)^\ell$$
$$\leq \left(\frac{1}{3}\right)^{\frac{1}{\varepsilon}-2} \leq \frac{1}{3},$$
where the first inequality uses (9) and $C_\varepsilon \geq 3$, the first equality follows from $C_\varepsilon = 3^{\frac{N}{\varepsilon^3}+1}$, and the last inequality holds because $\varepsilon \leq \frac{1}{3}$. Thus, together with (8), we can bound the number of recursive calls during the $A_{\text{thin}}$-heavy iterations by

$$\frac{1}{3}C_\varepsilon^\zeta D_\varepsilon^s.$$

Finally, if $s > 0$, we apply a $A_{\text{thin}}$-light iteration on $S$ and reduce $s$ by one. Each pair $(v, \ell)$ of $S$, has to be in one of the following 4 sets: (i) $A_{\text{big}}^{\text{close}}$, (ii) $A_{\text{small}}$, (iii) $A_{\text{sep}}^D$, or (iv) $A_{\text{DP}}$. In the first case we can additionally add $(v, \ell)$ to $A_{\text{big}}^C$ with a choice of $f(v, \ell) \in \{1, ..., \kappa_2\}$, which adds another $\kappa_2$ choices. We perform at most $\mathcal{P}(\zeta, s-1)$ recursive calls to EFFICIENT-THINNED-EXPLORATION-SNUkC in line 22 of Algorithm 5 for each combination of choices. Hence, we can bound the total number of recursive calls in this $V_{\text{thin}}$-light iteration by $(\kappa_2 + 4)^{|S|}\mathcal{P}(\zeta, s-1)$. By Observation 6.23, we have $|S| \leq a_\varepsilon L$. Combining this with the induction hypothesis, we get

$$(\kappa_2 + 4)^{|S|}\mathcal{P}(\zeta, s-1) \leq (\kappa_2 + 4)^{|S|} \cdot C_\varepsilon^\zeta D_\varepsilon^{s-1} \leq (\kappa_2 + 4)^{a_\varepsilon L} \cdot C_\varepsilon^\zeta D_\varepsilon^{s-1} = \frac{C_\varepsilon^\zeta D_\varepsilon^s}{\kappa_2 + 4},$$

by choice of $D_\varepsilon$.

Combining both cases, and adding the initial call to EFFICIENT-THINNED-EXPLORATION-SNUkC, we have shown that

$$\mathcal{P}(\zeta, s) \leq \frac{1}{3}C_\varepsilon^\zeta D_\varepsilon^s + \frac{1}{\kappa_2 + 4}C_\varepsilon^\zeta D_\varepsilon^s + 1 \leq C_\varepsilon^\zeta D_\varepsilon^s$$

as desired, where the second inequality holds because $C_\varepsilon \geq 2$, $D_\varepsilon \geq 2$, and $\zeta + s \geq 1$. □

# 7 Compression

In this section we prove the reductions to 1-feasible $\varepsilon$-compressed SRMFC-T and to $(1, 1)$-feasible $\varepsilon$-compressed SNUkC.

## 7.1 Compression for SRMFC-T

We start by proving the compression result for SRMFC-T.

**Theorem 7.1.** *Let $(G, r, B)$ be an SRMFC-T instance with $B_1 \geq 1$. Let $\varepsilon > 0$ be given. Then we can efficiently determine another SRMFC-T instance $(G', r', B')$ satisfying:*
- *$G'$ has $L' = \left\lceil \log_{1+\varepsilon}\left(\sum_{\ell=1}^L B_\ell\right)\right\rceil + 1$ many levels, where $L$ denotes the number of levels of $G$,*
- *the budget $B'_{\leq \ell}$ up to level $\ell$ is given by $B'_{\leq \ell} = (1 + \varepsilon)^{\ell - 1}$ for $\ell \in [L']$,*
- *for every $\alpha \in \mathbb{R}_{\geq 0}$, if $(G, r, B)$ is $\alpha$-feasible, then so is $(G', r', B')$, and*
- *for every $\alpha \in \mathbb{R}_{\geq 0}$, an $\alpha$-feasible solution of $(G', r', B')$ can be transformed efficiently into a $(1 + \varepsilon)\alpha$-feasible solution of $(G, r, B)$.*

*Proof of Theorem 7.1.* We first introduce the operations from [1] that we need in the proof. Given an instance $(G, r, B)$ of SRMFC-T, a *down-push* at level $\ell$ is obtained by setting the budget of level $\ell - 1$ to $B_\ell + B_{\ell-1}$ and setting the budget of level $\ell$ to 0. Observe that, as we move budget downwards, an $\alpha$-feasible solution to $(G, r, B)$ is also $\alpha$-feasible for the down-pushed instance. Note that in [1], this operation was called an *up-push*, as they considered trees growing downwards from the root.

Next, consider an instance $(G, r, B)$ of SRMFC-T that has a level $\ell$ with $B_\ell = 0$. The *contraction* of the instance at level $\ell$ is obtained by applying the following operation to every level-$\ell$-vertex $v$: Let $u$ be the predecessor of $v$ in the tree and let $w_1, ..., w_j$ be the successors of $v$ in the tree. If $v$ is a leaf, we remove all vertices from the subtree $T_u$ rooted at $u$, making $u$ a leaf. Else, we remove $v$ from the tree and add an edge between $u$ and each $w_i$. After processing every vertex, we shift the budgets starting from $\ell + 1$ down by one level to account for the index shift due to removing level $\ell$. Observe that every solution to the original instance gives rise to a solution of the contraction by identifying the vertices $u$ and $v$. Here we crucially use that $u$ is a leaf in the contracted instance, if $v$ was a leaf in the original instance. Furthermore, we use that $B_\ell = 0$, thus the budgets impose the same constraints on the vertices $u$ and $v$. Vice versa, every solution to the contraction gives rise to a solution of the original instance.

Additionally, we will need the following, new operation, which can be seen as a partial revert of a contraction. Consider any instance $(G, r, B)$ of SRMFC-T. A *splitting* is defined with respect to a level $\ell \in [L]$ and a new budget $B'_\ell$ with $0 \leq B'_\ell \leq B_\ell$. The splitting is obtained by splitting every vertex $v$ of level $\ell$ into two vertices $v_1$ and $v_2$, connected by an edge. The predecessor of $v_1$ is the predecessor of $v$ and the successors of $v_2$ are the successors of $v$. Note that this introduces a new level $\ell + 1$ and shifts all levels $\ell' > \ell$ up by one level. For this reason, we shift all budgets starting from $\ell + 1$ up by one level. Finally, we set the budget of level $\ell$ to $B'_\ell$ and the budget of level $\ell + 1$ to $B_\ell - B'_\ell$. Observe that the splitting of an SRMFC-T instance is equivalent to the original instance. This is due to the fact that for every vertex $v$ in level $\ell$ that is protected by a solution of the original instance, we can choose to protect $v_2$ in the split instance. Vice versa, if $v_1$ or $v_2$ is protected in the split instance, then we can choose to protect $v$ in the original instance.

Equipped with these operations, we are ready to prove the compression result. We will construct the instance $(G', r', B')$ from $(G, r, B)$ using the three operations described above, together with a slight modification of the budgets. Let $L' = \left\lceil \log_{1+\varepsilon} \sum_{\ell=1}^{L} B_\ell \right\rceil + 1$ denote the desired number of levels, where $L$ denotes the number of levels of $G$.

First, we apply splitting operations to obtain an instance $(G^{\text{split}}, r, B^{\text{split}})$ with $L^{\text{split}}$ levels such that for every $\ell \in [L']$ there is a level $k_\ell \in [L^{\text{split}}]$ such that $\sum_{j=1}^{k_\ell} B_j^{\text{split}} = (1+\varepsilon)^{\ell-1}$. This can be done the following way. First, we increase the budget on the final level by $(1+\varepsilon)^{L'-1} - \sum_{\ell=1}^{L} B_\ell$. Then, for every $\ell \in [L'-1]$ in descending order, we apply a splitting operation on level $k_\ell$ where $k_\ell$ is the smallest $k$ with $\sum_{j=1}^{k} B_j \geq (1+\varepsilon)^{\ell-1}$. The value of the split is chosen as $(1+\varepsilon)^{\ell-1} - \sum_{j=1}^{k_\ell - 1} B_j$. Note that $B_1 \geq 1$ implies $k_1 = 1$.

Next, we perform down-pushes from all levels $i \in \{k_{\ell-1} + 1, ..., k_\ell\}$ to level $k_{\ell-1} + 1$, for every $\ell \in [L']$ (using $k_0 := 0$). This gives us an instance $(G^{\text{split}}, r, B^{\text{push}})$ with the following properties:

- There are levels $k_1 < ... < k_{L'}$ such that $\sum_{i=1}^{k_{\ell-1}+1} B_i^{\text{push}} = (1+\varepsilon)^{\ell-1}$ for $\ell \in [L']$.
- All levels $\ell \notin \{k_{\ell-1} + 1 : \ell \in [L']\}$ satisfy $B_\ell^{\text{push}} = 0$.

Now contracting all levels with $B_\ell^{\text{push}} = 0$ gives us the desired instance $(G', r, B')$. We have to check the four properties.

The first and second property follow by construction. Let $\alpha \in \mathbb{R}_{\geq 0}$. To see the third property, note that increasing the budget of a level preserves $\alpha$-feasibility, as the budget constraints are looser. Additionally, splits and contractions yield equivalent instances and down-pushes of an $\alpha$-feasible instance are still $\alpha$-feasible.

Finally, consider the last property. We will show that any $\alpha$-feasible solution of $(G^{\text{split}}, r, B^{\text{push}})$ is $(1+\varepsilon)\alpha$-feasible for $(G^{\text{split}}, r, B^{\text{split}})$, even when not using the budget of the last level of the latter. Assuming the claim, the property follows. The former instance is equivalent to $(G', r, B')$. On the other hand, every $\alpha$-feasible solution to the latter instance that does not use the budget of the last level gives rise to an $\alpha$-feasible solution to $(G, r, B)$. This holds, as we only increase the budget on the last level, and the budget of this level is not used by the solution. All remaining transformations from $(G, r, B)$ to $(G^{\text{split}}, r, B^{\text{split}})$ give equivalent instances.

To show the claim, let $R$ be an $\alpha$-feasible solution to $(G^{\text{split}}, r, B^{\text{push}})$. For $\ell \in [L'-1]$ and $k_\ell \leq k < k_{\ell+1}$ observe that

$$(1+\varepsilon)\sum_{j=1}^{k} B_j^{\text{split}} \geq (1+\varepsilon)\sum_{j=1}^{k_\ell} B_j^{\text{split}} \geq (1+\varepsilon)^\ell = \sum_{j=1}^{k_\ell+1} B_j^{\text{push}} = \sum_{j=1}^{k+1} B_j^{\text{push}}.$$

As observed above, $B_1 \geq 1$ implies $k_1 = 1$. Moreover, by construction, $k_{L'}$ equals the number of levels of $G^{\text{split}}$. Thus, the above argumentation holds for all levels $k \in [k_{L'} - 1]$. Hence, $R$ is $(1+\varepsilon)\alpha$-feasible for $(G^{\text{split}}, r, B^{\text{split}})$ even when not using the budget of the last level. □

Now, we are ready to prove the reduction to 1-feasible $\varepsilon$-compressed instances.

*Proof of Theorem 2.4.* Given an instance $(G, r, B)$ of SRMFC-T with $L$ levels and $n$ vertices, let $\alpha_{\text{OPT}}$ denotes its optimal stretch. Note that for at least one level $\ell \in [L]$, $\alpha_{\text{OPT}} B_\ell$ is an integer, by optimality of $\alpha_{\text{OPT}}$. Moreover, $\alpha_{\text{OPT}} B_\ell \leq n$. Thus, we can guess $\alpha_{\text{OPT}}$ by trying at most $nL$ different values. Now, consider the instance $\left(G, r, \frac{B}{\alpha_{\text{OPT}}}\right)$, which is 1-feasible by construction. As long as $B_1 < 1$, push this budget to the second level and contract the first level into the root. (See the proof of Theorem 7.1 for a formal description of this operation.)

To the resulting 1-feasible instance $(G', r, B')$, apply Theorem 7.1 to obtain a 1-feasible $\varepsilon$-compressed instance of SRMFC-T. Let $R_\varepsilon$ denote the solution to this instance computed by the $\alpha$-approximation algorithm. By, Theorem 7.1, we can efficiently transform $R_\varepsilon$ into a $((1+\varepsilon)\alpha)$-feasible solution $R$ of $(G', r, B')$. Note that this solution is $((1+\varepsilon)\alpha\alpha_{\text{OPT}})$-feasible for $(G, r, B)$. □

## 7.2 Compression for SNUkC

Next, we prove the compression result for SNUkC.

**Theorem 7.2.** *Let $(V, d, k, r)$ be an instance of SNUkC with $L$ levels for which $k_1 \geq 1$. Let $\varepsilon > 0$ and $h \in \mathbb{Z}_{\geq 0}$ be given. Then we can efficiently determine an instance $(V, d, k', r')$ of SNUkC such that:*
- $k' \in \mathbb{R}_{>0}^{L'}$ *and* $r' \in \mathbb{R}_{\geq 0}^{L'}$ *with* $L' \leq \left\lceil \log_{1+\varepsilon} \sum_{\ell=1}^{L} k_\ell \right\rceil + 1$,
- *the budget $k'_{\leq \ell}$ up to level $\ell$ is given by $k'_{\leq \ell} = (1+\varepsilon)^j$ for some $j \in \mathbb{Z}_{\geq 0}$,*
- *for $\ell \in [h-1]$, we have $r'_{\ell+1} \geq (1+\varepsilon)r'_\ell$,*
- *for every $\alpha, \beta \in \mathbb{R}_{\geq 0}$, if the instance $(V, d, k, r)$ is $(\alpha, \beta)$-feasible then so is $(V, d, k', r')$,*
- *for every $\alpha, \beta \in \mathbb{R}_{\geq 0}$, an $(\alpha, \beta)$-feasible solution of $(V, d, k', r')$ can be transformed efficiently into an $((1+\varepsilon)\alpha, (1+\varepsilon)\beta)$-feasible solution of $(V, d, k, r)$.*

Whereas most parts of the proof closely follow the proof of Theorem 7.1, we start by aggregating radii to ensure that in the compressed instance, the radii differ by at least a factor $1+\varepsilon$. We exploit this property in the analysis of our final algorithm to keep the approximation factor we lose on the size of the radii small. With this property, we cannot guarantee $k'_{\leq \ell} = (1+\varepsilon)^{\ell-1}$ for each $\ell \in [L']$. Instead, we guarantee that $k'_{\leq \ell+1} \geq (1+\varepsilon)k'_{\leq \ell}$ for each $\ell \in [L'-1]$, and $k'_1 \geq 1$.

*Proof of Theorem 7.2.* We first introduce operations that we need in the proof. Given an instance $(V, d, k, r)$ of SNUkC, a *down-push* at level $\ell$ is obtained by setting the budget of level $\ell - 1$ to $k_\ell + k_{\ell-1}$ and setting the budget of level $\ell$ to 0. Observe that, as we move budget downwards, a feasible solution to $(V, d, k, r)$ is also feasible for the down-pushed instance.

Next, consider an instance $(V, d, k, r)$ of SNUkC that has a level $\ell$ with $k_\ell = 0$. The *contraction* of the instance at level $\ell$ is obtained by shifting the budgets and the radii starting from $\ell + 1$ down by one level to account for the index shift due to removing the level $\ell$. Observe that this operation results in an equivalent instance.

Additionally, we will need the following operation, which can be seen as a partial revert of a contraction. Consider any instance $(V, d, k, r)$ of SNUkC. A *splitting* is defined with respect to a level $\ell \in [L]$ and a new budget $k'_\ell$ with $0 \leq k'_\ell \leq k_\ell$. The splitting is obtained by replacing $k_\ell$ by two budgets $k'_\ell$ and $k_\ell - k'_\ell$ with the same radius and adapting the budget vector and radius vector accordingly. Observe that the resulting instance is equivalent to the original instance.

Equipped with these operations, we are ready to prove the compression result. We will construct the instance $(V, d, k', r')$ from $(V, d, k, r)$ using the operations described above, together with a slight modification of the budgets and the radii. Let $\hat{L} := \left\lceil \log_{1+\varepsilon} \sum_{\ell=1}^{L} k_\ell \right\rceil + 1$ denote the desired bound on the number of levels, where $L$ denotes the number of levels of the original instance.

First, increase each non-zero radius $r_\ell$ with $\ell \in [h]$ by at most a factor $1 + \varepsilon$, so that afterwards it equals $(1+\varepsilon)^j$ for some $j \in \mathbb{Z}$. Let $(V, d, k, \tilde{r})$ denote the resulting instance of SNUkC.

Afterwards, we apply splitting operations to obtain an instance $(V, d, k^{\text{split}}, r^{\text{split}})$ with $L^{\text{split}}$ levels such that for every $\ell \in [\hat{L}]$ there is a level $i_\ell \in [L^{\text{split}}]$ such that $\sum_{j=1}^{i_\ell} k_j^{\text{split}} = (1+\varepsilon)^{\ell-1}$. This can be done the following way. First, we increase the budget on the final level by $(1+\varepsilon)^{\hat{L}-1} - \sum_{\ell=1}^{L} k_\ell$. Then, for every $\ell \in [\hat{L}-1]$ in descending order, we apply a splitting operation on level $i_\ell$ where $i_\ell$ is smallest $i$ with $\sum_{j=1}^{i} k_j \geq (1+\varepsilon)^{\ell-1}$. The value of the split is chosen as $(1+\varepsilon)^{\ell-1} - \sum_{j=1}^{i_\ell - 1} k_j$. Note that $k_1 \geq 1$ implies $i_1 = 1$.

Next, we perform down-pushes from all levels $j \in \{i_{\ell-1}+1, ..., i_\ell\}$ to level $i_{\ell-1}+1$, for every $\ell \in [\hat{L}]$ (using $i_0 := 0$). This gives us an instance $(V, d, k^{\text{push}}, r^{\text{push}})$ with the following properties:
- There are levels $i_1 < ... < i_{\hat{L}}$ such that $\sum_{j=1}^{i_{\ell-1}+1} k_j^{\text{push}} = (1+\varepsilon)^{\ell-1}$ for $\ell \in [\hat{L}]$.
- All levels $\ell \notin \{i_{\ell-1}+1 : \ell \in [\hat{L}]\}$ satisfy $k_\ell^{\text{push}} = 0$.

Then, for all levels with the same radius, move all their budget to the smallest index with this radius. Now, contracting all levels with $k_\ell^{\text{push}} = 0$ gives us the desired instance $(V, d, k', r')$ with $L' \leq \hat{L}$ levels. We have to check the five properties.

The first three properties follow by construction. Let $\alpha, \beta \in \mathbb{R}_{\geq 0}$. To see the fourth property, note that increasing the budget and the radius of a level preserves $(\alpha, \beta)$-feasibility, as the budget constraints are looser. Additionally, splits and contractions yield equivalent instances and down-pushes of an $(\alpha, \beta)$-feasible instance are still $(\alpha, \beta)$-feasible.

Finally, consider the last property. We will show that any $(\alpha, \beta)$-feasible solution of $(V, d, k^{\text{split}}, r^{\text{push}})$ is $((1+\varepsilon)\alpha, \beta)$-feasible for $(V, d, k^{\text{split}}, r^{\text{split}})$, even when not using the budget of the last level of the latter. Assuming the claim, the property follows. Any solution to the former instance is also $(\alpha, (1+\varepsilon)\beta)$-feasible for $(V, d, k, r)$, as we only increased the radii by at most a factor $1 + \varepsilon$ and perform equivalent transformations. On the other hand, every $(\alpha, \beta)$-feasible solution to the latter instance that does not use the budget of the last level gives rise to an $(\alpha, \beta)$-feasible solution to $(V, d, k, \tilde{r})$ (which is $(\alpha, (1+\varepsilon)\beta)$-feasible to $(V, d, k, r)$). This holds, as we only increase the budget on the last level, and the budget of this level is not used by the solution. All remaining transformations from $(V, d, k, \tilde{r})$ to $(V, d, k^{\text{split}}, r^{\text{split}})$ give equivalent instances.

To show the claim, let $C$ be an $(\alpha, \beta)$-feasible solution to $(V, d, k^{\text{split}}, r^{\text{push}})$. For $\ell \in [L'-1]$ and $i_\ell \leq i < i_{\ell+1}$ observe that

$$(1+\varepsilon) \sum_{j=1}^{i} k_j^{\text{split}} \geq (1+\varepsilon) \sum_{j=1}^{i_\ell} k_j^{\text{split}} \geq (1+\varepsilon)^\ell = \sum_{j=1}^{i_{\ell+1}} k_j^{\text{push}} = \sum_{j=1}^{i+1} k_j^{\text{push}}.$$

As observed above, $k_1 \geq 1$ implies $i_1 = 1$. Moreover, by construction, $i_{L'}$ equals the number of entries in $k^{\text{split}}$. Thus, the above argumentation holds for all levels $i \in [i_{L'} - 1]$. Hence, $C$ is $((1+\varepsilon)\alpha, \beta)$-feasible to $(V, d, k^{\text{split}}, r^{\text{split}})$ even when not using the budget of the last level. □

Using this theorem, we show the reduction to $(1, 1)$-feasible $\varepsilon$-compressed instances.

*Proof of Theorem 6.4.* Given an instance $(V, d, k, r)$ of SNUkC with $L$ levels and $n$ vertices, let $\beta_{\text{OPT}}$ denotes its optimal dilation. Note that for the optimal dilation there are vertices $u, v \in V$ and $\ell \in [L]$ such that $\beta_{\text{OPT}} d(u, v) = r_\ell$. Thus, we can guess $\beta_{\text{OPT}}$ by trying at most $\frac{n(n-1)}{2} L$ different values. Now, consider the instance $(V, d, k, \beta_{\text{OPT}} r)$, which is $(1, 1)$-feasible by construction.

As long as $k_1 < 1$, push this budget to the second level and delete the first level. This yields an equivalent instance, as no feasible solution can use the first level if $k_1 < 1$.

To the resulting $(1, 1)$-feasible instance $(V, d, k', r')$, apply Theorem 7.2 to obtain a $(1, 1)$-feasible $\varepsilon$-compressed instance of SNUkC. Let $C_\varepsilon$ denote the solution to this instance computed by the $(\alpha, \beta)$-approximation algorithm. By, Theorem 7.2, we can efficiently transform $C_\varepsilon$ into a $((1+\varepsilon)\alpha, (1+\varepsilon)\beta)$-feasible solution $C$ of $(V, d, k', r')$. Note that this solution is $((1+\varepsilon)\alpha, (1+\varepsilon)\beta\beta_{\text{OPT}})$-feasible for $(V, d, k, r)$. □

## 8 Dynamic program

In this section, we prove Theorem 2.6 and Theorem 6.10.

*Proof of Theorem 2.6.* We build a dynamic programming table top-down. Starting from $L$ and going to $0$ we consider every level $\ell \in [L] \cup \{0\}$ (using $V_0 := \{r\}$). For every vertex $v$ on level $\ell$ we store all budget vectors $C = (c_\ell, ..., c_L) \in \mathbb{Z}_{\geq 0}^{L-\ell+1}$, with $c_i \leq \left\lfloor \sum_{j=1}^{i} B_j \right\rfloor - \left\lfloor \sum_{j=1}^{i-1} B_j \right\rfloor$ for $\ell \leq i \leq L$ (in which case we call $C$ *feasible* with respect to budgets $B$), and for which there is a feasible solution to the SRMFC-T problem on the sub-tree $T_v$ rooted at $v$ with budget vector $C$. Here, we allow to fireproof $v$ if $c_\ell > 0$. In particular, we store at most $(B_{\max} + 1)^L$ many budget combinations for each vertex $v$, where $B_{\max} := \max_{\ell \in [L]} B_\ell$.

In order to compute the entry for a vertex $v$ on level $\ell$, we start with a list $L_v^0$ containing only the all-zero vector and iterate over the successors $u_1, ..., u_k$ of $v$. We maintain the invariant that after processing the first $i$ successors, $L_v^i$ contains all budget combinations that correspond to sets that protect all vertices in the sub-trees $T_{u_1}, ..., T_{u_i}$, by fireproofing vertices in these sub-trees. For a successor $u_i$ we have to show how to update $L_v^{i-1}$ to $L_v^i$. Let $L_{u_i}$ be the list of budget combinations for $u_i$. Then, for each pair $(C_1, C_2)$ of budget combinations in $L_v^{i-1}$ and $L_{u_i}$, we add the combination $C_1 + C_2$ to $L_v^i$ if it is feasible with respect to the budgets $B$. This budget combination corresponds to combining a solution for $T_{u_1}, ..., T_{u_{i-1}}$ using budget $C_1$ with a solution for $T_{u_i}$ using budget $C_2$. This update can be done in time $O(|L_v| \cdot |L_{u_i}|) = O((B_{\max} + 1)^{2L})$. Set $L_v = L_v^k$. Additionally, if $\left\lfloor \sum_{j=1}^{\ell} B_j \right\rfloor - \left\lfloor \sum_{j=1}^{\ell-1} B_j \right\rfloor > 0$, we consider the case that we protect $v$ itself, using a single unit of budget on level $\ell$, and add the corresponding budget combination to $L_v$. From the invariant of the merging step it follows that $L_v$ now contains all budget combinations that allow to protect all vertices in $T_v$.

Any entry in $L_r$ gives rise to a feasible solution via backtracking. The merging step is performed once for each vertex $u$ in the tree, namely when processing its predecessor. Thus, the total runtime is bounded by $O(n(B_{\max} + 1)^{2L})$. □

Now, we prove Theorem 6.10. Note that for SRMFC-T, we heavily exploited the tree structure in the dynamic program. In particular, we used the divide-and-conquer approach to compute the budget combinations for the successors of a vertex. Since we do not have this tree structure anymore, we get worse running time guarantees. In Section 6, we compensated for this by drastically restricting the vertex set on which we run the dynamic program.

*Proof of Theorem 6.10.* For a set $C \subseteq V \times [L]$, we define its coverage $\text{cov}(C)$ as

$$\text{cov}(C) = \{(v, \ell) \in \mathcal{S} : \exists (v', \ell') \in C : \ell' \leq \ell \text{ and } d(v, v') \leq \beta r_{\ell'}\}.$$

Based on this, we set for $\ell \in [L]$

$$\mathcal{M}_\ell = \left\{ \text{cov}(C) : C \subseteq V \times \{\ell, ..., L\}, |C_{\ell'}| \leq \left\lfloor \sum_{j=1}^{\ell'} k_j \right\rfloor - \left\lfloor \sum_{j=1}^{\ell'-1} k_j \right\rfloor \forall \ell' \in \{\ell, ..., L\} \right\}$$

to be the set of subsets of $\mathcal{S}$ that can be covered by some $C \subseteq V \times [L]$ that does not use the budget of level below $\ell$.

Additionally, we maintain for each set $M \in \mathcal{M}_\ell$ a set $C_M$ of pairs $(v, \ell')$ such that $M = \text{cov}(C_M)$. Note that if different choices of $C_M$ are possible, we only keep one of them.

We can compute $\mathcal{M}_\ell$ for each $\ell \in [L]$ by a top-down dynamic programming approach. We start with $\mathcal{M}_{L+1} = \{\emptyset\}$. For $\ell$ going from $L$ down to $1$, we compute $\mathcal{M}_\ell$ as follows: For each $C_\ell \subseteq V \times \{\ell\}$ with $|C_\ell| \leq \left\lfloor \sum_{j=1}^{\ell} k_j \right\rfloor - \left\lfloor \sum_{j=1}^{\ell-1} k_j \right\rfloor$ and each $M \in \mathcal{M}_{\ell+1}$, we add $M' = M \cup \text{cov}(C_\ell)$ to $\mathcal{M}_\ell$, and set $C_{M'} = C_M \cup C_\ell$.

If $\mathcal{M}_1$ contains the set $\mathcal{S}$, output its corresponding set $C_\mathcal{S}$. Else, output that there is no such set $C$.

It remains to consider the runtime of the algorithm. We perform $L$ iterations. When computing $\mathcal{M}_\ell$, we iterate over all sets in $\mathcal{M}_{\ell+1}$ and each choice of $C_\ell$. Note that $|\mathcal{M}_{\ell+1}| \leq 2^n$, and there are at most $2^n$ choices for $C_\ell$. This gives a runtime of $O(L 4^n)$. □

# 9 Rounding fractional solutions

Most results proved in this section are based on sparsity of vertex solutions to a natural polytope relaxation of SRMFC-T. By leveraging an LP-aware reduction from NUkC to RMFC, we can construct sparse solutions to a natural polytope relaxation of the SNUkC problem as well.

We close this section by a short discussion of the integrality gap of the corresponding linear programming relaxation for SRMFC-T.

## 9.1 Rounding fractional solutions of SRMFC-T

We start by proving the results for SRMFC-T, namely Lemma 2.5, Lemma 3.1, and a general version of Lemma 3.3. To prove all these statements, we start by generalizing the sparsity result of [1] to fractional coverage requirements.

Let $(G, r, B)$ be an instance of SRMFC-T with fractional coverage requirements $\delta_t \in [0, 1]$ for $t \in \Gamma$. We are interested in fractional solutions with small stretch $\alpha \in \mathbb{R}_{\geq 0}$, which are described by the polytope

$$Q_{\alpha, \delta} := \left\{ x \in \mathbb{R}_{\geq 0}^{V \setminus \{r\}} : x(V_{\leq \ell}) \leq \alpha \sum_{i=1}^{\ell} B_i \ \forall \ell \in [L] \text{ and } x(P_t) \geq \delta_t \ \forall t \in \Gamma \right\}.$$

For $x \in Q_{\alpha, \delta}$, we call a support vertex $v$ of $x$ *loose* if $x(P_v) < \delta_t$ for every leaf $t$ in $T_v$. Analogously, we call a support vertex $v$ of $x$ *tight* if $x(P_v) = \delta_t$ for some leaf $t$ in $T_v$. Note that if $x$ is a vertex of $Q_{\alpha, \delta}$, then every support vertex of $x$ is either loose or tight. If $\delta_t = 1$ for every $t \in \Gamma$ it was shown in [1] that a vertex of $Q_{\alpha, \delta}$ has a bounded number of loose vertices. Using the same arguments, we can immediately extend this result to fractional coverage requirements. We want to stress that the proof of this result is up to minor adaptations and reformulations the same as the proof of [1:Lemma 6] for the firefighting problem. We repeat it here for the sake of completeness.

**Lemma 9.1.** *Let $(G, r, B)$ be an instance of SRMFC-T with fractional coverage requirements $\delta \in [0, 1]^\Gamma$ and $L$ levels. Let $\alpha \in \mathbb{R}_{\geq 0}$ and let $x$ be a vertex of $Q_{\alpha, \delta}$. Then, at most $L$ support vertices of $x$ are loose.*

*Proof.* Let $x$ be a vertex of $Q_{\alpha,\delta}$. We first partition the tight constraints of $x$ into three sets:
(i) Tight nonnegativity constraints, one for each vertex in $\mathcal{F}_1 := \{u \in V \setminus \{r\} : x_u = 0\}$.
(ii) Tight budget constraints, one for each level $\ell \in [L]$ in $\mathcal{F}_2 := \{\ell \in [L] : x(V_{\leq \ell}) = \alpha \sum_{i=1}^{\ell} B_i\}$.
(iii) Tight coverage constraints, one for each leaf $t$ in $\mathcal{F}_3 := \{t \in \Gamma : x(P_t) = \delta_t\}$.

A vertex $x$ is uniquely defined by any $|V| - 1$ linearly independent tight constraints. As the first set of constraints is linearly independent, we can pick a full-rank subsystem that corresponds to $\mathcal{F}_1$ together with $\mathcal{F}_2' \subseteq \mathcal{F}_2$ and $\mathcal{F}_3' \subseteq \mathcal{F}_3$ such that $\mathcal{F}_1 \cup \mathcal{F}_2' \cup \mathcal{F}_3'$ corresponds to a full-rank system defining $x$. Then we have $|V| - 1 = |\mathcal{F}_1| + |\mathcal{F}_2'| + |\mathcal{F}_3'|$.

Let $V^{\mathcal{L}} \subseteq \operatorname{supp}(x)$ and $V^{\mathcal{T}} \subseteq \operatorname{supp}(x)$ denote the set of loose and tight support vertices of $x$, respectively. Note that $|V^{\mathcal{L}}| + |V^{\mathcal{T}}| = |\operatorname{supp}(x)| = |V| - 1 - |\mathcal{F}_1|$. We claim that $|\mathcal{F}_3'| \leq |V^{\mathcal{T}}|$. Assuming this we have

$$|V^{\mathcal{L}}| \leq |V^{\mathcal{L}}| + |V^{\mathcal{T}}| - |\mathcal{F}_3'| = |V| - 1 - |\mathcal{F}_1| - |\mathcal{F}_3'| = |\mathcal{F}_2'| \leq L.$$

So, it remains to show that $|\mathcal{F}_3'| \leq |V^{\mathcal{T}}|$. To do so, we define an injective mapping $f : \mathcal{F}_3' \to V^{\mathcal{T}}$ as follows. For a tight coverage constraint $x(P_t) = \delta_t$ corresponding to a leaf $t \in \Gamma$, we define $f(t)$ to be the first support vertex on the path $P_t$ when starting in $t$. Then, by definition, $f(t)$ is tight, as $x(P_{f(t)}) = x(P_t) = \delta_t$. Furthermore, we claim that $f$ is injective. Assume that we had $f(t_1) = f(t_2)$ for two leaves $t_1, t_2$. Then, $x$ vanishes on $P_{t_1} \Delta P_{t_2}$, as $f(t_1)$ is the first support vertex on both paths. In particular, $\chi^{P_{t_1}} - \chi^{P_{t_2}}$ is in $\operatorname{span}(\{\chi^v : v \in \mathcal{F}_1\})$, contradicting the fact that we chose a full-rank subsystem of tight constraints. Thus, $f$ is injective and we have $|\mathcal{F}_3'| \leq |V^{\mathcal{T}}|$, finishing the proof. □

Note that Lemma 2.5 directly follows from Lemma 9.1. Indeed, compute a vertex $x \in Q_\alpha(\Gamma)$. (Recall that in Lemma 2.5, we assume $Q_\alpha(\Gamma) \neq \emptyset$.) Now, take all tight vertices $v$ with $x_v = 1$ and additionally all loose vertices. Note that all leaves are protected by the resulting vertex set, as every leaf $t$ either has a tight predecessor $v$ with $x_v = 1$ or the topmost support vertex on $P_t$ is loose. We add at most $L$ loose vertices in total, hence we in particular add at most $L$ vertices on each level. As $B_1 \geq \frac{L}{\varepsilon}$, this gives an $(\alpha + \varepsilon)$-feasible solution.

With Lemma 9.1, we can also prove Lemma 3.1. We prove the following generalization of Lemma 3.1 which we will also use for proving our sparsity result for SNUkC.

**Lemma 9.2.** *Let $0 < \varepsilon \leq \frac{1}{7}$ and $\gamma \in (0, 1]$, and let $(G, r, B)$ be an SRMFC-T instance with $L$ levels and fractional coverage requirements $\delta$. Let $h_1, h_2 \in [L]$ with $h_1 \geq h_2$, so that $h_1 = L$ or $B_{h_1+1} \geq \frac{L}{\varepsilon}$, and $h_2 = h_1$ or $B_{h_2+1} \geq \frac{h_1}{\varepsilon}$. For $x \in Q_{1,\delta}$, we can efficiently compute a point $y \in Q_{1+\varepsilon^2, \delta'}$ with $\delta_t' = \delta_t - 2\varepsilon\gamma$ for each $t \in \Gamma$ and*

- $\operatorname{supp}(y) \subseteq \operatorname{supp}(x)$,
- $y_v \geq \frac{\varepsilon\gamma}{h_2}$ *for each $v \in \operatorname{supp}(y) \cap V_{\leq h_2}$, and*
- $y_v \geq \varepsilon\gamma$ *for each $v \in \operatorname{supp}(y) \cap V_{>h_2} \cap V_{\leq h_1}$.*

Indeed, Lemma 3.1 is a consequence of Lemma 9.2 by choosing $\gamma = 1$, $h_1 = \hat{h}$, $h_2 = \check{h}$, and scaling the resulting point $y \in Q_{1+\varepsilon^2,\delta'}$ with a factor of $\frac{1}{1-2\varepsilon}$. Using $\varepsilon \leq \frac{1}{7}$, the resulting vector lies in $Q_{1+3\varepsilon}(\Gamma)$ as desired.

*Proof of Lemma 9.2.* Create a new instance of SRMFC-T by removing all levels $\ell \in \{h_1 + 1, ..., L\}$ and contracting the first $h_2$ levels to a new root $r'$. Then, the level-$h_1$-vertices of $(G, r, B)$ are the leaves of the new instance $(G', r', B')$. For $t \in V_{h_1}$, we set

$$\tilde{\delta}_t = \varepsilon\gamma \left\lfloor \frac{x\left(P_t \cap V_{\leq h_1} \cap V_{>h_2}\right)}{\varepsilon\gamma} \right\rfloor.$$

In other words, the coverage constraint of $t$ is $x\left(P_t \cap V_{\leq h_1} \cap V_{>h_2}\right)$ rounded down to the next multiple of $\varepsilon\gamma$. Then, compute a vertex $x'$ of the polytope $Q_{1,\tilde{\delta}}$ for $(G', r', B')$ with coverage requirements $\tilde{\delta}$, such that $\mathrm{supp}(x') \subseteq \mathrm{supp}(x)$. This exists, as $x \in Q_{1,\tilde{\delta}}$. Round up the loose vertices of $x'$ to the next multiple of $\varepsilon\gamma$, and round down the tight vertices of $x'$ to the next multiple of $\varepsilon\gamma$. Let $x''$ denote the resulting vector.

First of all, we claim that $x''$ satisfies the fractional coverage requirements. Note that for each tight vertex $v$, we have that $x'(P_v) = \tilde{\delta}_t$ for some leaf $t$. In particular, $x'(P_v)$ is a multiple of $\varepsilon\gamma$. We show that $x''(P_v) \geq x'(P_v)$ by induction on the number of support vertices on $P_v$. This directly shows $x''(P_t) \geq x'(P_t)$ for each $t \in V_{h_1}$. If $v$ is the only support vertex on $P_v$, then $x'_v$ is a multiple of $\varepsilon\gamma$ and $x''(P_v) = x''_v = x'_v = x'(P_v)$.

So, assume that $v$ has at least one predecessor in the support of $x'$. Furthermore, let $w$ be the first tight predecessor of $v$ in the support of $x'$, or $r$, if no tight predecessor exists. Note that $x'(P_v \setminus P_w)$ is a multiple of $\varepsilon\gamma$ since both $x'(P_v)$ and $x'(P_w)$ are a multiple of $\varepsilon\gamma$. Furthermore, $v$ is the only vertex on $P_v \setminus P_w$ for which $x''_v$ was rounded down to the next multiple of $\varepsilon\gamma$. For every other vertex $u$ on $P_v \setminus P_w$, $x''_u$ was rounded up to the next multiple of $\varepsilon\gamma$. This implies that $x''(P_v \setminus P_w) > x'(P_v \setminus P_w) - \varepsilon\gamma$. Since both $x''(P_v \setminus P_w)$ and $x'(P_v \setminus P_w)$ are a multiple of $\varepsilon\gamma$, this yields $x''(P_v \setminus P_w) \geq x'(P_v \setminus P_w)$. By the induction hypothesis, we have $x''(P_w) \geq x'(P_w)$. Combining these two inequalities, we obtain $x''(P_v) \geq x'(P_v)$, as desired.

By Lemma 9.1, $x'$ has at most $h_1$ loose vertices. Hence, $x''(V_{\leq \ell}) \leq x'(V_{\leq \ell}) + \varepsilon\gamma h_1$ for each level $\ell \in [L']$, where $L'$ denotes the number of levels in $G'$. Note that if $h_1 > h_2$, then $B_{h_2+1} \geq \frac{h_1}{\varepsilon}$. Thus, replacing $x$ on $V_{\leq h_1} \cap V_{>h_2}$ by $x''$ we violate the budget constraints by at most factor $1 + \varepsilon^2\gamma$. Note that we rounded the coverage constraints down to the next multiple of $\varepsilon\gamma$, which reduces the coverage we obtain by up to $\varepsilon\gamma$. As an additional step, we round down the value of $x$ on each vertex in $V_{\leq h_2}$ to a multiple of $\frac{\varepsilon\gamma}{h_2}$. This reduces the total coverage of a leaf by at most an additional $\varepsilon\gamma$, as on at most $h_2$ many vertices we lose at most $\frac{\varepsilon\gamma}{h_2}$. In total this yields a vector $\tilde{x}$ with

- $\mathrm{supp}(\tilde{x}) \subseteq \mathrm{supp}(x)$,
- $\tilde{x}_v \geq \frac{\varepsilon\gamma}{h_2}$ for each $v \in \mathrm{supp}(\tilde{x}) \cap V_{\leq h_2}$,
- $\tilde{x}_v \geq \varepsilon\gamma$ for each $v \in \mathrm{supp}(\tilde{x}) \cap V_{>h_2} \cap V_{\leq h_1}$,
- $\tilde{x}(V_{\leq \ell}) \leq (1+\varepsilon^2)B_{\leq \ell}$ for each $\ell \in [L]$, and
- $\tilde{x}(P_t) \geq \delta_t - 2\varepsilon\gamma$ for each $t \in \Gamma$,

as desired. □

Finally, using Lemma 2.5, we obtain the following generalization of Lemma 3.3.

**Lemma 9.3.** *Define $h_0 := L$ and for $i \in \mathbb{Z}_{\geq 1}$, let $h_i := \left\lceil \log_{1+\varepsilon}\left(\frac{h_{i-1}}{\varepsilon^2}\right) \right\rceil + 1$. Consider an $\varepsilon$-compressed SRMFC-T instance with $L$ levels, let $k \in \mathbb{Z}_{\geq 1}$, and let $\alpha \in \mathbb{R}_{>0}$. Let $y \in Q_\alpha(\Gamma)$ with $\mathrm{supp}(y) \cap V_{\leq h_k} = \emptyset$. Then we can efficiently compute a $(k\alpha + \varepsilon)$-feasible solution that does not fireproof any vertex in $V_{\leq h_k}$.*

*Proof.* We separate $(G, r, B)$ into $k$ parts, on each of which we can apply Lemma 2.5. To do so, define $\Gamma_i := \left\{ t \in \Gamma : y\left(P_t \cap V_{>h_i} \cap V_{\leq h_{i-1}}\right) \geq \frac{1}{k} \right\}$, for $i \in [k]$. Note that $\bigcup_{i \in [k]} \Gamma_i = \Gamma$ since $\mathrm{supp}(y) \cap V_{\leq h_k} = \emptyset$. Furthermore, let $y_i$ be the restriction of $y$ to the vertices in $V_{>h_i} \cap V_{\leq h_{i-1}}$. For $i \in [k]$, we construct a new instance of SRMFC-T by deleting the levels $\ell > h_i$ and contracting the first $h_{i-1}$ levels into the root. For the resulting tree, we use $B(i) := \left(B_{h_i+1}, ..., B_{h_{i-1}}\right)$ as budgets.

Note that $y_i \in Q_{k\alpha}(\Gamma_i)$ for the resulting SRMFC-T instance $(G_i, r, B(i))$. Moreover, we have $B(i)_1 \geq \frac{L_i}{\varepsilon}$, where $L_i$ denotes the number of levels of the tree $G_i$. Thus, using Lemma 2.5, we can efficiently compute a $(k\alpha + \varepsilon)$-feasible solution to $(G_i, r, B(i))$. Using $\bigcup_{i \in [k]} \Gamma_i = \Gamma$, the union of all these solutions is $(k\alpha + \varepsilon)$-feasible for $(G, r, B)$. □

## 9.2 LP-aware reduction from SNUkC to SRMFC-T

Next, we prove the corresponding statements for SNUkC. We start by proving Lemma 6.5, using the LP-aware reduction from NUkC to RMFC-T in [7].

*Proof of Lemma 6.5.* Let $y \in Q_{\alpha,\beta}(V)$. Consider the NUkC instance $(V, d, k', r)$ with $k'_\ell = \lceil k_\ell \rceil$ for each $\ell \in [L]$. Note that we can modify $y$ to obtain a point in the polytope

$$\left\{ x \in [0,1]^{V \times [L]} \;\middle|\; \begin{array}{l} \sum_{\ell=1}^{L} \sum_{u \in \text{Ball}(v, \beta r_\ell)} x_{u,\ell} \geq 1 \quad \forall v \in V \\[6pt] \sum_{v \in V} x_{v,\ell} \leq \alpha k'_\ell \, \forall \ell \in [L] \end{array} \right\},$$

which is referred to as the NUkC LP in [7]. Using the LP-aware reduction from NUkC to RMFC-T in [7], namely Theorem 3.4 in [7], we get an instance $(G, r, k')$ of RMFC-T, which admits a point in the polytope

$$Q := \left\{ x \in \mathbb{R}_{\geq 0}^{V \setminus \{r\}} : x(V_\ell) \leq \alpha k'_\ell \; \forall \ell \in [L] \; \text{ and } \; x(P_t) \geq 1 \; \forall t \in \Gamma \right\},$$

which is referred to as the RMFC-T LP in [7] and as $\text{LP}_{\text{FF}}$ in [1]. We can round this fractional solution using Lemma 3.1 in [1]. More precisely, we compute a vertex in $Q$ and round up all loose vertices. This gives an integral solution $R$. Since there are at most $L$ loose vertices as guaranteed by Lemma 3.1 in [1], we have $|R \cap V_\ell| \leq \alpha k'_\ell + L \leq (\alpha + \varepsilon) k'_\ell$. Thus, using the guarantees of Theorem 3.4 in [7], we can compute a solution $C$ to the NUkC instance $(V, d, k', r)$ with
- $|C_\ell| \leq (\alpha + \varepsilon) k'_\ell$ for each $\ell \in [L]$, and
- $\bigcup_{(v,\ell) \in C} \text{Ball}(v, 8\beta r_\ell) = V$.

Note that this solution is $((1+\varepsilon)(\alpha + \varepsilon), 8\beta)$ feasible for the SNUkC instance $(V, d, k, r)$. $\square$

Next, we present a different LP-aware reduction from NUkC to RMFC. While our reduction gives slightly worse guarantees than the one used in [7], it also works for fractional coverage requirements, which we crucially exploit when proving Lemma 6.7.

Let $(V, d, k, r)$ be an instance of SNUkC with fractional coverage requirements $\delta_v \in [0, 1]$ for $v \in V$. We are interested in fractional solutions with small stretch $\alpha, \beta \in \mathbb{R}_{\geq 0}$, which are described by the polytope

$$Q_{\alpha,\beta,\delta}^{\text{NUkC}} := \left\{ x \in [0,1]^{V \times [L]} \;\middle|\; \begin{array}{l} \sum_{\ell=1}^{L} \sum_{u \in \text{Ball}(v, \beta r_\ell)} x_{u,\ell} \geq \delta_v \quad \forall v \in V \\[6pt] \sum_{v \in V} \sum_{\ell'=1}^{\ell} x_{v,\ell'} \leq \alpha k_{\leq \ell} \, \forall \ell \in [L] \end{array} \right\}$$

To avoid confusion, we refer to the polytope $Q_{\alpha,\delta}$ defined in the previous subsection as $Q_{\alpha,\delta}^{\text{RMFC}}$. For ease of notation, for some vector $\delta \in \mathbb{R}^U$ on some ground set $U$ and some value $\kappa \in \mathbb{R}$, we denote by $\delta + \kappa$ the vector given by $\delta_u + \kappa$ for each $u \in U$.

**Lemma 9.4.** *Let $(V, d, k, r)$ be an instance of SNUkC with $L$ levels and fractional coverage requirements $\delta \in [0, 1]^V$. Let $\eta \in \mathbb{R}_{>2}$ and assume $r_\ell \geq \eta r_{\ell+1}$ for each $\ell \in [L-1]$. Let $\alpha, \beta \in \mathbb{R}_{\geq 0}$ and let $x \in Q_{\alpha,\beta,\delta}^{\text{NUkC}}$. Then, we can efficiently construct an instance $(G, \hat{r}, k)$ of SRMFC-T with $L$ levels, leaves $\Gamma$ and fractional coverage requirements $\delta' \in [0, 1]^\Gamma$, together with an assignment $\psi : V' \to V$ of the vertices $V'$ of $G$ to vertices in $V$ such that*
- *the point $y^x$ defined by*

$$y_v^x := \sum_{u \in \text{Ball}(\psi(v), \beta r_\ell)} x_{u,\ell}$$

*for every* $\ell \in [L]$ *and every level-$\ell$-vertex $v$ of $G$, satisfies* $y^x \in Q^{\text{RMFC}}_{\alpha,\delta'}$, *and*

- *for every* $\alpha', \kappa \in \mathbb{R}_{\geq 0}$ *and every* $y \in Q^{\text{RMFC}}_{\alpha',\delta'+\kappa}$, *the point $x^y$ defined by*

$$x^y_{v,\ell} := \sum_{v' \in V'_\ell : \psi(v') = v} y_{v'}$$

*satisfies* $x^y \in Q^{\text{NUkC}}_{\alpha', \frac{2\eta}{\eta-2}\beta, \delta+\kappa}$. *Moreover, if* $\text{supp}(y) \subseteq \text{supp}(y^x)$, *we have*

$$\text{supp}(x^y) \subseteq \bigcup_{(v,\ell) \in \text{supp}(x)} \{(u, \ell) : u \in \text{Ball}(v, \beta r_\ell)\} \ . \tag{10}$$

*Proof.* We build the SRMFC-T instance level by level, starting with the leaves on level $L+1$. On this level we add one vertex for each element in $V$. We set $\psi(v) = v$ for each $v \in V$.

Assume that we already constructed the level-$(\ell+1)$-vertices $V_{\ell+1}$ for $\ell \in [L]$. We construct the level-$\ell$-vertices $V_\ell$ as follows. As long as there is a vertex in $V_{L+1} = V$ with no ancestor on level $\ell$, let $v \in V$ be such a vertex with $\sum_{u \in \text{Ball}(v,\beta r_\ell)} x_{u,\ell}$ maximal. We add a vertex $v'$ to level $\ell$. Then, we set $y^x_{v'} := \sum_{u \in \text{Ball}(v,\beta r_\ell)} x_{u,\ell}$ and $\psi(v') := v$. Finally, we add edges from $v'$ to vertices in level $\ell+1$ so that $v'$ is ancestor of each vertex $u \in V$ which does not yet have a predecessor in level $\ell$ and for which $d(u,v) \leq 2\beta r_\ell$.

Finally, insert a root $\hat{r}$, and connect all level-$1$-vertices to the root. Let $G$ denote the resulting tree with $L+1$ levels, and let $V'$ denote its vertex set. We prove all statements for the SRMFC-T instance $(G, \hat{r}, k)$ (using $k_{L+1} := 0$) with fractional coverage requirements $\delta'_{v'} = \delta_{\psi(v')}$ for each leaf $v'$ of $G$. We then obtain an equivalent SRMFC-T instance $(G', \hat{r}, k)$ with the desired number of levels and fractional coverage requirements $\delta''$ by deleting level $L+1$ and setting $\delta''_v = \max_{t \in \Gamma_v} \delta'_t$ for each level-$L$-vertex $v$, where $\Gamma_v$ denotes the leaves of the subtree of $G$ rooted at $v$.

To show the first property, note that $y^x(P_v) \geq \delta'_v$ for each leaf $v$ of $G$. Indeed, for each level $\ell \in [L]$, we have for the level $\ell$ ancestor $v'$ of $v$ in $G$ that $y^x_{v'} \geq \sum_{u \in \text{Ball}(v,\beta r_\ell)} x_{u,\ell}$ by construction. Moreover, for two level-$\ell$-vertices of $G$, the corresponding vertices in $V$ have distance $> 2\beta r_\ell$ with respect to $d$. Thus, we have

$$y^x(V'_{\leq \ell}) \leq \sum_{v \in V} \sum_{\ell'=1}^{\ell} x_{v,\ell'} \leq \alpha k_{\leq \ell}$$

as desired.

To show the second property, let $\alpha', \kappa \in \mathbb{R}_{\geq 0}$ and $y \in Q^{\text{RMFC}}_{\alpha',\delta'+\kappa}$. Then, for each $\ell \in [L]$,

$$\sum_{v \in V} \sum_{\ell' \leq \ell} x^y_{v,\ell'} = y(V'_{\leq \ell}) \leq \alpha' k_{\leq \ell} \ .$$

We have to show that $\sum_{\ell=1}^{L} \sum_{u \in \text{Ball}(v,(\frac{2\eta}{\eta-2})\beta r_\ell)} x^y_{u,\ell} \geq \delta_v + \kappa$ for each $v \in V$. To do so, consider a level-$\ell$-vertex $u'$ in $G$ and a vertex $v' \in T_{u'}$. We show by backward induction on $\ell$ that $d(\psi(u'), \psi(v')) \leq \frac{2\eta}{\eta-2}\beta r_\ell$. Clearly, this holds for $\ell = L$ by construction. For $\ell < L$, let $w'$ be the level-$(\ell+1)$-vertex on the $u'$-$v'$-path in $G$. Note there must be some leaf $\tilde{v} \in T_{w'}$ with $d(\psi(u'), \psi(\tilde{v})) \leq 2\beta r_\ell$ by construction. Thus, we have

$$d(\psi(u'), \psi(v')) \leq d(\psi(u'), \psi(\tilde{v})) + d(\psi(\tilde{v}), \psi(w')) + d(\psi(w'), \psi(v'))$$

$$\leq 2\beta r_\ell + \frac{4\eta}{\eta - 2}\beta r_{\ell+1}$$

$$\leq 2\beta r_\ell + \frac{4}{\eta - 2}\beta r_\ell$$

$$= \frac{2\eta}{\eta - 2}\beta r_\ell,$$

where we used the induction hypothesis twice in the second inequality. Now, for $v \in V$, let $v'$ be the leaf of $G$ corresponding to $v$, i.e., $\psi(v') = v$. The above argument shows that if $u'$ is a level-$\ell$-vertex on $P_{v'}$, then $\psi(u')$ is in $\text{Ball}\left(v, \left(\frac{2\eta}{\eta-2}\right)\beta r_\ell\right)$. This gives

$$\delta_v + \kappa = \delta'_{\psi(v')} + \kappa \leq \sum_{u' \in P_v} y_{u'} \leq \sum_{\ell=1}^{L} \sum_{u \in \text{Ball}\left(v, \left(\frac{2\eta}{\eta-2}\right)\beta r_\ell\right)} x_{u,\ell}^y.$$

It remains to analyze the support of $x^y$. Let $v \in V$ and $\ell \in [L]$ with $x_{v,\ell}^y > 0$. In particular, there is some $v' \in V'_\ell$ with $\psi(v') = v$ and $y_{v'} > 0$. If $\text{supp}(y) \subseteq \text{supp}(y^x)$ this implies $y_{v'}^x > 0$. By construction of $y^x$, there is some $u \in \text{Ball}(v, \beta r_\ell)$ with $x_{u,\ell} > 0$. □

With this reduction, we are able to prove Lemma 6.7.

*Proof of Lemma 6.7.* We start by restricting the instance to the first $\hat{h}$ levels. To be precise, let $(V, d, k', r')$ denote the instance with $\hat{h}$ levels with $k'_\ell = k_\ell$ and $r'_\ell = r_\ell$ for each $\ell \in [\hat{h}]$.

For $m \in [\lambda]$, we consider the *slice* $(V, d, k(m), r(m))$ of $(V, d, k', r')$ containing all levels $\ell$ with $\ell \equiv m \mod \lambda$. More precisely, let $(V, d, k(m), r(m))$ denote the instance with $L_m$ levels defined by $k(m)_i = k'_{(i-1)\lambda+m}$ and $r(m)_i = r'_{(i-1)\lambda+m}$ for each $i \leq \frac{\hat{h}+\lambda-m}{\lambda}$. Define fractional coverage requirements $\delta(m) \in [0, 1]^V$ as

$$\delta(m)_v := \frac{\varepsilon}{\lambda} \left\lfloor \frac{\lambda}{\varepsilon} \sum_{\ell \in [L_m]: \ell \equiv m \mod \lambda} \sum_{u \in \text{Ball}(v, \beta r_\ell)} x_{u,\ell} \right\rfloor.$$

In other words, we set $\delta_v$ to the coverage of $v$ on the considered slice rounded down to the next multiple of $\frac{\varepsilon}{\lambda}$.

For the slice $(V, d, k(m), r(m))$, let $(G, \hat{r}, k(m))$ denote the SRMFC-T instance with fractional coverage requirements $\delta'$ obtained by applying Lemma 9.4 to the restriction $x(m)$ of $x$ to the slice with $\eta = (1+\varepsilon)^\lambda$. Indeed, since the given SNUkC instance is $\varepsilon$-compressed, two radii in the same slice differ by at least factor $(1+\varepsilon)^\lambda$. Furthermore, let $y^{x(m)} \in Q^{\text{RMFC}}_{1,\delta'}$ be the fractional point for this instance as defined in the first part of Lemma 9.4. Now, let $y \in Q^{\text{RMFC}}_{1+\varepsilon^2, \delta' - \frac{2\varepsilon}{\lambda}}$ be the point obtained from $y^{x(m)}$ by applying Lemma 9.2 with $\gamma = \frac{1}{\lambda}$, $h_1 = L_m$, and $h_2$ maximal with $(h_2 - 1)\lambda + m \leq \check{h}$. In particular, $y$ satisfies

- $y_v \geq \frac{\varepsilon}{\lambda \check{h}}$ for each $v \in \text{supp}(y) \cap V_{\leq h_2}$, and
- $y_v \geq \frac{\check{\varepsilon}}{\lambda}$ for each $v \in \text{supp}(y) \cap V_{>h_2} \cap V_{\leq h_1}$.

Lemma 9.4 guarantees that $x^y \in Q^{\text{NUkC}}_{1+\varepsilon^2, \frac{2}{1-2(1+\varepsilon)^{-\lambda}}, \delta(m)-\frac{2\varepsilon}{\lambda}}$. Additionally, by construction of $x^y$, we have

$$x_{v,\ell}^y = \sum_{v' \in V'_\ell: \psi(v')=v} y_{v'}.$$

In particular, we can use the lower bounds on $y$ to obtain the following lower bounds on $x^y$.

- $x_{v,\ell}^y \geq \frac{\varepsilon}{\lambda \check{h}}$ for each $(v, \ell) \in \text{supp}(x^y)$ with $\ell \leq h_2$,

- $x^y_{v,\ell} \geq \frac{\varepsilon}{\lambda}$ for each $(v,\ell) \in \mathrm{supp}(x^y)$ with $\ell \in [h_2, L_m]$.

Additionally, (10) in Lemma 9.4 guarantees that for each $(v,\ell) \in \mathrm{supp}(x^y)$, there is some $(u,\ell) \in \mathrm{supp}(x)$ with $d(u,v) \leq r_\ell$. In particular, replacing each such $(v,\ell)$ by a corresponding $(u,\ell)$ and increasing the radius by an additional $r_\ell$, we can find a point $x(m) \in Q^{\mathrm{NUkC}}_{1+\varepsilon^2, 1+\frac{2}{1-2(1+\varepsilon)^{-\lambda}}, \delta(m) - \frac{2\varepsilon}{\lambda}}$ satisfying the same lower bounds and $\mathrm{supp}(x(m)) \subseteq \mathrm{supp}(x)$.

We can finally combine this to a point $x'$ with $\frac{1}{1-3\varepsilon}x' \in Q_{1+7\varepsilon, \beta(\lambda)}(U)$ that satisfies the properties of Lemma 6.7. To do so, we set

$$x'_{v,\ell} := \begin{cases} x(m)_{v, \lceil \frac{\ell}{\lambda} \rceil} & \text{if } \ell \leq \hat{h} \text{ and } \ell \equiv m \bmod \lambda \\ x_{v,\ell} & \text{if } \ell > \hat{h} \end{cases}$$

for each $v \in V$ and each $\ell \in [L]$. So, in other words, on levels $1, \ldots, \hat{h}$, we set $x'$ to the value of the corresponding slice. On levels $\hat{h}+1, \ldots, L$, we keep the value of $x$. Since $x(m)$ satisfies the lower bounds for each $m \in [\lambda]$, we have

- $x'_{v,\ell} \geq \frac{\varepsilon}{\lambda \check{h}}$ for each $(v,\ell) \in \mathrm{supp}(x')$ with $\ell \leq \check{h}$,
- $x'_{v,\ell} \geq \frac{\varepsilon}{\lambda}$ for each $(v,\ell) \in \mathrm{supp}(x')$ with $\ell \in [\check{h}+1, \hat{h}]$.

Furthermore, since $\mathrm{supp}(x(m)) \subseteq \mathrm{supp}(x)$ for each $m \in [\lambda]$, we have $\mathrm{supp}(x') \subseteq \mathrm{supp}(x)$.

We will show that $\frac{1}{1-3\varepsilon}x' \in Q_{1+7\varepsilon, \beta(\lambda)}(U)$. Note that for each $v \in U$, using the fractional coverage requirements of $x(m)$, we have

$$\sum_{\ell=1}^{L} \sum_{u \in \mathrm{Ball}(v, \beta(\lambda)_\ell r_\ell)} x'_{u,\ell} \geq \sum_{m=1}^{\lambda} \left( \delta(m)_v - \frac{2\varepsilon}{\lambda} \right) + \sum_{\ell=\hat{h}+1}^{L} \sum_{u \in \mathrm{Ball}(v, \beta r_\ell)} x_{u,\ell}$$

$$\geq \sum_{m=1}^{\lambda} \left( \sum_{\ell \in [L']: \ell \equiv m \bmod \lambda} \sum_{u \in \mathrm{Ball}(v, \beta r_\ell)} x_{u,\ell} - \frac{3\varepsilon}{\lambda} \right) + \sum_{\ell=\hat{h}+1}^{L} \sum_{u \in \mathrm{Ball}(v, \beta r_\ell)} x_{u,\ell}$$

$$\geq \left( \sum_{\ell=1}^{L} \sum_{u \in \mathrm{Ball}(v, \beta r_\ell)} x_{u,\ell} \right) - \lambda \left( \frac{3\varepsilon}{\lambda} \right) \geq 1 - 3\varepsilon .$$

Here, we used that the fractional coverage requirements $\delta(m)$ are at most $\frac{\varepsilon}{\lambda}$ smaller than the coverage of $v$ on the corresponding slice, combined with $x$ completely covering $v$. Moreover, using $\varepsilon \leq \frac{1}{7}$, we have $\frac{1+\varepsilon^2}{1-3\varepsilon} \leq 1 + 7\varepsilon$. Thus, $\frac{1}{1-3\varepsilon}x' \in Q_{1+7\varepsilon, \beta(\lambda)}(U)$ as desired. □

## 9.3 Integrality gap of the natural LP relaxation for SRMFC-T

We show that the integrality gap of the natural LP relaxation $\min\{\alpha : x \in Q_\alpha(\Gamma)\}$ for SRMFC-T is at least $\frac{\log^* n}{2} - 1$, where $n$ denotes the number of vertices. This follows immediately when using the same construction as in [8], who established a lower bound of $\frac{\log^* n}{2}$ for the natural LP relaxation for RMFC-T.

**Lemma 9.5.** *For each $M \in \mathbb{Z}_{>0}$, there is an instance of $(G, r, B)$ of SRMFC-T with $n$ vertices so that (i) $M \geq \frac{\log^* n}{2}$, (ii) $Q_1(\Gamma) \neq \emptyset$, and (iii) the optimal stretch is larger than $M - 1$.*

*Proof.* Given an instance $(G, r)$ of RMFC-T with $L$ levels, consider the natural linear programming relaxation $\min\{\alpha : x \in Q'_\alpha\}$ with

$$Q'_\alpha := \left\{ x \in \mathbb{R}^{V \setminus \{r\}}_{\geq 0} : x(V_\ell) \leq \alpha \ \forall \ell \in [L] \text{ and } x(P_t) \geq 1 \ \forall t \in \Gamma \right\}.$$

In their paper, Chalermsook and Chuzhoy [8] construct an instance $(G, r)$ of RMFC-T with $L$ levels and $n$ vertices so that $M \geq \frac{\log^* n}{2}$, $Q'_1 \neq \emptyset$, and there is no feasible integral solution using less than $M$ firefighters. Now, define budgets $B \in \mathbb{R}_{\geq 0}^L$ by $B_\ell = 1$ for each $\ell \in L$. Note that the resulting instance $(G, r, B)$ of SMRF-T satisfies the desired properties. Indeed, $Q_1' \subseteq Q_1(\Gamma)$, and every integral solution with stretch $M - 1$ yields a solution to $(G, r)$ using at most $M - 1$ firefighters. □

## 10 Additional proofs

*Proof of Theorem 2.1.* Given an instance $G = (V, E)$ with root $r$ of RMFC-T, we guess the optimum number $B_{\mathrm{OPT}}$ of firefighters for this instance, using $B_{\mathrm{OPT}} \leq |V|$. Let $L$ denote the depth of $G$. For the instance $(G, r, B)$ of SRMFC-T with $B_\ell = B_{\mathrm{OPT}}$ for each $\ell \in [L]$, we call the $\alpha$-approximation algorithm for SRMFC-T. The solution $R$ computed by this algorithm satisfies

$$|R \cap V_{\leq \ell}| \leq \alpha \sum_{j=1}^{\ell} B_j = \alpha \ell B_{\mathrm{OPT}} \qquad \forall \ell \in [L].$$

As long as there is some level $\ell \in [L]$ with $|R \cap V_\ell| > \lceil \alpha B_{\mathrm{OPT}} \rceil$, there has to be a level $\ell' < \ell$ with $|R \cap V_{\ell'}| < \lceil \alpha B_{\mathrm{OPT}} \rceil$. Replace any vertex in $R \cap V_\ell$ by its unique ancestor in $V_{\ell'}$. In the end, we have $|R \cap V_\ell| \leq \lceil \alpha B_{\mathrm{OPT}} \rceil$ for each $\ell \in [L]$ as desired. □

*Proof of Theorem 2.2.* Let $0 < \varepsilon < 1$ be given. We will show how to compute a $(1 + 17\varepsilon)$-approximation for SRMFC-T, which implies the existence of a PTAS. By Theorem 2.4, it suffices to give a $(1 + 8\varepsilon)$-approximation for 1-feasible $\varepsilon$-compressed SRMFC-T instances.

Hence, let $(G, r, B)$ be a 1-feasible $\varepsilon$-compressed SRMFC-T instance. Applying Theorem 2.7, we can efficiently compute polynomially many partitions of the leaves $\Gamma$ such that for one partition $(\Gamma_{\mathrm{bot}}, \Gamma_{\mathrm{top}})$ both the bottom instance and the top instance are $(1 + 7\varepsilon)$-feasible. For each of these partitions, we can check existence of a $(1 + 7\varepsilon)$-feasible fractional solution for the top instance, and, if it exists, we can compute a $(1 + 8\varepsilon)$-feasible solution using Lemma 2.5. Additionally, using dynamic programming (Theorem 2.6), we can compute an optimal solution for the bottom instance. Combining both approaches and iterating over all partitions and returning the best solution found this way, we obtain a $(1 + 8\varepsilon)$-feasible solution for the original instance, finishing the proof. □